\documentclass[11pt,letterpaper]{article}
\usepackage{jheppub}
\usepackage{style}
\usepackage{verbatim}

\begin{document}

\title{Continuous Measurement of Quantum Axion Dark Matter}

\author[1,2,3,4]{Yunjia Bao,}
\author[1,2,3,4]{Dhong Yeon Cheong,}
\author[5,6]{Nicholas L.~Rodd,}
\author[5,6]{Joey Takach,}
\author[1,2,3,4]{Lian-Tao Wang,}
\author[5,6]{and Kevin Zhou}

\affiliation[1]{\footnotesize Enrico Fermi Institute, The University of Chicago, Chicago, IL 60637, USA}
\affiliation[2]{\footnotesize Department of Physics, The University of Chicago, Chicago, IL 60637, USA}
\affiliation[3]{\footnotesize Leinweber Institute for Theoretical Physics, The University of Chicago, Chicago,
IL 60637, USA}
\affiliation[4]{\footnotesize Kavli Institute for Cosmological Physics, The University of Chicago, Chicago,
IL 60637, USA}
\affiliation[5]{\footnotesize Theory Group, Lawrence Berkeley National Laboratory, Berkeley, CA 94720, USA}
\affiliation[6]{\footnotesize Leinweber Institute for Theoretical Physics, University of California, Berkeley, CA 94720, USA}

\emailAdd{yunjia.bao@uchicago.edu}
\emailAdd{dycheong@uchicago.edu}
\emailAdd{nrodd@lbl.gov}
\emailAdd{joseph\_takach@berkeley.edu}
\emailAdd{liantaow@uchicago.edu}
\emailAdd{kzhou7@berkeley.edu}

\abstract{
We provide a fully quantum-mechanical description of the continuous measurement of wavelike dark matter.
Our focus is cavity haloscopes, although our results inform a broad range of axion searches.
We extend the input-output formalism by treating the axion as a quantum field, finding how its quantum state is imprinted onto higher-order moments of observables of the outgoing transmission-line modes.
We further develop a novel measurement-record formalism, which computes the joint probability distribution of all measurement outcomes and is well suited to studying the axion in the time domain.
With it, we compute the exact axion-induced number distribution for the first time, demonstrating that it is not Poisson.
We argue that virialized axion dark matter is expected to be in a Gaussian quantum state, and show that number or quadrature measurements can distinguish this from the predictions of a coherent state or even a Bose--Einstein condensate.
Finally, we show that intrinsically quantum effects, such as those from Fock or squeezed axion states, are far too weak to detect.
} 

\maketitle
\setcounter{page}{2}

\vspace{-6mm}
\paragraph{Conventions.} We adopt similar conventions to our previous work~\cite{Bao:2025nsd,suppression_paper}.
We use natural units, $\hbar = c = k_B = 1$, and a mostly-negative spacetime metric.
Integrals and delta functions over coherent states are written as $d\alpha = d \, \text{Re}(\alpha) \, d \, \text{Im}(\alpha)$ and $\delta(\alpha) = \delta(\text{Re}(\alpha)) \, \delta(\text{Im}(\alpha))$.
A ``Gaussian state'' denotes either the thermal Gaussian in Eq.~\eqref{eq:gaussian_state_def} or its multimode extension in Eq.~\eqref{eq:VGH}, while ``generalized Gaussian'' states are discussed in App.~\ref{app:QuantumStates}.
We use $p$ for the probability (density) of measurement outcomes and reserve $P$ for the Glauber--Sudarshan $P$-function.
We treat all fields as quantum, but generally suppress operator hats.

\section{Introduction}
\label{sec:intro}
%!TEX root = main.tex

The search for axion dark matter (DM) would once have sounded like science fiction: macroscopic devices operating beyond the quantum limit to search for an invisible wave that fills the universe.
Instruments can operate down to the standard quantum limit, and the residual quantum noise floor can be beaten down with squeezing~\cite{Malnou:2018dxn,HAYSTAC:2020kwv}, number measurements~\cite{Lamoreaux:2013koa,Dixit:2020ymh,Braggio:2024xed}, and other techniques~\cite{Fang:2024ple,Chou:2023hcc}.
The axion has yet to appear, but with its mass unknown, DM could hide at any frequency, and its discovery may require even larger next-generation instruments.

Despite such high stakes, the theoretical description of the axion DM signal has not been developed with similar sophistication.
The conventional starting point, as described in influential reviews~\cite{Marsh:2015xka,Irastorza:2018dyq,Hui:2021tkt}, is that ultralight axion DM can be treated as a classical wave because it has large occupancy.
But large occupation does not ensure classicality; for instance, a Fock state does not become more classical as its occupancy increases.
Fundamentally, the axion is quantum mechanical, and it can only be treated as classical if its measured effects match that of a classical-wave model.
Conversely, one could consider searching for \textit{intrinsically} quantum effects, i.e.~those which cannot arise in any classical model for the axion.

In Refs.~\cite{Bao:2025nsd,suppression_paper}, we established two severe obstructions to observing intrinsically quantum effects.
The first is that realistic detectors coarse-grain the axion signal, potentially erasing these effects, and the second is that these effects are suppressed by the axion's weak coupling, but \textit{not} enhanced by its large occupation.
This justifies the classical approximation but does not imply the axion must be thought of as a classical wave.\footnote{As emphasized in Ref.~\cite{suppression_paper}, these conclusions hold more broadly for weakly coupled waves including gravitational waves; see also Refs.~\cite{Carney:2023nzz,Carney:2024dsj,Miyauchi:2026cnt}.}
One is always free to choose the basis most convenient for the calculation; with Fock states, the particle description is forefront, whilst with coherent states, the wave picture is more apparent.
For axions, the coherent-state basis is especially convenient, as one can then rigorously map the predictions of the full quantum theory to a classical-wave description that can be derived from first principles rather than heuristics.

In those works we considered a simplified discrete measurement model, in which one repeatedly prepares a given detector state, allows it to evolve for some time while interacting with the axion, and then projectively measures the final detector state.
However, the vast majority of existing axion experiments operate using continuous measurement.
For example, in a typical cavity haloscope~\cite{Sikivie:1983ip} -- realized by ADMX~\cite{ADMX:2001dbg,ADMX:2003rdr}, HAYSTAC~\cite{Brubaker:2017ohw}, and CAPP/IBS-DMAG~\cite{CAPP:2020utb} -- the cavity is not directly measured.
Instead, the cavity is coupled to a transmission line or waveguide, and the axion signal is continuously imprinted on the outgoing transmission-line modes.
The HAYSTAC experiment evades the standard quantum limit by injecting a squeezed state into the transmission line~\cite{Malnou:2018dxn,HAYSTAC:2020kwv,HAYSTAC:2023cam,HAYSTAC:2024jch}.
Other efforts seek to avoid vacuum quadrature noise entirely by monitoring the output with microwave photon counters~\cite{Lamoreaux:2013koa,Lescanne:2020awk,Balembois:2023bvs,Braggio:2024xed,Pankratov:2022fnh,Pankratov:2024kdv,Pankratov:2025cby,Kuo:2024duq}.
Discrete projective or single-shot measurements are used only for a small subset of existing axion experiments, and there is no existing framework for continuous measurement with axions in an arbitrary quantum state.

\begin{figure}[t]
\centering
\includegraphics[width=1\linewidth]{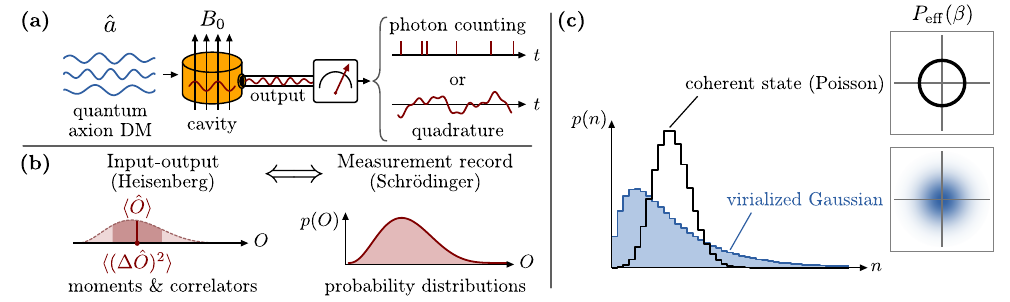}
\vspace{-0.7cm}
\caption{Overview of this work.
\textbf{(a)} The axion-induced signal in a haloscope is continuously monitored at the end of a transmission line by a photon counting or quadrature measurement.
Our formalism can track an \emph{arbitrary} axion quantum state through the detection pipeline to its impact on either observable.
\textbf{(b)} The setup can be equivalently described using the standard input-output formalism or the measurement record formalism.
The former involves operators in Heisenberg picture, naturally yielding moments and correlators, whilst the latter tracks states in Schrödinger picture, yielding the joint probability distribution of the measurement record.
\textbf{(c)} As an example, we derive new results for dark matter number statistics.
A (phase-averaged) coherent state, whose Glauber--Sudarshan $P$-function is shown at right, yields Poisson statistics.
But a more realistic model of the quantum state of virialized DM is a Gaussian, which also features amplitude fluctuations.
Its number statistics considerably deviate from Poisson and approach a negative binomial distribution in the long-integration-time limit.
}
\label{fig:overview}
\end{figure}

In this work, we construct an end-to-end quantum-mechanical description of axion detection in continuously monitored experiments, flexible enough to trace the dependence on the DM quantum state through to a variety of realistic observables.
A schematic summary of our approach is provided in Fig.~\ref{fig:overview}.

We begin in Sec.~\ref{sec:review} by reviewing relevant results from Refs.~\cite{Bao:2025nsd,suppression_paper}, such as the notion of effective modes, which make precise the idea that a detector only couples to certain coarse-grained combinations of the fundamental axion plane-wave modes.
We also discuss what we believe is the most likely candidate for the quantum state of axion DM, which we refer to as the virialized Gaussian hypothesis, and describe alternative hypotheses which lead to distinct experimental predictions.

In Sec.~\ref{sec:inout}, we construct a model of a cavity haloscope where the entire detection chain -- axion, cavity, transmission line -- is treated quantum mechanically.
This treatment builds on the well-established input-output formalism~\cite{Collett:1984ulf,Gardiner:1985uof}, which phrases the problem in terms of how ``inputs'' sent into the transmission line map to ``outputs'' propagating out; the state of the cavity is removed using the Heisenberg equations of motion.
We first review input-output theory for a particle-theory audience, highlighting relevant subtleties, such as the use of negative-frequency modes.
Our primary addition is that we treat the axion as quantum and trace the impact of its quantum state on transmission-line observables, relevant for quadrature and number measurements.
Higher moments of these observables encode the fluctuations of the DM state.
To illustrate this, we show that the variance of number measurements can be used to distinguish whether the axion is in a virialized state or instead behaves as we might expect for a coherent state.
(Distinguishing these is possible since the axion does not exhibit nonclassical behavior in either case.)
The virialized state statistics are a distinctive fingerprint of axion DM that could be tested by an experiment shortly after discovery.

In Sec.~\ref{sec:measurement} we introduce the \textit{measurement-record formalism}, which is to the best of our knowledge a new approach for the problem.
The underlying physics is equivalent to that of Sec.~\ref{sec:inout}, which is built on the Heisenberg picture.
Here we move to the Schr\"odinger picture, tracking the evolution of the entire system in a joint coherent-state basis; more precisely, we work in terms of the joint Glauber--Sudarshan $P$-function~\cite{Glauber:1963tx,Sudarshan:1963ts}, reviewed in Sec.~\ref{sec:review}.
These calculations are tractable due to the close connection between coherent states and classical waves, which renders the calculations no more difficult than a fully classical description of the problem.
Arbitrary measurements are accounted for by repeatedly conditioning the state (i.e.~through projection) on the outcomes of said measurements.

The measurement-record formalism is inspired by classic models of continuous position measurement~\cite{PhysRevA.36.5543,barchielli1982model} and more modern approaches, such as stochastic master equations and Fokker--Planck equations, that model the non-unitary evolution of a state subject to continuous measurements~\cite{Jacobs2014QuantumMeasurement,carmichael2013statistical}.
However, rather than focus on the state of the system, we compute the probability distribution of the measurement record, the set of \textit{all} measurement outcomes recorded by the experiment.
This distribution contains all information extractable from a series of measurements, as opposed to only moments, which are more naturally computed in conventional input-output theory.
We exemplify the power of this approach by computing the exact number and quadrature statistics expected of axion DM.
For quadrature, we recover known results~\cite{Foster:2017hbq,Derevianko:2016vpm,Centers:2019dyn,Kim:2021yyo,Dror:2022xpi,Cheong:2024ose}.
The number distribution is much less well studied, and has been heuristically treated as Poisson~\cite{Lamoreaux:2013koa,Braggio:2024xed}.
However, we show that the number distribution is never Poisson; instead, depending on the length of time the number measurements are collected over, it interpolates from a geometric to a negative binomial distribution.

Using our two approaches, we establish that the no-go results of Refs.~\cite{Bao:2025nsd,suppression_paper} generalize to continuous measurement.
Whilst this conclusion implies there should always be an equivalent classical description one can use for the axion, it does not imply a classical heuristic is always more useful.
Indeed, the quantum description is often no harder than the classical, and it has the enormous benefit of being the fundamental description.
Both continuous-measurement formalisms can be useful in their own contexts; input-output can be more convenient when describing frequency-domain observables, whereas the measurement record allows for a clear description in the time domain.
The formalism also enables us to straightforwardly study properties of the axion in the time domain and compare the behavior of virialized DM with that of single-mode DM, which is a model for a Bose--Einstein condensate~\cite{Semikoz:1995rd,Sikivie:2009qn,Guth:2014hsa}.
Broadly, we establish a framework that can be used for precise continuous-measurement computations to replace heuristics.
Finally, in Sec.~\ref{sec:discussion} we conclude with a discussion of the implications of our results for the observability of nonclassical effects, a comparison between our approaches and other continuous-measurement schemes, and a list of open directions raised by our work.

\section{Nonclassicality and the State of Dark Matter}
\label{sec:review}
%!TEX root = main.tex

Here we establish fundamental results used throughout this work.
In Sec.~\ref{subsec:review}, we review the formalism and results from our companion works, Refs.~\cite{Bao:2025nsd,suppression_paper}, needed for our study of continuous measurement.
In Sec.~\ref{subsec:DMQstate}, we discuss the DM quantum state in detail, and explicitly define the associated assumptions we adopt throughout.
Lastly, in Sec.~\ref{subsec:twopowers} we comment on the different notions of power one can define for quantum measurement.

%%%%%%%%%%%%%%%%%%%%%%%%%%%%%%%%%
\subsection{Review: Effective Modes and the Quantum Central Limit Theorem}
\label{subsec:review}
%%%%%%%%%%%%%%%%%%%%%%%%%%%%%%%%%

This subsection reviews the relevant results in Refs.~\cite{Bao:2025nsd,suppression_paper}, including the Glauber--Sudarshan $P$-function; the reduction of the axion-photon interaction in a cavity haloscope to a two-mode problem by the introduction of an \textit{effective mode}; the exact solution of that two-mode problem; and the quantum central limit theorem (CLT), which underpins a general expectation that the axion state is Gaussian.
Throughout, we simply state results; see Ref.~\cite{suppression_paper} for detailed derivations and further discussion.

%%%%%%%%%%%%%%%%%%%%%%%%%%%%%%%%%
\paragraph{The $P$-Function.}
%%%%%%%%%%%%%%%%%%%%%%%%%%%%%%%%%
%
The Glauber--Sudarshan $P$-function~\cite{Glauber:1963tx,Sudarshan:1963ts} is both a powerful computational tool for solving the dynamics of general quantum systems -- as we exploit to the fullest in Sec.~\ref{sec:measurement} -- and a rigorous indicator of which states are \textit{intrinsically} quantum.

The $P$-function is defined in terms of coherent states, which are eigenstates of a bosonic annihilation operator, satisfying $a \ket{\alpha} = \alpha \ket{\alpha}$ and labeled by a complex eigenvalue $\alpha$.
The coherent states are pure states that form an overcomplete basis of the single-mode bosonic Hilbert space.
The states are not orthogonal and have overlap $|\la \alpha | \beta \ra|^2 = e^{-|\alpha - \beta|^2}$.
If we perform number measurements of a coherent state $\ket{\alpha}$ with the number operator $N = a^\dag a$, the results will be described by a Poisson distribution with mean $|\alpha|^2$.
If we instead measure dimensionless quadratures of a coherent state, defined by
\be
X = \frac{1}{\sqrt{2}} \big(a + a^\dag\big),\qquad 
Y = \frac{-i}{\sqrt{2}} \big(a - a^\dag\big),
\ee
then the results will be Gaussian distributed with means $(\la X \ra, \la Y \ra) = \sqrt{2}\,(\text{Re}(\alpha), \text{Im}(\alpha))$ and variances $\var(X) = \var(Y) = 1/2$.

The density matrix of a bosonic mode can always be written in the form
\be \label{eq:P_def}
\rho = \int d\alpha \, P(\alpha) \ket{\alpha}\bra{\alpha},
\ee
where the integral is taken over the full complex plane.
This expression defines the $P$-function.
From this, we can compute expectation values of any operator $\mathcal{O}$ by
\be
\la \mathcal{O} \ra = \tr{\rho \, \mathcal{O}} = \int d\alpha \, P(\alpha) \bra{\alpha}\mathcal{O} \ket{\alpha}.
\ee
In the case where $\mathcal{O}$ is normally ordered, we have the further simplification $\bra{\alpha}\mathcal{O}(a^\dag, a) \ket{\alpha} = \mathcal{O}(\alpha^*,\alpha)$, a result known as the optical equivalence theorem~\cite{mandel1995optical}.
In addition, for any $\mathcal{O}$ with nondegenerate eigenstates $\ket{\mathcal{O}}$, we can compute the probability of observing a value $\mathcal{O}$ as
\be \label{eq:P_probability}
p(\mathcal{O}) = \tr\!\big\{\rho \ketbra{\mathcal{O}}{\mathcal{O}} \!\big\} = \int d\alpha \, P(\alpha) |\!\braket{\mathcal{O}}{\alpha}\!|^2.
\ee

A coherent state $\ket{\alpha_0}$ is the quantum analog of a classical wave of known amplitude and phase set by $\alpha_0$.
It has the simple $P$-function $P(\alpha) = \delta(\alpha-\alpha_0)$.
The quantum state for a system where we know the amplitude but not the phase of the wave is the mixed state
\be \label{eq:phaseMixtureP}
P(\alpha) = \frac{1}{\pi} \, \delta \Big(|\alpha|^2 - |\alpha_0|^2 \Big).
\ee
In both of these states, the mean occupancy is $\la N \ra = |\alpha_0|^2$.
Finally, a state where both the amplitude and phase are uncertain is the (thermal) Gaussian state with mean occupancy $N$,\footnote{To avoid a potential confusion, although the Gaussian form in Eq.~\eqref{eq:gaussian_state_def} is sometimes described as a ``thermal'' Gaussian, a system need not be thermalized in order to have modes described by this $P$-function.
In particular, we \textit{do not} assume in this work that the local DM is thermalized.}
\be \label{eq:gaussian_state_def}
P(\alpha) = \frac{1}{\pi N} \, \exp{-|\alpha|^2/N}.
\ee
Further properties and examples are discussed in App.~\ref{app:QuantumStates}.

As shown below in Eq.~\eqref{eq:joint_coherent}, a coherent state acts on a weakly coupled detector like a definite classical field value.
Since the above examples all have $P(\alpha) \geq 0$, they act like ensembles of classical field values with probability density $P(\alpha)$.
However, other quantum states can have $P$-functions that achieve negative values, or are more singular than a delta function.
Examples include Fock states, squeezed states, or ``cat state'' superpositions $\propto \ket{\alpha} + \ket{-\alpha}$.
These states are intrinsically quantum: we cannot reproduce all their predictions with a probabilistic ensemble of classical waves.

As the $P$-function of nonclassical states can be highly singular, it is often easier to work with the characteristic function.
In fact, we can define this for a family of quasiprobability distributions, of which the $P$-function is just one example,
\be \label{eq:s_ordered}
\tilde{P}^{(s)} (\lambda) = e^{s|\lambda|^2/2} \tr\!\Big[ \rho \, e^{\lambda a^{\dag}-\lambda^* a} \Big] = \int d\alpha\, P^{(s)}(\alpha) e^{\lambda \alpha^*-\lambda^* \alpha},
\ee
which is defined for complex $\lambda$ and $s \in [-1,1]$.
Inverting the result with $s=1$, one finds that $P^{(1)}(\alpha)$ corresponds exactly to the Glauber--Sudarshan function of Eq.~\eqref{eq:P_def}.
Likewise, $s=0$ and $s=-1$ yield the Wigner function and Husimi $Q$-function.

%%%%%%%%%%%%%%%%%%%%%%%%%%%%%%%%%
\paragraph{Effective Modes.}
%%%%%%%%%%%%%%%%%%%%%%%%%%%%%%%%%
%
We now review the effective-mode formalism of Refs.~\cite{Bao:2025nsd,suppression_paper}, which greatly simplifies the description of how quantum waves couple to a quantum detector.
Here we focus on axion DM interacting with a cavity haloscope through $\Lag \supset - \ga \phi F_{\mu\nu} \tilde{F}^{\mu\nu}/4$.

A single cavity mode $c$ couples to an approximate continuum of axion plane-wave modes $a_\bk$ with 3-momentum $\bk$, weighted by their overlap with the cavity.
For a cavity of volume $V_c$ in a background magnetic field $B_0 \hat{\vb{z}}$, this overlap is quantified by a dimensionless form factor
\be \label{eq:formfactor}
C(\bk) = \frac{\sqrt{2}\, \omega_c}{\sqrt{\omega_\bk\,V_c}}
\int_{V_c} \hspace{-0.1cm} d^3 \bx\, u(\bx) e^{i \bk \cdot \bx},
\ee
where $\omega_c$ and $u(\bx)$ are the angular frequency and spatial profile of the cavity mode, and $\omega_\bk^2 = m_a^2 + |\bk|^2$.
For a cylindrical cavity, $u(\bx)$ involves Bessel functions, with the explicit form given in Ref.~\cite{suppression_paper}.
Using these weightings, one can define the effective mode as the linear combination of plane-wave axion modes the detector couples to,
\be \label{eq:effective_mode_def}
a_\eff = \frac{1}{2} \sqrt{\frac{V_c}{\Omega}} \int\! \frac{d^3\bk}{(2\pi)^3} \, C(\bk)\,a_\bk, \qquad 
\Omega = \frac{V_c}{4}\int \!\frac{d^3\bk}{(2\pi)^3} \, |C(\bk)|^2.
\ee
The dimensionless normalizing factor $\Omega$, typically $\mathcal{O}(1)$, enforces the canonical commutation relation $[a_\eff, a_\eff^\dag] = 1$.
The axion-photon interaction Hamiltonian then takes the simple form
\be \label{eq:Hint_simplified}
H_\text{int} = i g \big( c^\dag a_\eff - c\, a_\eff^\dag \big), 
\qquad 
g = \ga B_0 \sqrt{\Omega}.
\ee
Here we used the rotating-wave approximation (RWA), i.e.~we discarded counter-rotating terms $c\,a_\eff$ and $c^\dag a_\eff^\dag$, which do not conserve the total excitation number; these terms are associated with effects that oscillate as $e^{\pm i(\omega_c + m_a)t}$ and average to zero over the relevant measurement times, unlike the resonantly enhanced terms retained above, which require $\omega_c \simeq m_a$.
For a resonant experiment with quality factor $Q_c$, the RWA is a good approximation, up to $\mathcal{O}(1/Q_c)$ corrections to the signal amplitude.
For broadband experiments, which we do not consider in detail, the RWA is not appropriate, and counter-rotating terms should be included.

All results above are specialized to a single cavity mode, e.g.~the TM$_{010}$ mode.
Analogous effective modes $a_\ell$ can be constructed for every cavity mode indexed by $\ell$, although they are not independent in the sense that $[a_{\ell},a^\dag_{\ell'}]$ can be nonzero for $\ell \neq \ell'$.
A detailed exploration of effective modes is given in Ref.~\cite{suppression_paper}, but here we recall two important features.
First, the occupation number of the effective mode for axion DM is
\be \label{eq:Neff-DM}
N_\eff = \la a^{\dag}_\eff a_\eff \ra \simeq \left(\frac{|C(\vb{0})|^2}{4 \Omega} \right) \bar{n} V_c,
\ee
where $\bar{n}$ is the local DM number density.
For low-lying cavity modes, the expression in parentheses is order one, so that $N_\eff$ is approximately the expected number of axions within the detector.
For the axion masses targeted by ADMX, we have $N_\eff \sim 10^{19}$, and more generally, for ultralight DM experiments we always expect $N_\eff \gg 1$.

Second, Eq.~\eqref{eq:effective_mode_def} shows that $a_\eff$ is assembled from a set of plane waves spanning a finite range of frequencies, so it does not evolve in time as a single free mode.
In the interaction picture, each fundamental mode in the sum carries a phase $e^{-i \omega_\bk t}$, and so the effective mode does not simply acquire an overall phase, but evolves into a different combination.
The dominantly occupied DM modes occupy a frequency band $\sim m_a v^2$ with $v \sim 10^{-3}$ the characteristic DM velocity, which implies the effective modes drift on a characteristic timescale $\tau_\DM \sim 1/(m_a v^2)$.
(This is essentially the DM coherence time; a rigorous definition of the coherence time, used throughout this work, is discussed in App.~\ref{app:coherence_time} and Ref.~\cite{Cheong:2024ose}.)

As such, Eq.~\eqref{eq:effective_mode_def} specifies an approximately fixed mode only over timescales $t \ll \tau_\DM$.
A common intuition is that measurements of DM remain correlated for $t \lesssim \tau_\DM$, but become roughly independent for $t \gtrsim \tau_\DM$, where one essentially probes an independent realization of the effective mode.
The frequency analog of this statement is that if we resolve the DM below its intrinsic bandwidth of $1/\tau_\DM$, we expect the separate frequency modes to be statistically independent.
This is an assumption and whether it holds depends on the underlying DM $P$-function, as we discuss further in Sec.~\ref{subsec:DMQstate}.
We adopt this assumption in Sec.~\ref{sec:inout}, although we interrogate it in far greater detail in Sec.~\ref{sec:measurement}, demonstrating that it need not always hold.

%%%%%%%%%%%%%%%%%%%%%%%%%%%%%%%%%
\paragraph{Cavity State Evolution.}
%%%%%%%%%%%%%%%%%%%%%%%%%%%%%%%%%
%
The Hamiltonian of Eq.~\eqref{eq:Hint_simplified} describes two linearly coupled oscillators, exchanging energy at a rate set by $g$, and can be solved exactly.
The interaction conserves the total number of axion plus cavity quanta while rotating the two modes into one another and therefore maps coherent states to coherent states; in the interaction picture,
\be \label{eq:joint_coherent}
U(t) \ket{\alpha, \gamma} = \big\vert \alpha \cos gt - \gamma \sin gt,\, \gamma \cos gt + \alpha \sin gt \big\rangle,
\ee
where $\alpha$ and $\gamma$ label the axion and cavity modes, respectively.
As we emphasized in Ref.~\cite{suppression_paper}, to leading order in $gt$, the evolution of the cavity state is exactly what one would find in a fully \textit{classical} treatment of the axion with the same interaction in Eq.~\eqref{eq:Hint_simplified}.

Equation~\eqref{eq:joint_coherent} was useful for treating discrete measurements in Ref.~\cite{suppression_paper}, and we greatly generalize it in Sec.~\ref{sec:measurement} to track the cavity state through continuous measurement.
It shows that as time progresses, the axion state is slowly rotated into the cavity state.
If the process could continue until $t \sim 1/g$, then the DM would deposit almost all its energy into the cavity; however, the validity of the simple model breaks down well before this due to, for example, cavity dissipation.
Instead, in the steady state, we show that for continuous measurements, the expected fraction of axions that convert to photons is given by the efficiency
\be \label{eq:eta_value}
\eta = (2g t_m)^2 \sim 10^{-22} \left( \frac{\ga}{10^{-15} \, \mathrm{GeV}^{-1}} \frac{B_0}{10 \, \mathrm{T}^{\vrule height 4.5pt width 0pt}} \frac{Q_c}{10^5} \frac{10^{-5} \, \mathrm{eV}}{m_a} \right)^2\!,
\ee
where $t_m = Q_c / \omega_c$ is the cavity's characteristic decay time.\footnote{Here $Q_c$ is the \textit{loaded} quality factor; this is defined carefully in Sec.~\ref{sec:inout}.}
This is very small for realistic parameter values, but axion DM remains detectable due to the correspondingly large value of $N_\eff$.
A key finding of Ref.~\cite{suppression_paper} was that for discrete measurements, intrinsically quantum effects are penalized by at least one extra factor of $\eta$, \textit{not} compensated by a factor of $N_\eff$.

Returning to the Hamiltonian, we assume the axion and cavity systems are initially independent, $P(\alpha,\gamma) = P_a(\alpha) P_c(\gamma)$.
Then, using Eq.~\eqref{eq:joint_coherent}, the cavity $P$-function evolves to
\be \label{eq:approx_cavity_P}
P_c(\gamma, t) \simeq \int d\alpha \, \frac{P_a(\alpha / g t)}{(g t)^2} \frac{P_c((\gamma - \alpha) / \sqrt{1-(g t)^2})}{1 - (g t)^2},
\ee
which is a weighted convolution of the two initial $P$-functions.
This result played a central role in Refs.~\cite{Bao:2025nsd,suppression_paper}, where we tracked the cavity state,\footnote{In those works, we treated projective measurement by ignoring dissipation and imagining that the projective measurements were spaced by $t_m$.
This is a rough treatment, only accurate at the $\mathcal{O}(1)$ level (sufficient for the parametric statements made in those works), and instead yields a natural efficiency $\eta = (g t_m)^2$.} but in continuous measurement one usually focuses on the electromagnetic radiation leaving the cavity, down a transmission line toward a measurement device.
In this view, the cavity is solely an intermediary between the DM signal and our measured output, and its state can be formally eliminated by working with a steady-state solution.
(The conditional evolution of the cavity under continuous measurement could be studied, but it is not the focus of our work.)

%%%%%%%%%%%%%%%%%%%%%%%%%%%%%%%%%
\paragraph{The Quantum Central Limit Theorem.}
%%%%%%%%%%%%%%%%%%%%%%%%%%%%%%%%%
%
The quantum CLT was one of the core obstructions to observing nonclassical effects outlined in Refs.~\cite{Bao:2025nsd,suppression_paper}.
Since the detector couples to the DM field only through $a_\eff$, all of the axion's observable consequences are encoded in the effective mode's $P$-function, obtained by tracing out the remaining DM modes,
\be \label{eq:Peff}
P_\eff(\beta) = \int d\bm{\alpha} \, P_a(\bm{\alpha}) \, \delta\big(\beta - \alpha_\eff(\bm{\alpha}) \big).
\ee
Here $\bm{\alpha} = \{\alpha_1, \alpha_2, \ldots\}$ enumerates the fundamental axion modes with joint $P$-function $P_a(\bm{\alpha})$, and $\alpha_\eff(\bm{\alpha})$ is the effective-mode eigenvalue, $a_\eff \ket{\bm{\alpha}} = \alpha_\eff(\bm{\alpha}) \ket{\bm{\alpha}}$.
Note that in general $P_\eff(\beta)$ can evolve nontrivially over timescales $t \gtrsim \tau_\DM$; we examine this issue in Sec.~\ref{sec:measurement}.

If we assume that the $P$-function factorizes over the axion modes, 
\be \label{eq:Pa}
P_a(\bm{\alpha}) = \prod_\bk P_\bk(\alpha_\bk),
\ee
then Eq.~\eqref{eq:Peff} becomes a weighted convolution of the individual mode $P$-functions.
By similar logic to the classical CLT, when many modes contribute, the convolution generically drives $P_\eff$ toward a generalized Gaussian state, as defined in App.~\ref{app:QuantumStates}, even if the fundamental modes are in highly nonclassical states.
This is exactly the quantum CLT.
If the generalized Gaussian is stationary, then the $P$-function of the effective mode takes the thermal Gaussian form in Eq.~\eqref{eq:gaussian_state_def}, which satisfies $P \geq 0$, implying that any intrinsically quantum information has been washed out.
Conditions for the quantum CLT to apply, the CLT's rate of convergence, and ways it can be evaded are discussed in Ref.~\cite{suppression_paper}.

%%%%%%%%%%%%%%%%%%%%%%%%%%%%%%%%%
\subsection{The Quantum State of Dark Matter}
\label{subsec:DMQstate}
%%%%%%%%%%%%%%%%%%%%%%%%%%%%%%%%%

In this work we develop the formalism to compute the output of a continuous-measurement experiment for an \textit{arbitrary} DM state.
In particular, in Sec.~\ref{sec:measurement} we show how the probability of the observed experimental data can be reduced to an integral over a general $P_a(\bm{\alpha})$.
Yet in order to show the formalism in action, we need to specify an explicit $P_a$.
This leads to a natural question: what is the quantum state of DM?
To the best of our knowledge, this question is currently unresolved.
In this subsection we speculate on the potential identity of the DM state.
In doing so, we introduce the various examples used throughout this work.

%%%%%%%%%%%%%%%%%%%%%%%%%%%%%%%%%
\paragraph{Homogeneity and Stationarity.}
%%%%%%%%%%%%%%%%%%%%%%%%%%%%%%%%%
%
Virialization drives DM toward an attractor distribution that has several of the properties of an equilibrium state; see, for instance, the classic treatment in Ref.~\cite{Lynden-Bell:1966zjv}.
Within the galactic frame, DM could have properties that slowly vary over galactic time and distance scales.
However, one might generically expect that over terrestrial times and distances the DM's statistical properties are approximately invariant.
Technically, this requires the DM correlation functions to be independent of any absolute position or time, implying DM is both homogeneous and stationary.
These assumptions considerably simplify the description of DM and we invoke them often in this work.
For instance, at the level of a single mode, stationarity requires $P(\alpha) = P(|\alpha|)$, which is satisfied by the examples in Eqs.~\eqref{eq:phaseMixtureP} and \eqref{eq:gaussian_state_def}.

DM need not satisfy either assumption.
The existence of localized DM structures such as miniclusters~\cite{Kolb:1993zz,Eggemeier:2022hqa} or the resolved structure of streams~\cite{OHare:2025jpr} would violate homogeneity.
Over longer measurement timescales in the frame of our experiments, daily and annual modulations introduce a violation of stationarity.
Further, simple forms of $P_a$, like that of a pure coherent state with a fixed amplitude and phase, are neither stationary nor homogeneous.

%%%%%%%%%%%%%%%%%%%%%%%%%%%%%%%%%
\paragraph{Virialized Gaussian Hypothesis.}
%%%%%%%%%%%%%%%%%%%%%%%%%%%%%%%%%
%
The existence of the quantum CLT already suggests that a generalized Gaussian form for $P_a$ has a privileged position amongst all possible quantum states.
Similar to the classical CLT, Gaussian states naturally emerge when an observation cannot resolve the scale on which statistically independent contributions occur.
If our detector coarse-grains over $\mathcal{N} \gg 1$ such contributions, the detected mode is rapidly driven toward a Gaussian as $\mathcal{N}$ increases.
Applying this mindset to DM, where there is an enormous mismatch between the galactic scales over which it is defined and the terrestrial scales where we search for it, we are motivated to introduce a \textit{virialized Gaussian hypothesis},
\be \label{eq:VGH}
P_a(\bm{\alpha}) = \prod_j \frac{1}{\pi N_j} e^{-|\alpha_j|^2/N_j}.
\ee
Here, $j$ indexes coarse-grained modes each with mean occupancy $N_j$, as we discuss below.

In Refs.~\cite{Bao:2025nsd,suppression_paper} we used the quantum CLT together with stationarity to advocate for a baseline expectation that effective DM modes exist in the Gaussian state of Eq.~\eqref{eq:gaussian_state_def}, with a mean occupancy set by $N_\eff$ in Eq.~\eqref{eq:Neff-DM}.
For simplicity, we assumed that for each projective measurement one obtained an independent realization of the Gaussian state.
This was reasonable since the projective measurements could be spaced arbitrarily far apart in time, but for continuous measurement, one must track correlations and their rate of decay.
Further, we can resolve the DM into distinct frequency bins, and the behavior within each bin and the correlations between them must be specified.
Here we provide an extended discussion reaffirming the Gaussian hypothesis as relevant for this work.

When discussing the quantum CLT, we noted that if the fundamental DM modes factorize as in Eq.~\eqref{eq:Pa}, then the effective mode the detector resolves is a sum of many independent modes and is therefore driven to a Gaussian.
The presentation there framed the essential factorization at the level of the fundamental axion modes, separated by $|\Delta \bk| \sim \Delta k_\text{fund}$.
This is unnecessary: the conclusion can still hold even if the fundamental DM modes are correlated for $|\Delta \bk| \lesssim \Delta k_\text{corr}$, where $\Delta k_\text{corr}$ could be comparable to or much larger than $\Delta k_\text{fund}$.
The reason for this is there can remain an intermediate scale, $\Delta k_j \gg \Delta k_\text{corr}$, over which we can form coarse-grained modes comprised of many independent modes, which are then in Gaussian states by the quantum CLT.
(In practice, we only need $\Delta k_j \gtrsim \Delta k_\text{corr}$.
As discussed in Ref.~\cite{suppression_paper}, the quantum CLT can be highly accurate even for relatively few modes.)
As long as the detector can only resolve $\Delta k_\text{det} \gtrsim \Delta k_j$, it sees DM in a virialized Gaussian state.

Although we are unaware of a quantitative calculation, virialization and phase mixing during the formation of the galaxy could render the DM modes effectively independent at some scale.
The classic picture of violent relaxation~\cite{Lynden-Bell:1966zjv} is that virialization results from coarse-graining over the strongly mixed phase space that arises early in the galaxy's formation as the DM falls into a rapidly varying gravitational potential.
The result is that the apparently smooth local DM emerges from an enormous number of filaments or streams~\cite{Vogelsberger:2010gd}.
This same process could strongly suppress large-scale correlations in the DM momentum space, yielding a scale $\Delta k_\text{corr}$ at which the DM $P$-function factorizes into independent momentum modes.
The scale $\Delta k_\text{corr}$ could be far larger than $\Delta k_\text{fund}$, but the CLT is still applicable if $\Delta k_\text{corr} \ll \Delta k_j$.
Combined with an assumption of stationarity, we arrive at Eq.~\eqref{eq:VGH}.\footnote{Technically, Eq.~\eqref{eq:VGH} is not stationary.
The modes indexed by $j$ coarse-grain over a finite momentum scale, and therefore evolve over a timescale $\sim\!1/\Delta \omega_j$, with $\Delta \omega_j$ the corresponding frequency bandwidth.
However, since $1/\Delta k_j \gg 1/\Delta k_\text{det}$, with $\Delta k_\text{det}$ the detected bandwidth, we take the evolution on the experimental timescale as negligible and therefore we can effectively treat Eq.~\eqref{eq:VGH} as stationary.}

Going forward, we assume $\Delta k_\text{det} \gg \Delta k_j \gg \Delta k_\text{corr} \geq \Delta k_\text{fund}$, a hierarchy which is possible due to the enormous mismatch of scales between our galaxy and our instruments.
(We return to provide a more quantitative discussion of the interplay between these various scales and the quantum CLT toward the end of Sec.~\ref{sec:measurement}.)
The exact values of $N_j$ that appear in Eq.~\eqref{eq:VGH} are set by the local DM phase-space density at a momentum $\bk_j$.
Axion haloscopes do not resolve the individual $j$ modes, but instead detect the frequency components of $a_\eff$, each of which is a linear combination of the multiple $j$ modes sharing that frequency.
Under the virialized Gaussian hypothesis, these frequency components are also Gaussian (being a sum of independent Gaussians), and as derived in Sec.~\ref{sec:inout} have an occupancy per unit frequency given by $n_\eff[\omega]$, which integrates to $N_\eff$ in Eq.~\eqref{eq:Neff-DM}.

For the remainder of this work we take the model in Eq.~\eqref{eq:VGH} as our default DM hypothesis, and call it the Gaussian state model.
Yet as DM need not follow this distribution, we consider additional models.

%%%%%%%%%%%%%%%%%%%%%%%%%%%%%%%%%
\paragraph{Coherent-State Hypothesis.}
%%%%%%%%%%%%%%%%%%%%%%%%%%%%%%%%%
%
Although we see the Gaussian state in Eq.~\eqref{eq:VGH} as the most likely description of local DM, axion DM is almost always described as a ``coherent field'' rather than a ``Gaussian field.''
This could simply be a result of imprecise language.\footnote{The phrase ``coherent field'' is also often used interchangeably with ``coherently oscillating,'' which indicates that the axion has a small fractional frequency width, $1/(m_a\tau_\DM) \sim v^2 \ll 1$.
Of course, this by itself does not determine the axion state.}
Indeed, statistical analyses of the axion field often use the random-plane-wave model constructed in Ref.~\cite{Foster:2017hbq}, which is the classical analog of Eq.~\eqref{eq:VGH}.
(For further discussion of this approach and the connection to the Gaussian model, see Refs.~\cite{Derevianko:2016vpm,Centers:2019dyn,Kim:2021yyo,Dror:2022xpi,Cheong:2024ose}.)

Nevertheless, as an alternative to the Gaussian model, one could ask whether there is an appropriate axion $P_a$ so that a detector sees something more akin to the phase-randomized coherent state model in Eq.~\eqref{eq:phaseMixtureP}, rather than a Gaussian.
There are significant obstacles to constructing any such $P_a$, as we now show explicitly.
These challenges appear to give further weight to the virialized Gaussian hypothesis.

The most direct path to realizing a coherent-state model for DM would be to take the system in a pure multimode coherent state at the level of the fundamental DM modes,
\be \label{eq:Pa_pure_coherent}
P_a(\bm{\alpha}) = \prod_{\bk} \delta(\alpha_{\bk}-\alpha_{\bk,0})
\ee
for some known constants $\alpha_{\bk,0}$.\footnote{This model, which has a $P$-function exactly in the form of Eq.~\eqref{eq:Pa}, provides a degenerate realization of the quantum CLT.
Any linear combination of pure coherent-state modes generates a mode in another pure coherent state.
As the coherent state is itself a generalized Gaussian state -- see App.~\ref{app:QuantumStates} -- this is then trivially consistent with the quantum CLT.}
This model corresponds to having complete information about the DM state.
It is unclear how one would obtain that information.
Physical mechanisms such as virialization may provide insight into the mean occupancy or energy density of DM as a function of $\bk$.
Even if one could then argue that certain interactions fix the magnitudes $|\alpha_{\bk,0}|$ to the mean expected from virialization, we know of no principle that could deterministically fix the individual phases.

The natural next attempt would be to assume that we have no knowledge of the phases of each $\alpha_{\bk,0}$, and moreover that all of these phases are completely independent.
This corresponds to taking a multimode generalization of the phase-averaged coherent state in Eq.~\eqref{eq:phaseMixtureP},
\be \label{eq:coherent_fundamental}
P_a(\bm{\alpha}) = \prod_\bk \frac{1}{\pi} \, \delta \Big(|\alpha_\bk|^2 - N_\bk \Big).
\ee
Yet this model still fails our requirement: it does not appear as a coherent state to a detector.
Following the virialized Gaussian hypothesis discussion, $\Delta k_\text{fund} \ll \Delta k_\text{det}$, so that the detector cannot resolve the fundamental and independent DM modes.
Accordingly, by the quantum CLT, to a detector this state is indistinguishable from the Gaussian state in Eq.~\eqref{eq:VGH}.\footnote{This argument is similar to the classical construction given for the Gaussian DM distribution in Ref.~\cite{Foster:2017hbq}.}

As a final attempt, we can simply posit -- without having any physical mechanism to achieve this in mind -- that at a momentum scale $\Delta k_j \gg \Delta k_\text{fund}$ we can write
\be \label{eq:coherent_j}
P_a(\bm{\alpha}) = \prod_j \frac{1}{\pi} \, \delta \Big(|\alpha_j|^2 - N_j \Big),
\ee
where $N_j$ should be interpreted as in Eq.~\eqref{eq:VGH}.
This model can only potentially achieve the desired behavior when $\Delta k_j \sim \Delta k_\text{det}$.
If $\Delta k_j \ll \Delta k_\text{det}$, the detector resolves multiple independent modes and the $P$-function is driven to Gaussian form by the quantum CLT, just as for Eq.~\eqref{eq:coherent_fundamental}.
If instead $\Delta k_j \gg \Delta k_\text{det}$, detected modes become strongly correlated, which would clash with our conventional expectation for the DM coherence time.
In particular, the standard intuition is that if we resolve the DM into frequency bins finer than $1/\tau_\DM$, those should be independent, and in this limit Eq.~\eqref{eq:coherent_j} would fail to satisfy this requirement.

Suppose we assume $\Delta k_j \sim \Delta k_\text{det}$, which requires the DM state to coincidentally involve scales that are tied to the DM detector's resolution.
Whilst this appears at best contrived, \textit{even this description fails}.
This is because axion detectors typically resolve the energy of the DM field and not its momentum; the detector thus cannot distinguish between the different modes with similar values of $|\bk|$.
Instead, the experiment resolves the combination of all modes with similar $|\bk|$, which again drives the observed mode away from a coherent state and toward a Gaussian, even if we tune $\Delta k_j \sim \Delta k_\text{det}$.\footnote{Interferometry techniques can give DM experiments sensitivity to the DM momenta rather than just energy~\cite{Foster:2020fln,Fukuda:2025zcf}.
In principle, this could revive the contrived model in Eq.~\eqref{eq:coherent_j}, although we do not pursue this further.}
To avoid this problem, one could construct a DM state that fixes the relative phase of all modes within each shell of fixed $|\bk|$.
However, this is even more contrived, as it gives the detector a privileged position; it is inconsistent with spatial homogeneity, as the relative phases between modes propagating in different directions change under spatial translations.

We conclude that it is not straightforward to define DM, at the level of $P_a$, such that the detected mode behaves as a coherent state with the expected DM coherence time.
Notwithstanding, in order to have a benchmark to compare the behavior of the virialized Gaussian model against, we define the \textit{coherent-state hypothesis} to be a model which behaves as we might naively expect a coherent state would from the perspective of the detector.
Given the challenges identified, we specify this hypothesis with respect to the modes a detector resolves (in both frequency and time), rather than at the level of the DM $P_a$.
We label the eigenvalues of the DM effective modes resolved by the detector as $\beta_k$.
The coherent-state hypothesis then asserts that each mode is drawn independently from a phase-averaged coherent state,
\be \label{eq:CSH}
P_\eff(\beta_k) = \frac{1}{\pi} \, \delta \Big( |\beta_k|^2 - N_k \Big),
\ee
where $N_k$ is the mean occupancy the virialized Gaussian hypothesis assigns to the same mode.
Matching the occupancies ensures the two hypotheses predict the same mean signal and will only deviate in higher moments.
To be explicit, we exclusively invoke this hypothesis when performing number measurements, and Eq.~\eqref{eq:CSH} implies that each measurement provides an independent draw from the Poisson distribution.\footnote{This is exactly how the coherent-state description of DM was invoked in Refs.~\cite{Bao:2025nsd,suppression_paper}.
Those works considered projective measurements and we simply posited that each measurement of number yielded an independent draw from the Poisson distribution for the coherent state model.}
Independent of the above discussion, one might have naively expected that the DM prediction for the counts is Poisson.
This is not what is predicted by the Gaussian model and we can use the coherent state hypothesis as a benchmark to establish that the differing predictions can be resolved experimentally.

%%%%%%%%%%%%%%%%%%%%%%%%%%%%%%%%%
\paragraph{More General Cases.}
%%%%%%%%%%%%%%%%%%%%%%%%%%%%%%%%%
%
One can envision scenarios where DM is in a quantum state not described by any of the models considered so far.
However, as emphasized in Refs.~\cite{Bao:2025nsd,suppression_paper}, such models will still often appear to the detector as a Gaussian.
The reason for this is similar to why the coherent state model repeatedly collapsed to a Gaussian: if the detected mode corresponds to multiple independent DM modes, then even if those modes have a $P$-function with no classical description, the quantum CLT applies and drives the observed mode to a Gaussian.
There are, however, potential exceptions.
If the local DM density is dominated by miniclusters, both homogeneity and stationarity could be badly violated, and a local Bose--Einstein condensate~\cite{Semikoz:1995rd,Sikivie:2009qn,Guth:2014hsa,Budker:2023sex,Paranjape:2026hkr} would have the DM dominantly occupying a single mode.
Both are examples where the assumptions underpinning the virialized Gaussian hypothesis fail.
In these cases it is even less clear what the quantum state of DM should be, although the formalism we develop allows us to study certain observational consequences of DM being in a condensate.
Nonetheless, we show that even in the most optimistic scenarios, establishing that DM exists in an intrinsically nonclassical state, such as a Fock state, is far out of reach of existing detectors.
Consistent with what we found for projective measurements in Refs.~\cite{Bao:2025nsd,suppression_paper}, such effects are suppressed by an additional power of the small efficiency $\eta$.
We also identify a further obstruction not discussed in those works: evading the quantum CLT by taking $P_a$ to have correlated modes can come at a cost, as those correlations can reduce the number of independent measurements with which to probe the quantum state.
We return to this final point in Sec.~\ref{subsec:nonclassical_singlemode}.

%%%%%%%%%%%%%%%%%%%%%%%%%%%%%%%%%
\subsection{Signal Power from Number and Quadrature}
\label{subsec:twopowers}
%%%%%%%%%%%%%%%%%%%%%%%%%%%%%%%%%

Throughout this work we focus on number and quadrature observables.\footnote{See Ref.~\cite{suppression_paper} for a discussion of more general observables, including entanglement and decoherence.}
Since we study measurements of the electromagnetic field exiting the cavity, the two correspond to measurements of photon number or the quadratures of the radiation field.
It is well known that depending on the relevant signal and backgrounds, there can be strong advantages for one measurement approach over another; see e.g.~Ref.~\cite{Lamoreaux:2013koa}.

Here we emphasize a different point: DM imprints itself distinctly on number and quadrature measurements.
For both types of measurements we can define a notion of ``signal power.''
For number measurements, one can construct a power by multiplying the photon count rate with the photon energy.
In the quadrature-measurement case, one can Fourier transform the measurement record and construct a power spectral density (PSD), which can be integrated to obtain signal power $P_{\mathrm{sig}}$.
However, in spite of the common name, these two notions of power probe different aspects of the DM state.

To exhibit the difference, consider an extremely idealized model in which DM is in a single-mode Gaussian state of mean occupancy $N_\eff$, and its properties can be measured directly (i.e.~without a penalty from the efficiency $\eta$).
Then by Eq.~\eqref{eq:P_probability}, the number distribution is
\be \label{eq:pN}
p(N) = \frac{N_\eff^N}{(1+N_\eff)^{N+1}},
\ee
which is the geometric distribution characteristic of the Bose--Einstein distribution.
The distribution of the power in this case has mean $N_\eff$ and variance $N_\eff(N_\eff+1)$.

Constructing the PSD distribution requires greater care.
From Eq.~\eqref{eq:P_probability}, the distribution of a single quadrature is
\be \label{eq:pX}
p(X) = \int d\alpha \, P_\eff(\alpha) |\!\braket{X}{\alpha}\!|^2
= \frac{1}{\sqrt{\pi(2N_\eff+1)}} \exp{ - \frac{X^2}{2N_\eff+1}},
\ee
so that the values are Gaussian distributed with mean zero and variance $N_\eff + 1/2$, where the $+1/2$ represents vacuum fluctuations.
However, the quadrature PSD depends on the time series $X(t) = X \cos(m_a t) + Y \sin(m_a t)$, and in general this requires measuring incompatible observables, $[X(t),X(t')] \neq 0$, since $[X,Y]=i$.
In the following sections, we encounter this issue in the context of continuous measurements.
Here we address the problem in a much simpler context: we consider simultaneously measuring $X$ and $Y$, and constructing a ``single-mode PSD'' from $X^2 + Y^2$.

To make this possible, even though $X$ and $Y$ do not commute, one can perform a heterodyne measurement.
We pass the DM through a 50:50 beamsplitter,\footnote{There are of course no such beamsplitters for DM.
In practice, the beamsplitter would be applied to the electromagnetic mode that has had DM imprinted on it.} which mixes the signal with a vacuum mode $v$ and creates two output modes
\be
h_{\pm} = \frac{1}{\sqrt{2}} \big(a \pm v\big).
\ee
The quadratures of the output modes can be written in terms of the quadratures of the axion and vacuum as $X_{\pm} = (X_a \pm X_v)/\sqrt{2}$ and $Y_\pm = (Y_a \pm Y_v)/\sqrt{2}$.
We can then measure $X_+$ and $Y_-$, i.e.~one quadrature from each output, as they are compatible,
\be
[X_+,Y_-] = \frac{1}{2} \big( [X_a,Y_a] - [X_v,Y_v] \big) = 0.
\ee
We then define the single-mode PSD as 
\be \label{eq:S_to_XpYm}
S = X_+^2 + Y_-^2.
\ee
To determine the distribution of $S$, first note that $X_a$ and $X_v$ are both Gaussian distributed with a mean of $0$, and variances $N_\eff + 1/2$ and $1/2$, respectively.
Since they are independent, $X_+$ is also Gaussian distributed, with
\be
p(X_+) = \frac{1}{\sqrt{\pi(N_\eff+1)}} \exp{-\frac{X_+^2}{N_\eff+1}}.
\ee
The distribution of $Y_-$ is identical and also independent from $X_+$.
Combining these results, we infer that $S$ is exponentially distributed,
\be \label{eq:pS}
p(S) = \frac{1}{N_\eff+1} \exp{-\frac{S}{N_\eff+1}},
\ee
with mean $N_\eff+1$ and variance $(N_\eff+1)^2$.

Note that the PSD mean involves an additive term of $+1$, which is double what one would have expected from the vacuum fluctuations in Eq.~\eqref{eq:pX}.\footnote{This is not simply due to the normalization of $S$, or the fact that both $X$ and $Y$ have vacuum fluctuations: within each quadrature, the ratio of thermal to vacuum noise is $2 N_\eff$, while here the ratio is only $N_\eff$.}
The extra $+1/2$ arises because in a heterodyne measurement, the vacuum fluctuations of both $a$ and $v$ contribute.
We can also quickly recover the $+1$ by defining 
\be \label{eq:defZ_heterodyne}
Z = \frac{1}{\sqrt{2}} \big( X_+ + i Y_- \big) = \frac{1}{\sqrt{2}} \big( a + v^\dag \big), 
\ee
and noting that 
\be \label{eq:S_Z_expec}
\la S \ra = 2 \la Z^\dag Z \ra = \la a^\dag a \ra + \la v v^\dag \ra = N_\eff+1.
\ee
An extra $+1/2$ arises generically whenever one measures conjugate observables~\cite{Arthurs:1965}.
For example, one could also have measured $X$ and $Y$ simultaneously by passing the signal through a phase-insensitive linear amplifier, after which the signal can be treated classically.
However, in that case the amplifier inevitably adds a half-quantum of noise~\cite{PhysRevD.26.1817}.
For instance, in a nondegenerate parametric amplifier, it comes from the vacuum fluctuations of the idler mode.

Comparing Eqs.~\eqref{eq:pN} and \eqref{eq:pS}, the two notions of power are manifestly distinct.
Not only do the means and variances disagree, but further $S$ is a continuous variable and $N$ is discrete.
However, as also discussed in Ref.~\cite{Cheong:2024ose}, in the limit $N_\eff \gg 1$ characteristic of wave DM,\footnote{That the two notions of power diverge in the particle DM limit is not immediately meaningful as all WIMP DM experiments measure number.
In particular, there are no experiments coupling to the quadrature of the WIMP field, just as there are no experiments looking to measure the quadrature of high-energy $\gamma$-ray photons.} all of these differences become unimportant, and the two notions of power coincide.
In Sec.~\ref{sec:inout}, we discuss this issue further in a full measurement model with the full DM mode structure, and show that many of the conclusions above continue to hold.

\section{Measurement in the Heisenberg Picture: Input-Output Theory}
\label{sec:inout}
%!TEX root = main.tex

\begin{figure}[t]
\centering
\includegraphics[width=0.7\linewidth]{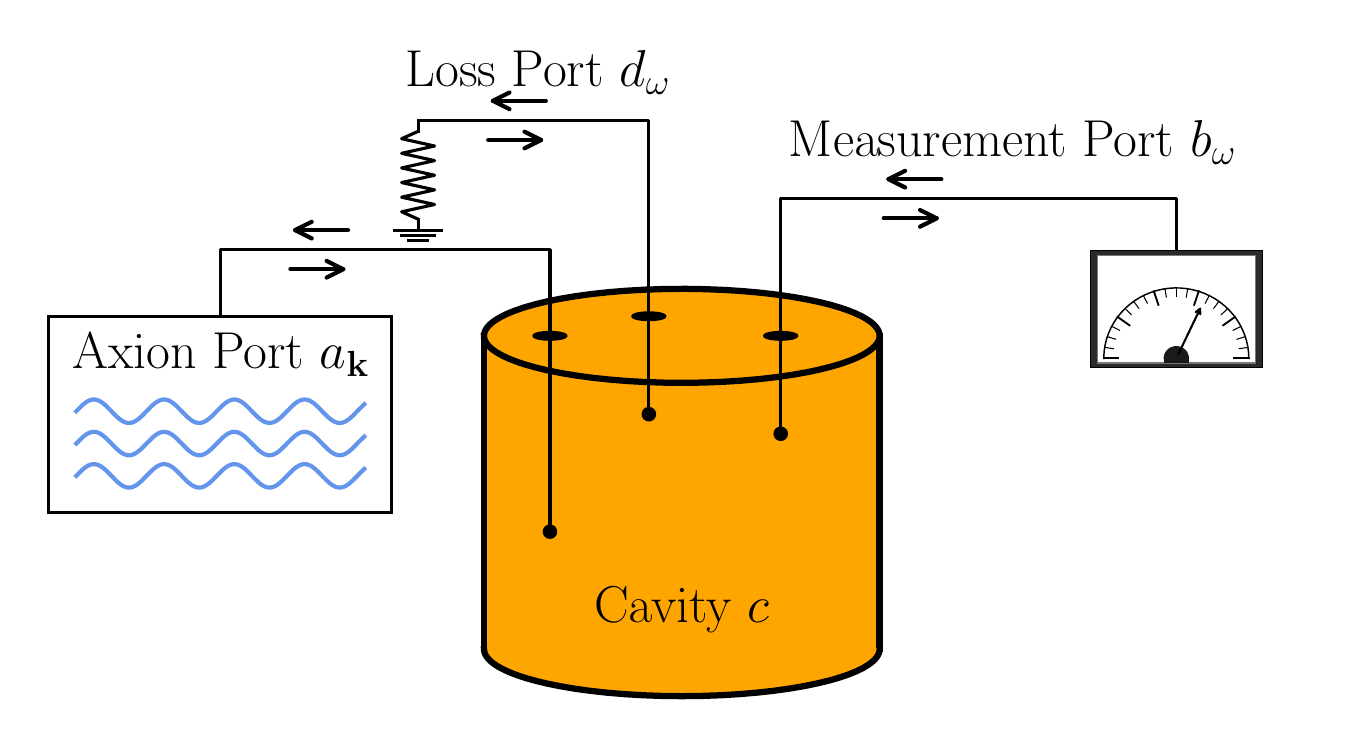}
\vspace{-0.6cm}
\caption{A schematic of the logic underpinning the input-output formalism: a single cavity mode $c$ is coupled to the axion, measurement, and loss ports.
The axion port provides the signal that drives the cavity, the measurement port carries the output signal that continuously monitors the cavity, and the loss port models additional dissipation in the cavity.
In input-output theory, each is viewed as a transmission line into the cavity.
Our depiction here is inspired by Ref.~\cite{Zheng:2016qjv}.
} 
\label{fig:cavity_io_scheme}
\end{figure}

A cavity haloscope which is continuously measured can be described by a cavity mode $c$ coupled to a set of transmission lines (see Fig.~\ref{fig:cavity_io_scheme}).
Measurements are performed on the field exiting through a physical transmission line, described with operators $b_\omega$, coupled to the cavity at a measurement port.
One can also send in squeezed states through this transmission line, to prepare a squeezed cavity state~\cite{Malnou:2018dxn,HAYSTAC:2020kwv,HAYSTAC:2023cam,HAYSTAC:2024jch}.
Intrinsic cavity dissipation is described with an effective transmission line with modes $d_\omega$, coupled at a decay (or ``loss'') port; this port also describes how thermal noise from the cavity walls enters the cavity state.
Finally, the axion signal enters through an effective transmission line associated with the continuum of axion modes, with operators $a^{(\dag)}_\bk$.
The full dynamics of the system are described by the Hamiltonian
\bea \label{eq:fullHamiltonian}
H &= \omega_c c^\dag c + \int \frac{d \omega}{2 \pi} \, \omega \qty(b_\omega^\dag b_\omega + d_\omega^\dag d_\omega) + \int \frac{d^3 \bk}{(2\pi)^3} \, \omega_\bk a_\bk^\dag a_\bk \\
&\quad + \Bigg[ i g c^\dag a_\eff + \int \frac{d \omega}{2\pi} \qty( i \sqrt{\kappa_m} \, c^\dag b_\omega + i \sqrt{\kappa_\ell} \, c^\dag d_\omega ) + \hc \Bigg],
\eea
where the cavity mode has angular frequency $\omega_c$, and $\kappa_m$ and $\kappa_\ell$ are the coupling rates to the measurement and loss ports, respectively.
Axion DM is described via an effective mode $a_\eff$ with coupling $g$, as given in Eqs.~\eqref{eq:effective_mode_def} and \eqref{eq:Hint_simplified}.
Though our discussion is general, for concreteness we note that in many haloscopes, $c$ corresponds to the TM$_{010}$ mode of a cylindrical cavity.

This system can be described with ``input-output theory,'' in which the outgoing waves along the transmission lines are related to the incoming waves by input-output relations.
In this section, we develop a version of input-output theory where all ingredients, including the axion, are treated quantum mechanically.
In Sec.~\ref{sec:string-OscillatorSystem}, we build intuition with an analogous mechanical model, in which a mass on a spring (the cavity) is coupled to a set of strings (the transmission lines).
Each small segment of an incoming waveform on the string momentarily interacts with the mass, before reflecting to become part of the outgoing waveform.
The example also illustrates how irreversible dissipation for the mass can emerge from coupling to the continuum of modes on a string.
In Sec.~\ref{subsec:io_intro} we review the standard input-output formalism for a cavity haloscope, which amounts to studying the dynamics of the Hamiltonian in Eq.~\eqref{eq:fullHamiltonian} in the Heisenberg picture.
In the process, we define discrete temporal modes -- the quantum analogs of the momentary interactions with the cavity -- relevant for Secs.~\ref{subsec:higherOrderIO} and \ref{sec:measurement}.

In Sec.~\ref{subsec:OutputObservables} we compute the expectation values of the PSD and total power.
These simple observables are easily accessible in the input-output formalism and can be computed for various DM and detector states.
However, in a haloscope experiment, once a (homogeneous) DM density and velocity distribution are taken, the mean DM occupation number is fixed, and only higher-order statistics can distinguish the underlying axion state.
As an explicit example, we compute the variance of number measurements in Sec.~\ref{subsec:higherOrderIO} and show it can be used to distinguish between the virialized Gaussian and coherent-state hypotheses from Sec.~\ref{subsec:DMQstate}.

%%%%%%%%%%%%%%%%%%%%%%%%%%%%%%%%%
\paragraph{Assumptions on the DM State.}
%%%%%%%%%%%%%%%%%%%%%%%%%%%%%%%%%
%
Our aim is to understand how the response of a detector to DM depends on the DM's quantum state.
Our discussion is completely general until Eq.~\eqref{eq:input_DM_number_mean}, where we assume the DM state is homogeneous.
Following this, in Secs.~\ref{subsec:OutputObservables} and \ref{subsec:higherOrderIO} we evaluate DM expectation values extensively; to make this tractable, we assume the DM state is both homogeneous and stationary, i.e., DM correlators only depend on differences of times and positions.
Homogeneity and stationarity are not assumed in Sec.~\ref{sec:measurement}.

Our ad hoc coherent-state hypothesis of Eq.~\eqref{eq:CSH} is neither homogeneous nor stationary; we verify results for its behavior separately where relevant.
At the end of Sec.~\ref{subsec:higherOrderIO} we further assume that if one makes two measurements of the DM field separated by at least a DM coherence time, then those measurements probe statistically independent realizations of the DM state, and use this to distinguish the virialized Gaussian and coherent-state hypotheses.
A more careful treatment of correlations as a function of time is provided in Sec.~\ref{sec:measurement}.

%%%%%%%%%%%%%%%%%%%%%%%%%%%%%%%%%
\paragraph{Comparison to Existing Literature.}
%%%%%%%%%%%%%%%%%%%%%%%%%%%%%%%%%
%
Input-output theory is a well-established formalism that has been widely applied to axion searches~\cite{Zheng:2016qjv,Malnou:2018dxn,Flower:2018qgb,Carney:2019cio,Lehnert:2021gbj,Wurtz:2021cnm,Brady:2022bus,Shi:2022wpf,Jiang:2022vpm,Beckey:2023shi,Chen:2023ryb,Chigusa:2023szl}; our central contribution is an end-to-end quantum description of axion detection.
We treat the axion quantum mechanically throughout, so that we can track how its quantum state is imprinted on the measured statistics.
We also provide a unified discussion of the expected first and second moments of the PSD and number operators appropriate for axion detection, which are presented in Secs.~\ref{subsec:OutputObservables} and \ref{subsec:higherOrderIO}.
Further, Sec.~\ref{subsec:io_intro} derives the well-known input-output relations in a way which highlights the connection to the simple dynamics of coupled oscillators exhibited in, for example, Eq.~\eqref{eq:joint_coherent}.

%%%%%%%%%%%%%%%%%%%%%%%%%%%%%%%%%
\subsection{Warm-Up: Classical String-Oscillator Model} 
\label{sec:string-OscillatorSystem}
%%%%%%%%%%%%%%%%%%%%%%%%%%%%%%%%%

\begin{figure}[!t]
\centering
\includegraphics[width=0.75\linewidth]{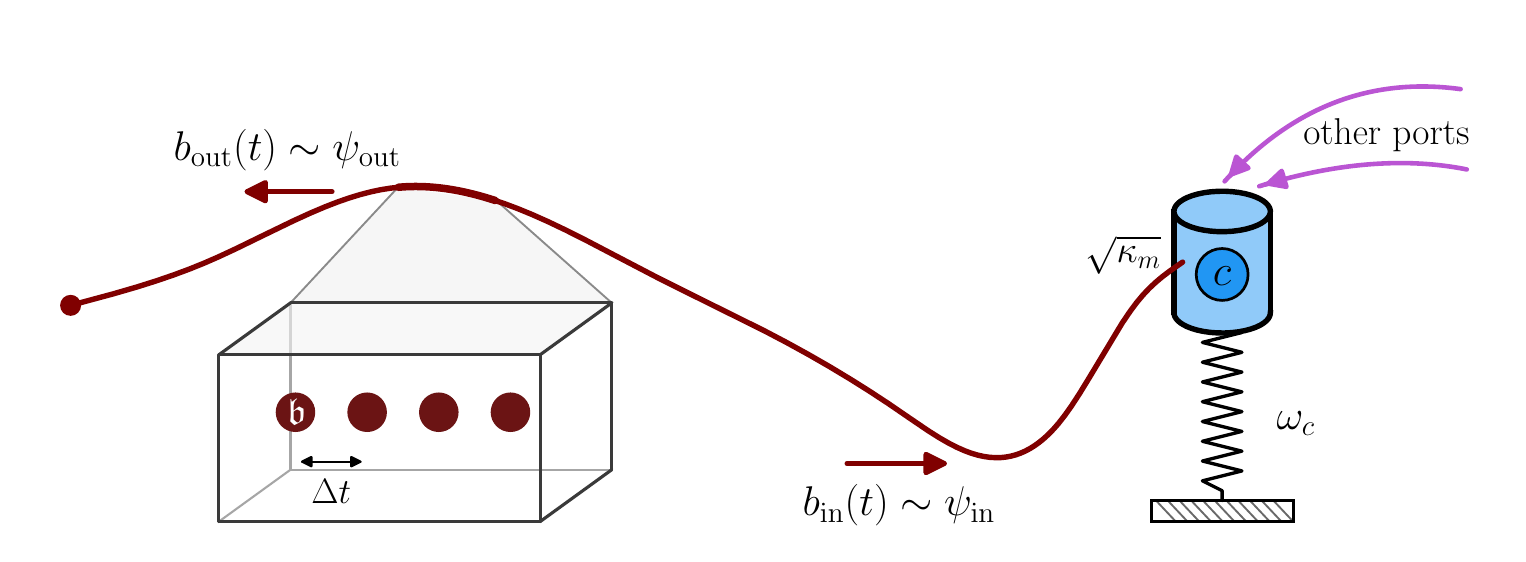}
\vspace{-0.3cm}
\caption{Toy model of a string (transmission line) coupled to a classical harmonic oscillator (cavity mode): the input field of an incoming waveform interacts with the oscillator, before producing an outgoing waveform that heads back down the line to be detected.
Once the system is quantized, the line can be described by a series of discrete modes $\db$ as defined in Eq.~\eqref{eq:discrete_b_in}.
}
\label{fig:string_oscillator_sketch}
\end{figure}

Consider the classical model depicted in Fig.~\ref{fig:string_oscillator_sketch}: a semi-infinite string linearly coupled at one end to a mass $m$ attached to a spring with spring constant $k$, so that it has a natural angular frequency $\omega_c = \sqrt{k/m}$.
In this analogy to a haloscope, the string plays the role of the transmission line, and the mass coupled to a spring represents the cavity.\footnote{This toy model has been widely used to describe the response of a damped harmonic oscillator to an incident field and its corresponding outgoing field -- for a classic review see Ref.~\cite{10.1119/1.13768} -- including in the context of input-output theory~\cite{drummond2013quantum}.}
This classical input-output model should be thought of as the continuous-measurement generalization of the two-coupled-mode system described by Eq.~\eqref{eq:Hint_simplified}, which we used to describe projective measurements in Refs.~\cite{Bao:2025nsd,suppression_paper}.

To probe the mass, we prepare an \textit{input} signal on the string in the infinite past $t \to -\infty$, described by a waveform $\psi_\tin(x,t)$,%
\footnote{We write $\psi_\tin(x,t)$ and $\psi_\tout(x,t)$ here as complex for convenience; the actual input and output waveforms are $\psi_\tin(x,t) + \hc$ and $\psi_\tout(x,t) + \hc$, cf.~Eq.~\eqref{appeq:IOWaveformDefine}.
}
specified by a set of Fourier coefficients $A_k$, 
\be \label{eq:psi_in_string}
\psi_\tin(x,t) = \int_0^\infty \frac{dk}{2\pi}\, A_k\, e^{i(kx-\omega_k t)}.
\ee
This wave propagates along the string until it briefly interacts with the mass, after which it is reflected back outward.
The \textit{output} signal $\psi_\tout(x,t)$ is described by Fourier coefficients $B_k$,
\be \label{eq:psi_out_def}
\psi_\tout(x,t) = \int_0^\infty \frac{dk}{2\pi}\, B_k\, e^{-i(kx+\omega_k t)}.
\ee
In App.~\ref{app:string-Oscillator}, we solve the dynamics of the coupled string-mass system, using a classical analog of Eq.~\eqref{eq:fullHamiltonian}, and show that in the steady state, the input and output coefficients are linearly related by a response function $\chi_k$,\footnote{There is a minus sign here and in Eq.~\eqref{eq:string_inout_w_DM} because far off-resonance, the mass acts like a hard boundary condition, flipping the sign of the wave profile.
Such signs can always be removed, e.g.~by using the conjugate variable or by absorbing them into the definition of the output field.}
\be \label{eq:trivial_string_inout}
B_k = -\chi_k A_k, \quad
\chi_k \equiv 1 - \frac{\kappa \omega_k}{\kappa \omega_k/2 - i (\omega_k^2 - \omega_c^2)/2} 
\simeq 1 - \frac{\kappa}{\kappa/2 - i (\omega_k - \omega_c)} 
\ee
where the final approximation for $\chi_k$ holds near the resonance, specifically when $|\omega_k-\omega_c| \ll \omega_c$.
(If the string has mass density $\mu$ and tension $\mathcal{T}$, then the coupling is $\kappa = \sqrt{\mathcal{T} \mu}/m$.)
Finally, once the reflected wave returns, we can measure it to deduce information about the mass's motion.
The output is fully determined by the input and the input-output relation Eq.~\eqref{eq:trivial_string_inout}.

In this simple case we have $\abs{\chi_k}^2 = 1$, i.e.~the scattering produces only a phase shift.
Consider next connecting an additional string, e.g.~representing the axion field.
If the incoming DM wave is described by Fourier coefficients $C_k$, a similar analysis reveals the following input-output relation
\be \label{eq:string_inout_w_DM}
B_k \simeq - \underbrace{\left( 1 - \frac{\kappa_{\scriptscriptstyle A}}{\kappa_\text{tot}/2 - i (\omega_k - \omega_c)} \right)}_{\equiv \chi_k^{\scriptscriptstyle A}} A_k
+ \underbrace{\left( \frac{\kappa_{\scriptscriptstyle C}}{\kappa_\text{tot}/2 - i (\omega_k - \omega_c)} \right)}_{\equiv \chi_k^{\scriptscriptstyle C}} C_k
\ee
where $\kappa_\text{tot} = \kappa_{\scriptscriptstyle A}+\kappa_{\scriptscriptstyle C}$.
Here $\chi_k^{\scriptscriptstyle A}$ has a comparable form to the susceptibility $\chi_k$ in Eq.~\eqref{eq:trivial_string_inout}.
A key difference is that while the numerator of the second term in $\chi_k^{\scriptscriptstyle A}$ is identical -- reflecting the strength of the coupling of the measurement string to the mass -- in the denominator we replace $\kappa$ by $\kappa_\text{tot}$, as this encodes the width of the oscillator that depends on all ports it can lose energy through.
With two strings, $\abs{\chi_k^{\scriptscriptstyle C}}^2$ is a Lorentzian bandpass with a resonance near $\omega_c$, while $\abs{\chi_k^{\scriptscriptstyle A}}^2$ is a bandstop.
For now, we see that through the final measurements of the output wave $\psi_\tout(x,t)$, so long as we know the input modes $A_k$, we can gain knowledge about the DM modes $C_k$.
Note also that while the mass's dynamics determine the output, the state of the mass does not appear explicitly in the input-output relations in steady state.

We emphasize that there is a key difference between the dynamics of the above string model and that of the coupled oscillators in Eq.~\eqref{eq:Hint_simplified} that was studied in Ref.~\cite{suppression_paper}.
There, the two modes periodically exchange energy indefinitely, without dissipation.
By contrast, a harmonic oscillator coupled to the continuum of modes in a semi-infinite string can permanently lose energy: the infinite nature of the string implies that energy transmitted down a line need never return, introducing a source of loss for the oscillator.

This classical string model is discussed in more detail in App.~\ref{app:string-Oscillator}.
Already though we can see that in classical input-output theory, with knowledge of the in and out modes we can infer properties of the DM string, whilst in a sense bypassing the state of the cavity.
In the following subsection we show how to lift this picture from classical variables to quantum mechanical operators.
Additional subtleties arise, but the conceptual picture is largely identical.

%%%%%%%%%%%%%%%%%%%%%%%%%%%%%%%%%
\subsection{Input-Output Formalism for Quantum Axion Detection}
\label{subsec:io_intro}
%%%%%%%%%%%%%%%%%%%%%%%%%%%%%%%%%

In this subsection, we introduce the terms of the Hamiltonian in Eq.~\eqref{eq:fullHamiltonian} sequentially, building up to the full dynamics.
First, consider the coupling to the measurement port: the cavity mode $c$ is measured by coupling to a semi-infinite transmission line with a continuum of modes $b_\omega$, normalized by $[b_\omega, b^\dag_{\omega'}]=2\pi \delta(\omega-\omega')$.
The corresponding Hamiltonian is
\be \label{eq:bare_inout_ham}
H = \omega_c c^\dag c + \int_{-\infty}^\infty \frac{d\omega}{2\pi} \, \omega b_\omega^\dag b_\omega + i\sqrt{\kappa_m}\int_{-\infty}^\infty \frac{d\omega}{2\pi} \, (c^\dag b_\omega - c b_\omega^\dag).
\ee
We have again applied the RWA, yielding an interaction analogous to Eq.~\eqref{eq:Hint_simplified}, but extended to a continuum of transmission-line modes.

In this subsection, we first derive the evolution of the mode operators by changing basis to \textit{temporal} modes of the transmission line.
We then derive an input-output relation comparable to Eq.~\eqref{eq:trivial_string_inout}, before repeating the procedure after adding ports to describe the axion signal and losses.
Lastly, we discuss the quantum states of the input modes.

%%%%%%%%%%%%%%%%%%%%%%%%%%%%%%%%%
\paragraph{Markov Approximation and Negative-Frequency Modes.}
%%%%%%%%%%%%%%%%%%%%%%%%%%%%%%%%%
%
We first note two subtleties that are implicit in Eq.~\eqref{eq:bare_inout_ham}.
First, we have taken the coupling between the cavity and all transmission-line modes to be $\sqrt{\kappa_m}$, independent of frequency.
Since this assumption implies the coupling is perfectly local in time, it is called the Markov approximation.
However, in reality this cannot be the case; for instance, for very low or high frequencies, the transmission line may not support modes at all.
The justification for this approximation is the same as for the RWA: since the cavity mode at $\omega_c$ is narrow, the only relevant transmission-line modes have $\omega \simeq \omega_c$.
For these $\omega$ the frequency variation of $\kappa_m$ is negligible, while for other $\omega$ the value of $\kappa_m$ is irrelevant, so we take it to be frequency independent for simplicity.

Second, we let the transmission line have modes with \textit{negative} frequencies.
This is very different from our classical treatment in Sec.~\ref{sec:string-OscillatorSystem}: a classical Fourier transform can have support at negative frequencies, but negative-frequency \textit{modes} correspond to quanta with negative energy, which are certainly unphysical.
Again, this extension is justified by the same logic as the RWA: all the negative-frequency modes are far from $\omega_c$, and hence have a negligible effect on the cavity dynamics.\footnote{Here it is essential that we actually applied the RWA, or else the mode at $-\omega_c$ would have an effect as well.}
As we see shortly, introducing the negative-frequency modes is mathematically convenient, as it allows us to construct commuting temporal modes.

In input-output theory, the introduction of negative-frequency modes is a standard (albeit often unemphasized) step; see Refs.~\cite{Gardiner:1985uof,Albarelli:2023vok,Clerk:2008tlb} for discussion.
From here on, we let all $\omega$ integrals implicitly range over $(-\infty, +\infty)$, unless shown otherwise.
Finally, while realistic transmission-line measurements only extract positive-frequency modes near $\omega_c$, some formal observables can depend on the negative-frequency modes; we address this subtlety in Sec.~\ref{subsec:OutputObservables}.

%%%%%%%%%%%%%%%%%%%%%%%%%%%%%%%%%
\paragraph{Temporal Input and Output Modes.}
%%%%%%%%%%%%%%%%%%%%%%%%%%%%%%%%%
%
The plane-wave modes $b_\omega$ diagonalize the free Hamiltonian of the transmission line, $H_0 = \int \tfrac{d\omega}{2\pi} \, \omega b_\omega^\dag b_\omega$, but are completely spatially and temporally delocalized.
Since the detector at the end of the transmission line interacts locally, it is useful to instead work with localized \textit{temporal modes}.
It is this set of modes that sequentially interact with the cavity, carrying away information that we ultimately measure.

These temporal modes can be defined in terms of the transmission-line modes as 
\be \label{eq:b_tin_def}
b_\tin(t) = U_0^\dag(t) \left( \int \frac{d\omega}{2\pi} \, b_\omega \right) U_0(t)
= \int \frac{d\omega}{2\pi} \, b_\omega e^{-i\omega t},
\ee
with $U_0(t) = e^{-i H_0 t}$.
The modes are \textit{local} in that they have the following commutation relation
\be \label{eq:b_tin_commutator}
[b_\tin(t), b^\dag_\tin(t')] = \delta\!\,(t-t'),
\ee
which results from the Markov approximation and extension to negative frequencies discussed above.
These commutation relations justify interpreting each $b_\tin(t)$ as an independent bosonic degree of freedom.
However, as is clear from Eq.~\eqref{eq:b_tin_def}, $b_\tin(t)$ does not represent an operator in the Heisenberg picture evolved to time $t$; instead, it is the interaction-picture operator obtained by \textit{freely} evolving the local transmission-line operator according to $H_0$, with $b_\omega$ corresponding to the Schr\"odinger-picture operators at the chosen reference time.

In this section, we are concerned with the Heisenberg-picture evolution of operators; however, the explicit use of these temporal modes requires the intermediate use of the interaction picture.
Concretely, we decompose the full time-evolution operator $U_\text{full}(t_f, t_i) = \exp{-i H(t_f-t_i)}$ as
\be \label{eq:interaction_U_split}
U_\text{full}(t_f, t_i) = U_0(t_f)U_I(t_f, t_i)U_0^\dag(t_i),
\ee
where $U_I$ is generated by the time-dependent Hamiltonian
\be \label{eq:HI(t)}
H_I(t) = \omega_c c^\dag c + i\sqrt{\kappa_m}\big(c^\dag b_\tin(t) - c\, b^\dag_\tin(t) \big).
\ee
To be explicit, here $c$ denotes the unevolved cavity operator.
Since $H_0$ contains only the free transmission-line Hamiltonian, $[H_0,c]=0$, the cavity operator has no explicit time dependence in the interaction picture.
When combined with the independence of the input modes from Eq.~\eqref{eq:b_tin_commutator}, the Hamiltonian in Eq.~\eqref{eq:HI(t)} describes the cavity interacting with each temporal mode for a brief instant, before moving on to the next mode, never interacting with any temporal mode more than once.
Consequently, the cavity never feels backaction again from the input modes it has already interacted with; it is a memoryless process.
This provides a further justification for describing Eq.~\eqref{eq:b_tin_commutator} as a Markov approximation.
From this perspective, $b_\tin^\dag(t) b_\tin(t)$ is a rate corresponding to the incoming flux of measurement-port quanta through a point on the transmission line at a time $t$.

To describe what our experiments measure, we need also to define the local temporal modes that couple to our detector, $b_\tout(t)$.
This output mode is defined similarly to $b_\tin$,\footnote{The explicit time dependence in $b_{\omega}(t)$ makes the definition of the output mode in Eq.~\eqref{eq:b_out_def} appear distinct from that commonly adopted in the literature, such as the classic formulation of Gardiner and Collett~\cite{Gardiner:1985uof}.
Nevertheless, as shown in App.~\ref{app:equivalentTemporalInputOutput}, our expression matches the classic result exactly in the limit $t_d \to 0$.
\label{footnote:output-GC}}
\be \label{eq:b_out_def}
b_\tout(t) \equiv \int\frac{d\omega}{2\pi}\, b_\omega(t) e^{i\omega t_d},
\ee
where $t_d$ is the time it takes an output signal to travel from the cavity to the detector.
Here, $b_\omega(t)$ is a Heisenberg-picture operator, evolved to time $t$ using the full time-evolution operator in Eq.~\eqref{eq:interaction_U_split}.
Explicitly, we can relate the output mode to either the input mode or the Schr\"odinger operator localized at the detector,
\be \label{eq:bout_to_bin}
b_\tout(t) = U_I^\dag (t, 0) b_\tin(t-t_d) U_I(t, 0) = U_\text{full}^\dag (t, 0) \left( \int \frac{d\omega}{2\pi} \, b_\omega e^{i\omega t_d} \right) U_\text{full}(t, 0).
\ee
This shows the output mode is the freely evolved input mode combined with the interaction with the cavity encoded in $U_I$.
The output mode is therefore the field operator that our detector along the transmission line has access to at any moment, and accordingly all output-field observables accessible to the detector can be constructed from $b_\tout(t)$.\footnote{Strictly speaking, the measured observables, such as a voltage, are constructed from the transmission line's canonical field operators, not $b_\tout(t)$.
The key distinction between the two is a factor of $\sqrt{\omega}$ in the integrand, not present in Eq.~\eqref{eq:b_out_def}, but essential in the field operators.
(This factor can be seen in the classical expressions in Eq.~\eqref{eq:app_canonical_string_mode_expansion}; the quantum forms are similar with $\beta_k \to b_\omega$.)
Yet for the narrow-bandwidth, near-resonance regime, $\sqrt{\omega}$ is well approximated by $\sqrt{\omega_c}$ over the relevant integration range, and the physical field operator becomes proportional to a quadrature of $b_\tout(t)$.}
As shown in App.~\ref{app:equivalentTemporalInputOutput}, one can use the above results to show that the output operator is also local,
\be \label{eq:b_out_commutator}
[b_\tout(t), b^\dag_\tout(t')] = \delta\!\,(t-t').
\ee
Further discussion can be found in Refs.~\cite{Gardiner:1985uof,Albarelli:2023vok,Clerk:2008tlb,Jacobs2014QuantumMeasurement}, which also justify the physical interpretation of $b_\tout(t)$ as an \textit{outgoing wave} analogous to Eq.~\eqref{eq:psi_out_def} in the string model.

Having introduced continuous input and output modes, it further proves useful to divide the full integration time into disjoint intervals of length $\Delta t$, and then define discrete temporal modes smeared over each interval (see Fig.~\ref{fig:string_oscillator_sketch}), 
\be \label{eq:discrete_b_in}
\db_\tin(t) \equiv \frac{1}{\sqrt{\Delta t}}\int_t^{t+\Delta t} \!dt'\, b_\tin(t').
\ee
The prefactor ensures the $\db_\tin(t)$ are dimensionless and canonically normalized, 
\be
[\db_\tin(t),\db^\dag_\tin(t')]=\delta_{t,t'}.
\ee
We define the discrete $\db_\tout(t)$ identically.
The introduction of $\Delta t$ is physically meaningful: any real measurement has a \textit{finite} sampling rate.
The finite time serves as a regulator in several of the calculations below, and final results are independent of $\Delta t$ if it is sufficiently small.
As such, we generically treat $\Delta t$ as a small parameter that we can expand our results in.
When performing such a power counting, we observe that for small $\Delta t$ we have $\db_\tin(t) \simeq b_\tin(t)\sqrt{\Delta t}$.
Since $b_\tin(t)$ is the physical measured variable that enters the Hamiltonian, we should regard any discrete mode $\db(t)$ as $\mathcal{O}(\sqrt{\Delta t})$.

To evaluate the interaction-picture time-evolution operator, given the Hamiltonian in Eq.~\eqref{eq:HI(t)} is time dependent, we need to compute a time-ordered exponential.
Taking advantage of the $\Delta t$ expansion, however, we can simplify the operator to
\bea \label{eq:time_ev_operator}
U_I(t+\Delta t,t) &\simeq \exp{-i\omega_c \Delta t\, c^\dag c}\exp{\sqrt{\kappa_m}\int_t^{t+\Delta t} \!dt' \left( c^\dag b_\tin(t') - c b_\tin^\dag (t')\right)} \\
&= \exp{-i\omega_c \Delta t\, c^\dag c}\exp{\sqrt{\kappa_m \Delta t} \left( c^\dag \db_\tin(t) - c \db^\dag_\tin(t) \right)},
\eea
where we introduced the discrete modes in the second line.
The result is approximate as we have neglected the term that arises from a commutator between the two exponents, which is formally $\mathcal{O}(\omega_c \Delta t \sqrt{\kappa_m \Delta t})$.
Here and throughout we assume a sampling rate such that these higher-order terms can be neglected; we further assume $\kappa_m \Delta t \ll 1$ as this combination appears a number of times in the results that follow.

Using the above formalism, we now study the Heisenberg-picture evolution of the cavity and output modes.
Consider first how the cavity mode $c$ evolves over a time $t$ to $t+\Delta t$.
Being explicit, we can use Eq.~\eqref{eq:interaction_U_split} to write
\bea
c(t+\Delta t) &= U_\text{full}^\dag(t+\Delta t, 0) \, c\, U_\text{full}(t+\Delta t, 0) \\
&= U_I^\dag(t,0)\big[U_I^\dag(t+\Delta t, t)\, c\, U_I(t+\Delta t, t) \big]U_I(t,0)
\eea
where we used $[H_0, c]=0$ to eliminate the factors of $U_0$.
Using Eq.~\eqref{eq:time_ev_operator} and the results from Sec.~2 of Ref.~\cite{suppression_paper}, we can evaluate the term in brackets as
\be
U_I^\dag(t+\Delta t, t)\, c\, U_I(t+\Delta t, t) \simeq c\,e^{-i\omega_c \Delta t}\cos(\sqrt{\kappa_m \Delta t}) + \db_\tin(t) \sin(\sqrt{\kappa_m\Delta t}).
\ee
The final $U_I(t,0)$ operator has no effect on $\db_\tin(t)$, since $U_I(t,0)$ contains only temporal modes active for times strictly less than $t$.
However, $U_I(t,0)$ does act on the cavity mode, evolving it exactly in the Heisenberg picture to time $t$.
We therefore arrive at 
\be \label{eq:cavity_mode_step}
c(t+\Delta t) \simeq c(t)e^{-i\omega_c \Delta t}\cos(\sqrt{\kappa_m \Delta t}) + \db_\tin(t) \sin(\sqrt{\kappa_m\Delta t}).
\ee
To evolve the output mode, which carries a portion of the cavity mode after it mixes with $\db_\tin(t)$, we apply an identical procedure to the first relation in Eq.~\eqref{eq:bout_to_bin} to obtain
\be \label{eq:b_out_step}
\db_\tout(t) \simeq \db_\tin(t-t_d) \cos(\sqrt{\kappa_m \Delta t}) - c(t-t_d)\sin(\sqrt{\kappa_m \Delta t}).
\ee
Going forward, we take $t_d \to 0$ for simplicity, as the location of our detector in a dispersionless transmission line has no effect on the physical outcomes.

%%%%%%%%%%%%%%%%%%%%%%%%%%%%%%%%%
\paragraph{Input-Output Relations.}
%%%%%%%%%%%%%%%%%%%%%%%%%%%%%%%%%
%
We now revert to continuum modes $b_\tin(t) \simeq \db_\tin(t)/\sqrt{\Delta t}$ and expand to first order in $\Delta t$.
By analogy, we identify the continuum output mode operator $b_\tout(t) \simeq \db_\tout(t)/\sqrt{\Delta t}$, which is what we measure at the end of the transmission line.
Equations~\eqref{eq:cavity_mode_step} and \eqref{eq:b_out_step} for small $\kappa_m \Delta t$ then reduce to
\begin{subequations}\label{eq:cont_inout_relations}
\begin{align}
\label{eq:cont_inout_relations_a}
b_\tout(t) &= b_\tin(t) - \sqrt{\kappa_m}\, c(t), \\
\label{eq:cont_inout_relations_c}
\dot c(t) &= -\left( i\omega_c + \frac{\kappa_m}{2}\right) c(t) + \sqrt{\kappa_m}\, b_\tin(t).
\end{align}
\end{subequations}
The first line is the well-known time-domain input-output relation: the difference between the input and output modes is dictated by the cavity.
The second line is the cavity equation of motion, which now contains a dissipative term $-\kappa_m c(t) /2$, so that the decay rate of cavity quanta is $\kappa_m$.
As discussed at the end of Sec.~\ref{sec:string-OscillatorSystem}, this irreversible dissipation emerges from coupling the cavity to a continuum of modes, even though the system as a whole is conservative.

Finally, by Fourier transforming both equations in Eq.~\eqref{eq:cont_inout_relations}, we can eliminate the cavity mode entirely.
We can then express the output purely in terms of the input,
\be \label{eq:first_bout_eq}
b_\tout[\omega] = \qty(1 - \frac{\kappa_m}{\kappa_m/2-i \Delta_{\omega}}) b_\tin[\omega], 
\ee
in which $\Delta_{\omega} \equiv \omega - \omega_c$ is the detuning.
This is a frequency-domain input-output relation, a quantum-mechanical analog of the classical result Eq.~\eqref{eq:trivial_string_inout}; note that we only recover the approximate classical expression that holds for $|\omega-\omega_c| \ll \omega_c$, since we applied the RWA.

%%%%%%%%%%%%%%%%%%%%%%%%%%%%%%%%%
\paragraph{Adding DM Interactions.}
%%%%%%%%%%%%%%%%%%%%%%%%%%%%%%%%%
%
We now add the interaction we hope to detect: an axion effective mode $a_\eff(t)$ that has a coupling $g$ to the cavity and an interaction Hamiltonian,
\be
H_\text{int}^\DM = i g \big( c^\dag a_\eff(t) - c\, a_\eff^\dag(t) \big).
\ee
For a given cavity mode, $g$ and $a_\eff(t)$ can be determined explicitly from Eqs.~\eqref{eq:effective_mode_def} and \eqref{eq:Hint_simplified}.

This interaction adds a factor $\exp\,\!\{g\Delta t ( c^\dag a_\eff(t) - c\, a_\eff^\dag(t))\}$ to the time-evolution operator in Eq.~\eqref{eq:time_ev_operator}.
The resulting input-output relation in Eq.~\eqref{eq:cont_inout_relations_a} is unchanged, but the cavity equation of motion gains an additional term,
\be \label{eq:cavity_EOM_DM}
\dot c(t) \simeq -\left( i\omega_c + \frac{\kappa_m}{2}\right) c(t) + g a_\eff(t) + \sqrt{\kappa_m} \, b_\tin(t).
\ee
This equation is approximate: here and in what follows we derive the cavity evolution and the input-output relation to leading order in $g$.
We do not, however, truncate observables constructed from these modes; indeed, the leading-order axion contribution to the observed power scales as $g^2 N_\eff$ and is retained.
We show explicitly in App.~\ref{app:weakAxionCoupling} that there are $\mathcal{O}(g^2)$ contributions to Eq.~\eqref{eq:cavity_EOM_DM}, such as an axion contribution to the cavity dissipation.
These contributions are negligible; as an example, the cavity decay rate into axions is parametrically $\kappa_a \sim g^2/\omega_c \ll \kappa_m$.
Neglecting these terms also implies that our truncated cavity and consequently output modes do not satisfy canonical commutation relations.
The violation arises from a contribution proportional to $g^2 [a_\eff,a_\eff^\dag]$ -- second order in $g$ and unenhanced by $N_\eff$ -- and would be exactly canceled by the second-order term absent in Eq.~\eqref{eq:cavity_EOM_DM}; it is thus consistent with our approximation to treat the modes as having canonical commutators.

In addition to the approximation just discussed, we neglect detector backaction on the axion state.\footnote{Neglecting the second-order correction to the cavity evolution in Eq.~\eqref{eq:cavity_EOM_DM} does not by itself imply we can disregard the cavity's impact on the axion's evolution.
For example, if the cavity were prepared with an enormous occupation, $\la c^\dag c \ra \sim N_\eff$, then the rate of power transfer from the cavity to the axion would not be negligible, and so we assume $\la c^\dag c \ra \ll N_\eff$.
Further, owing to the measurement and loss ports, the cavity dissipates power at a rate $\kappa_\text{tot}$ defined below in Eq.~\eqref{eq:freq_output_noise_DM}, and if evolved indefinitely would eventually dissipate all power in the axion field.
We therefore also assume that the system is only evolved for a timescale $T \ll \kappa_\text{tot}/g^2$, which for the parameters in Eq.~\eqref{eq:eta_value} is many orders of magnitude longer than any experimental integration time.
Similar considerations arise when deriving a steady-state regime in our Schr\"odinger-picture construction (see the discussion in Sec.~\ref{subsec:complete_model}).
\label{footnote:cavitybackaction}}
Consequently, $a_\eff(t)$ simply obeys its free evolution, making it an oscillator-like, \textit{operator-valued} driving term,
\be \label{eq:free_effective_mode}
a_\eff(t) = \frac{1}{2}\sqrt{\frac{V_c}{\Omega}}\int \frac{d^3\bk}{(2\pi)^3}\, C(\bk)\,a_\bk\,e^{-i\omega_\bk t}, 
\ee
which should be compared with the form in Eq.~\eqref{eq:effective_mode_def}.

By repeating the steps that led up to Eq.~\eqref{eq:first_bout_eq}, we find
\be
b_\tout[\omega] = \left(1-\frac{\kappa_m}{\kappa_m/2-i\Delta_{\omega}}\right)b_\tin[\omega] - \frac{g\sqrt{\kappa_m}}{\kappa_m/2 -i\Delta_{\omega}}a_\eff[\omega].
\ee
This is the quantum analog of Eq.~\eqref{eq:string_inout_w_DM}, derived for the classical string model.
The width of the cavity response is controlled by the measurement port coupling $\kappa_m$ alone as we neglect the axion contribution, parametrically $\kappa_a \sim g^2/\omega_c$.
As discussed further in App.~\ref{app:weakAxionCoupling}, our treatment of the axion effective mode is not Markovian.
For the transmission lines, the Markov approximation is a statement about the coupling: we take $\sqrt{\kappa_m}$ to be frequency independent, and this renders $b_\tin(t)$ local as in Eq.~\eqref{eq:b_tin_commutator}.
Neglecting backaction leaves $a_\eff(t)$ freely evolving, so the DM state is undisturbed through the evolution and retains its full set of correlations encoded in its $P$-function, $P_a$.\footnote{Non-Markovian DM dynamics can also be included using the Schwinger--Keldysh formalism; see Ref.~\cite{Badurina:2026owr}.}

%%%%%%%%%%%%%%%%%%%%%%%%%%%%%%%%%
\paragraph{Adding Cavity Losses.}
%%%%%%%%%%%%%%%%%%%%%%%%%%%%%%%%%
%
Finally, we add a transmission line coupled with rate $\kappa_\ell$ and with associated modes $d_\omega$, which allows for dissipation in the cavity walls together with thermal noise in the system,
\be
H_\text{int}^\text{loss} = i \sqrt{\kappa_\ell}\int \frac{d\omega}{2\pi} \,\big(c^\dag d_\omega - c d_\omega^\dag\big).
\ee
With this additional line, we obtain the final form of the frequency-domain input-output relation,
\be \label{eq:freq_output_noise_DM}
b_\tout[\omega] = \left(1-\frac{\kappa_m}{\kappa_\text{tot}/2-i\Delta_{\omega}}\right)b_\tin[\omega] - \frac{\sqrt{\kappa_m\kappa_\ell}}{\kappa_\text{tot}/2 -i\Delta_{\omega}}d_\tin[\omega]- \frac{g\sqrt{\kappa_m}}{\kappa_\text{tot}/2 -i\Delta_{\omega}}a_\eff[\omega],
\ee
where the total cavity decay rate $\kappa_\text{tot} = \kappa_m + \kappa_\ell$ controls the width of the cavity response.

Decay rates are also commonly expressed in terms of dimensionless coupling quality factors; see e.g.~Ref.~\cite{Brubaker:2017ohw}.
The intrinsic quality factor is $Q_{\ell} = \omega_c/\kappa_{\ell}$, the measurement port's coupling quality factor is $Q_m = \omega_c/\kappa_m$, and the loaded quality factor is $Q_c = \omega_c/\kappa_\text{tot} = (1/Q_{\ell} + 1/Q_m)^{-1}$.
Here, $\kappa_{\ell}$ is determined by the cavity material and geometry, while $\kappa_m$ depends on the adjustable strength of the transmission line and cavity coupling.
Defining $\beta = \kappa_m/\kappa_{\ell}$, cavities commonly operate near $\beta = 2$, which maximizes the scan rate at the standard quantum limit.
The cavity ring-up time $t_m$ used in Eq.~\eqref{eq:eta_value} is related to the decay rate by $t_m = 1/\kappa_\text{tot}$.

%%%%%%%%%%%%%%%%%%%%%%%%%%%%%%%%%
\paragraph{Transmission-Line Input States.}
%%%%%%%%%%%%%%%%%%%%%%%%%%%%%%%%%
%
To compute the expected distribution of observables, we must specify the states of the input modes.
The input modes in the loss port represent thermal noise from the cavity walls, so we take them to be thermal states.
Their second moments are\footnote{Our convention for the Hermitian conjugate of a Fourier transform of an operator is
\[ 
\mathcal{O}^\dag[\omega] \equiv \qty[ \int d t\, \mathcal{O}(t) e^{i \omega t} ]^\dag = \int d t\, \mathcal{O}^\dag(t) e^{-i \omega t}.
\]
Consequently, the Fourier transform of $\mathcal{O}^\dag(t)$ is $\mathcal{O}^\dag[-\omega]$.}
\be\label{eq:loss_input_noise_number_mean}
\la d^\dag_\tin[\omega] d_\tin[\omega']\ra = n_\ell[\omega]\,2\pi \deltafunc(\omega-\omega'),
\ee
where the number flux per frequency $n_\ell[\omega]$ vanishes at fictitious negative frequencies, and follows a Bose--Einstein distribution at positive frequencies~\cite{Clerk:2008tlb,Beckey:2023shi}, 
\be \label{eq:n_Bose_Einstein}
n_\ell[\omega] = \frac{\Theta(\omega)}{e^{\omega/T_{\ell}} - 1},
\ee
with $\Theta(\omega)$ the Heaviside step function.

The input modes for the measurement port can in principle be chosen, e.g.~one could take a squeezed state to achieve sensitivity beyond the standard quantum limit.
For simplicity, we take the measurement port input modes to also be in a thermal state,
\be \label{eq:input_noise_number_mean}
\la b^\dag_\tin[\omega] b_\tin[\omega']\ra = n_m[\omega]\,2\pi \deltafunc(\omega-\omega'), 
\ee
with $n_m[\omega]$ again given by the Bose--Einstein distribution Eq.~\eqref{eq:n_Bose_Einstein} with temperature $T_m$.
(When $T_m \to 0$, the measurement port input modes are in vacuum states.)
Equivalently, we are taking the input modes in the measurement and loss ports to have thermal, Gaussian $P$-functions that factorize in frequency space.

%%%%%%%%%%%%%%%%%%%%%%%%%%%%%%%%%
\paragraph{Axion Input State.}
%%%%%%%%%%%%%%%%%%%%%%%%%%%%%%%%%
%
Applying the definition of $a_\eff(t)$ in Eq.~\eqref{eq:free_effective_mode} we can compute
\begin{align}
\la a^\dag_\eff[\omega] a_\eff[\omega']\ra &= \int dt \, e^{-i\omega t}\int dt'\, e^{i\omega't'} \la a^\dag_\eff(t)a_\eff(t')\ra \\
&= \frac{V_c}{4\Omega}\int \frac{d^3\bk}{(2\pi)^3}\int \frac{d^3\bq}{(2\pi)^3}\, C^*(\bk)C(\bq)\la a^\dag_\bk a_\bq\ra \int dt\, e^{-i(\omega-\omega_\bk)t}\int dt'\, e^{i(\omega'-\omega_\bq)t'}.\nonumber
\end{align}
To simplify this expression we invoke our first assumption regarding the DM state; specifically we take the axion to be homogeneous.
This implies momentum conservation, which diagonalizes the correlators, allowing us to write $\la a^\dag_\bk a_\bq\ra = N_\bk\,(2\pi)^3\delta^{(3)}(\bk-\bq)$; see Ref.~\cite{suppression_paper} for further discussion of this point.
The correlator then simplifies to
\be \label{eq:input_DM_number_mean}
\la a^\dag_\eff[\omega] a_\eff[\omega']\ra = \frac{V_c}{4\Omega}\int \frac{d^3\bk}{(2\pi)^3}|C(\bk)|^2 N_\bk \, 2\pi\delta(\omega-\omega_\bk)2\pi\delta(\omega-\omega').
\ee
As expected, no negative-$\omega$ modes are occupied, since $\omega_\bk\geq m_a$.

Since the input modes of the three ports (loss, measurement, and axion) have independent origins, we assume throughout that they are uncorrelated (i.e.~in a product state).

%%%%%%%%%%%%%%%%%%%%%%%%%%%%%%%%%
\subsection{Quadrature and Number Observables}
\label{subsec:OutputObservables}
%%%%%%%%%%%%%%%%%%%%%%%%%%%%%%%%%

We now turn to measurement.
In this subsection we analyze quadrature and number (photon counting) measurements.
We construct the signal power from both the quadrature PSD and the photon count rate, in analogy to Sec.~\ref{subsec:twopowers}, and evaluate them in the standard halo model.

%%%%%%%%%%%%%%%%%%%%%%%%%%%%%%%%%
\paragraph{Idealized Measurements.} 
%%%%%%%%%%%%%%%%%%%%%%%%%%%%%%%%%
%
To fully describe an experiment's observed results, we must model the detector at the end of the transmission line.
In practice this could be, e.g., a microwave amplifier, single-photon counter, or bolometer.
However, explicitly modeling the detector would require introducing additional parameters involving, e.g., the gain, bandwidth, and noise temperature of an amplifier.

Thus, for simplicity we idealize the detector as performing projective measurements on the temporal modes, which arrive at the detector one after the other.
For quadrature measurements, we consider measuring 
\be
X_\tout(t) = \frac{1}{\sqrt{2}}\Bigl( b_\tout^\dag(t)+b_\tout(t)\Bigr),
\ee
while for number measurements, we consider measuring the output number flux 
\be
n_\tout(t) = b_\tout^\dag(t)b_\tout(t).
\ee
Using the locality of the output operators from Eq.~\eqref{eq:b_out_commutator}, it follows that
\be \label{eq:X_n_commutators}
[X_\tout(t),X_\tout(t')] = [n_\tout(t),n_\tout(t')] = 0,
\ee
and thus measurement outcomes of either $X_\tout$ or $n_\tout$ at successive times are all simultaneously observable.
We can then take arbitrary functions of the raw data $X_\tout(t)$ or $n_\tout(t)$ to form macroscopic observables, like the Fourier transform, PSD, and total averaged power.

However, this idealization comes at a price: the unphysical negative-frequency modes, which do not affect real detectors, can contribute to these idealized measurements.
Despite this, in what follows we show that our formalism is still a reasonable model for certain types of detector.

%%%%%%%%%%%%%%%%%%%%%%%%%%%%%%%%%
\paragraph{Quadrature and PSD Measurements.}
%%%%%%%%%%%%%%%%%%%%%%%%%%%%%%%%%
%
Equation~\eqref{eq:X_n_commutators} justifies that we can measure the output quadrature at different times and construct various derived observables from these measurements.
As a first example, if we continuously monitor the system for an integration time $T$, we can define the windowed Fourier transform by\footnote{Although $X_T[\omega] \neq (X_T[\omega])^\dag$, the Fourier transform is a valid observable as it is \textit{constructed} from Hermitian observables.
The non-Hermiticity here only represents an experimentalist multiplying each (quantum) measurement outcome with a (classical) phase, and then adding the results up.}
\be \label{eq:XT_def}
X_T[\omega] \equiv \int_0^T \!dt\, X_\tout(t)\, e^{i\omega t}.
\ee
From this, we define the windowed PSD of the quadrature as
\be
S[\omega] \equiv \frac{1}{T} \, X^\dag_T[\omega] X_T[\omega],
\ee
where again $X^\dag_T[\omega] = (X_T[\omega])^\dag$.

In terms of the time-domain observables, the mean observed PSD is
\begin{align}
\la S[\omega]\ra 
&= \frac{1}{T}\int_0^T\!dt\int_0^T\!dt'\, \la X_\tout(t)X_\tout(t')\ra e^{i\omega(t'-t)} \\
&= \frac{1}{T}\int_0^T\!dt\int_{-t}^{T-t}\!d\tau\, \la X_\tout(t)X_\tout(t+\tau)\ra e^{i\omega \tau},
\end{align}
where in the second step we changed variables to $\tau = t' - t$.
Now, the thermal states we adopted for the measurement and loss ports are stationary.
With all input modes independent, if we further assume that the axion field is stationary, then the same becomes true of the output modes, implying we can rewrite the correlation function in the integrand using
\be
\la X_\tout(t)X_\tout(t+\tau)\ra = \la X_\tout(0)X_\tout(\tau)\ra .
\ee
From here, the expression can be rearranged as
\bea \label{eq:partial_FT}
\la S[\omega]\ra &= \frac{1}{T}\int_0^T\!dt \int_{-t}^{T-t}\!d\tau\, \la X_\tout(0)X_\tout(\tau)\ra \,e^{i\omega\tau} \\
&=\int_{-T}^T\! d\tau \left(\frac{1}{T}\int_0^T\! dt \, \Theta(t+\tau)\Theta(T-\tau-t) \right)\la X_\tout(0)X_\tout(\tau)\ra \,e^{i\omega\tau} \\
&=\int_{-T}^T\!d\tau \left(1-\frac{|\tau|}{T} \right) \la X_\tout(0)X_\tout(\tau)\ra \,e^{i\omega\tau}.
\eea
If the signal has a finite linewidth, then it has a finite coherence time $\tau_\tout$.
The autocorrelation function will then eventually decay rapidly as $\tau$ becomes large; see App.~\ref{app:coherence_time} and Ref.~\cite{Cheong:2024ose} for further discussion.
In the limit $T \gg \tau_\tout$, the autocorrelation function decays faster than the kernel $(1-|\tau|/T)$, and we can approximate the expression by a proper Fourier transform, 
\be\label{eq:PSDasFourier}
\la S[\omega]\ra = \int_{-\infty}^{\infty} \!d\tau \, \la X_\tout(0)X_\tout(\tau)\ra \, e^{i\omega\tau}.
\ee
Therefore, for $T \gg \tau_\tout$, the mean PSD is the Fourier transform of the autocorrelation function of the individual quadrature measurements.
This is exactly the Wiener--Khinchin theorem, with the above derivation demonstrating we can use this result when computing the PSD for a sufficiently long continuous measurement.

This result is crucial, as the Wiener--Khinchin theorem also implies the relation
\be
2\pi \deltafunc(\omega-\omega') \la S[\omega]\ra = \la X^\dag_\tout[\omega'] X_\tout[\omega]\ra,
\ee
where $X_\tout[\omega]$ is the Fourier transform of its time-domain counterpart (cf.~Eq.~\eqref{eq:XT_def}),
\be
X_\tout[\omega] = \int_{-\infty}^{\infty}\! dt \, X_\tout(t) \, e^{i\omega t}.
\ee
We can express this result in terms of $b_\tout[\omega]$,
\bea \label{eq:PSD_prestationarity}
2\pi \deltafunc(\omega-\omega')\la S[\omega]\ra = \frac{1}{2}&\Bigl\la b_\tout^\dag[\omega']b_\tout[\omega] + b_\tout^\dag[\omega']b_\tout^\dag[-\omega] \\
&\quad + b_\tout[-\omega']b_\tout[\omega] +b_\tout[-\omega']b_\tout^\dag[-\omega]\Bigr\ra.
\eea
We can then use Eq.~\eqref{eq:freq_output_noise_DM} to relate the output modes to the input modes.

%%%%%%%%%%%%%%%%%%%%%%%%%%%%%%%%%
\paragraph{Evaluating the PSD.}
%%%%%%%%%%%%%%%%%%%%%%%%%%%%%%%%%
%
To evaluate the PSD, we first note that our assumptions on the input modes imply that the middle two terms in Eq.~\eqref{eq:PSD_prestationarity} vanish.
To see this, our independence assumption implies that expectation values such as $\la b_\tin[-\omega'] d_\tin[\omega] \ra$ vanish.
Next, the measurement and loss ports are assumed to be in thermal Gaussian states.
They are diagonal in the Fock basis, which implies $\la b_\tin[-\omega'] b_\tin[\omega] \ra = \la d_\tin[-\omega'] d_\tin[\omega] \ra = 0$.
Finally, for DM the condition of stationarity implies energy conservation, which for a nonrelativistic state can only be satisfied with an equal number of creation and annihilation operators.\footnote{As an explicit example, stationarity applied to $\la a_1^\dag a_2 a_3 \ra$ requires either the correlator vanishes or $\omega_1 = \omega_2 + \omega_3$.
The latter condition cannot be satisfied with a nonrelativistic dispersion, implying all such correlators vanish for stationary DM.
Note the coherent-state hypothesis in Eq.~\eqref{eq:CSH} is not stationary, although as each mode has a separate random phase, unbalanced correlators also vanish.}
As such, expectation values with an unequal number of creation and annihilation operators vanish.
Discarding these terms and using $[b_\tout[\omega],b^\dag_\tout[\omega']]=2\pi \delta(\omega-\omega')$, we have
\be \label{eq:PSD_penultimate}
2\pi \deltafunc(\omega-\omega')\la S[\omega]\ra = \frac{1}{2}\Bigl\la b_\tout^\dag[\omega']b_\tout[\omega] + b^\dag_\tout[-\omega]b_\tout[-\omega']\Bigr\ra + \pi\deltafunc(\omega-\omega').
\ee
Using the input-output relation in Eq.~\eqref{eq:freq_output_noise_DM}, we find
\begin{align}
2\pi \deltafunc(\omega-\omega')\la S[\omega]\ra &= \pi\deltafunc(\omega-\omega') \nonumber \\
&+\frac{1}{2}\left[ \left(1-\frac{\kappa_m\kappa_\ell}{(\kappa_\text{tot}/2)^2 + (\omega-\omega_c)^2}\right)\Bigl\la b^\dag_\tin[\omega'] b_\tin[\omega]\Bigr\ra + (\omega^{(\prime)} \!\to {-}\omega^{(\prime)}) \right] \nonumber \\
& +\frac{1}{2}\left[ \frac{\kappa_m \kappa_\ell}{(\kappa_\text{tot}/2)^2 + (\omega - \omega_c)^2}\Bigl\la d^\dag_\tin[\omega'] d_\tin[\omega]\Bigr\ra + (\omega^{(\prime)} \!\to {-}\omega^{(\prime)})\right] \\
& +\frac{1}{2}\left[ \frac{g^2\kappa_m}{(\kappa_\text{tot}/2)^2 + (\omega - \omega_c)^2}\Bigl\la a^\dag_\eff[\omega'] a_\eff[\omega]\Bigr\ra + (\omega^{(\prime)} \!\to {-}\omega^{(\prime)})\right]\!.\nonumber
\end{align}
The final expression in each bracket indicates the same expression but with the sign of $\omega$ and $\omega'$ inverted.
We also simplified the prefactors using the fact that each expectation value on the right-hand side is proportional to $\delta(\omega-\omega')$.
Finally, by integrating both sides against $\omega'$ and applying the mode two-point functions in Eqs.~\eqref{eq:loss_input_noise_number_mean},~\eqref{eq:input_noise_number_mean}, and \eqref{eq:input_DM_number_mean}, we find the PSD\footnote{Note that this PSD is two-sided and symmetric.
This choice is purely conventional; see Ref.~\cite{Cheong:2024ose} for more discussion on one-sided versus two-sided PSDs.
Asymmetric PSDs can arise in quantum measurement theory~\cite{Clerk:2008tlb}, but do not appear here because our PSD is constructed from commuting observables $X_\tout(t)$.}
\bea \label{eq:PSD_mean}
\la S[\omega]\ra = \frac{1}{2} &+\frac{1}{2}\left[ \left(1-\frac{\kappa_m\kappa_\ell}{(\kappa_\text{tot}/2)^2 + (\omega-\omega_c)^2}\right) n_m[\omega] +(\omega \to -\omega)\right] \\
&+\frac{1}{2}\left[ \frac{\kappa_\ell \kappa_m}{(\kappa_\text{tot}/2)^2 + (\omega-\omega_c)^2} n_\ell[\omega] + (\omega \to -\omega)\right] \\
&+\frac{1}{2}\left[ \frac{g^2\kappa_m}{(\kappa_\text{tot}/2)^2 + (\omega-\omega_c)^2}\neff[\omega] + (\omega \to -\omega)\right]\!,
\eea
in which we defined the DM effective mode's occupation number per angular frequency
\be
\label{eq:n_eff}
\neff[\omega] \equiv \frac{V_c}{4\Omega}\int \frac{d^3\bk}{(2\pi)^3}\, |C(\bk)|^2 N_\bk\, 2\pi\delta(\omega-\omega_\bk).
\ee
The DM occupancy satisfies $\int \frac{d\omega}{2\pi}\, \neff[\omega] = N_\eff$, consistent with Eq.~\eqref{eq:Neff-DM} and the exact form provided in Ref.~\cite{suppression_paper}.
If the effective mode seen by the detector follows the virialized Gaussian hypothesis of Eq.~\eqref{eq:VGH}, then dividing the frequency range into bins, the mode in each bin is Gaussian with an occupancy set by integrating $\neff[\omega]$ over the bin.
Summing the occupancies over each bin then recovers $N_\eff$.

In the PSD, the measurement, loss, and signal ports each contribute via the mean PSD of each port weighted by an appropriate susceptibility factor.
This is a known result, matching Eq.~(A12) of Ref.~\cite{Malnou:2018dxn} up to conventions.
What is novel is that we have treated DM quantum mechanically, and shown how it enters through the properties of its effective mode.
Note that the ratio of thermal to vacuum noise is the thermal occupancy $n_{\mathrm{th}} = (e^{\omega_c / T} - 1)^{-1}$, rather than the expected $2 n_{\mathrm{th}}$.
This is due to negative-frequency modes, as we discuss below.

\begin{figure}[!t]
\centering
\includegraphics[width=0.47\linewidth]{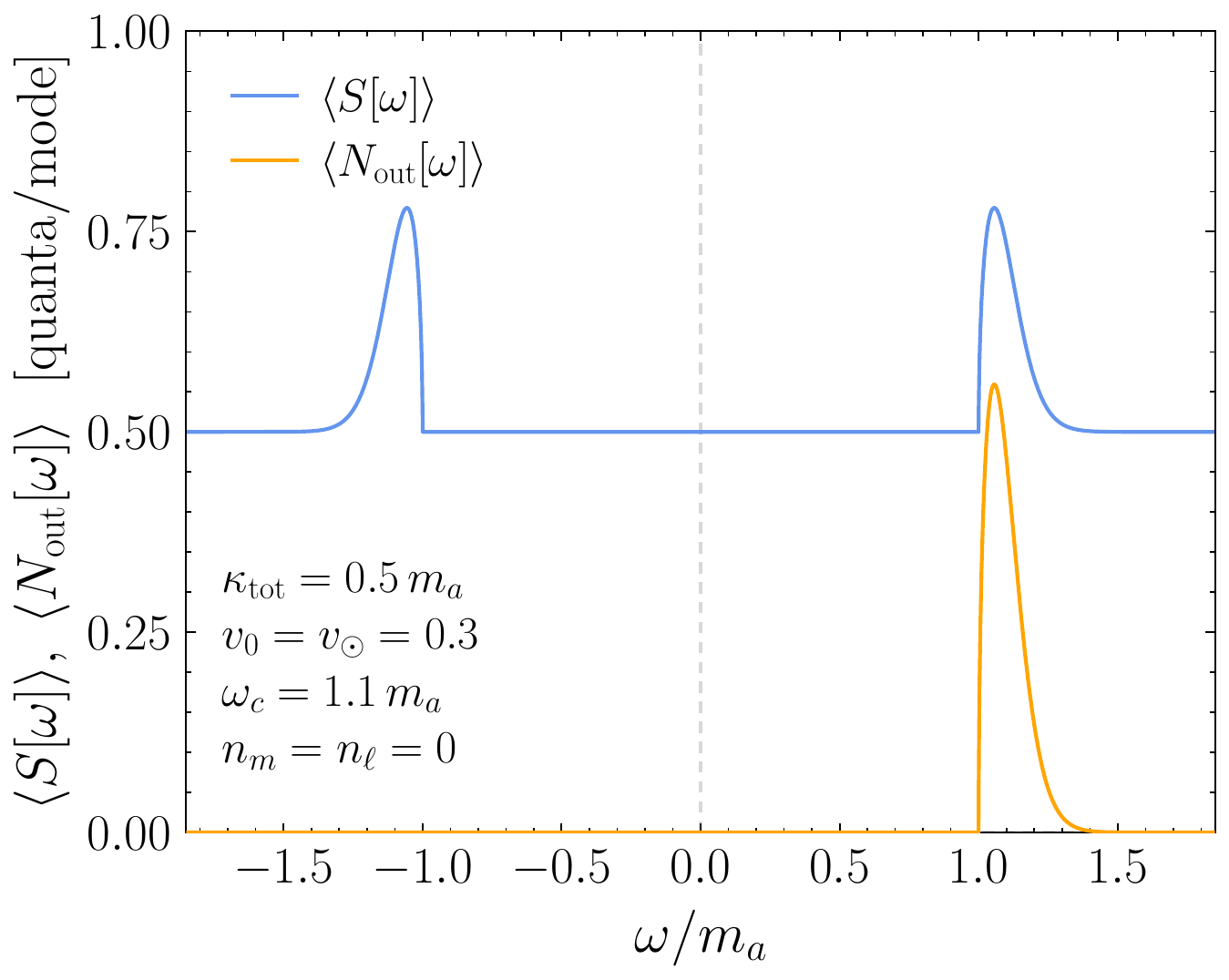}
\vspace{-0.3cm}
\caption{Schematic comparison of the expectation values of the quadrature PSD (blue) and the differential photon-number spectrum (orange) assuming the cavity is driven solely by the axion field, with no thermal noise.
Though the PSD $S[\omega]$ is two-sided and has a noise floor due to vacuum fluctuations, its expectation value carries the same information as $\la N_\tout[\omega] \ra$.
Yet as we discuss in Sec.~\ref{subsec:higherOrderIO}, the fluctuations of these observables can differ significantly.}
\label{fig:SvsN_Input_Output}
\end{figure}

%%%%%%%%%%%%%%%%%%%%%%%%%%%%%%%%%
\paragraph{Photon-Counting Measurement.}
%%%%%%%%%%%%%%%%%%%%%%%%%%%%%%%%%
%
We can also measure number instead of quadrature.
In particular, the total photon count collected in a time $T$ is described by the operator
\be
N_\tout \equiv \int_0^T \!d t \, b_\tout^\dag(t) b_\tout(t),
\ee
whose expectation value is 
\be
\la N_\tout \ra = \int_0^T \!dt\, \left\la\left( \int\frac{d\omega}{2\pi} \, b^\dag_\tout[\omega] e^{i\omega t}\right)\left( \int\frac{d\omega'}{2\pi} \, b_\tout[\omega'] e^{-i\omega' t}\right)\right\ra\!.
\ee
We evaluate this expectation value in a similar manner to how we arrived at Eq.~\eqref{eq:PSD_mean}.
Equations~\eqref{eq:loss_input_noise_number_mean}, \eqref{eq:input_noise_number_mean} and \eqref{eq:input_DM_number_mean} yield factors of $\delta(\omega - \omega')$, which we use to evaluate the $\omega'$ integral, leaving a trivial integral over time.
The final result is
\begin{align} \label{eq:mean_avg_count_rate}
\la N_\tout \ra = T \int\frac{d\omega}{2\pi}\Biggl[\left(1-\frac{\kappa_m\kappa_\ell}{(\kappa_\text{tot}/2)^2 + (\omega-\omega_c)^2}\right)&n_m[\omega] + \frac{\kappa_m\kappa_\ell}{(\kappa_\text{tot}/2)^2 + (\omega-\omega_c)^2} n_\ell[\omega] \nonumber \\
+\frac{g^2\kappa_m}{(\kappa_\text{tot}/2)^2 + (\omega-\omega_c)^2}&\neff[\omega]\Biggr].
\end{align}
Defining the quantity in brackets as $\la N_\tout[\omega] \ra$, we have
\be \label{eq:psd_discussion}
2 \la S[\omega] \ra = \la N_\tout[\omega] \ra + \la N_\tout[-\omega] \ra + 1
\ee
as we show explicitly in Fig.~\ref{fig:SvsN_Input_Output}, in the idealized case of no thermal noise.
This confirms the classical intuition, discussed in Sec.~\ref{subsec:twopowers}, that the mean quadrature PSD and the mean number spectrum contain equivalent information.
However, $S[\omega]$ and $N_\tout[\omega]$ correspond to distinct observables with distinct fluctuations, as we discuss further in Sec.~\ref{subsec:higherOrderIO}.

%%%%%%%%%%%%%%%%%%%%%%%%%%%%%%%%%
\paragraph{Effect of Negative-Frequency Modes.}
%%%%%%%%%%%%%%%%%%%%%%%%%%%%%%%%%
%
If the PSD of Eq.~\eqref{eq:PSD_mean} is folded to a one-sided PSD, then the quantum noise floor is $+1$ at all frequencies, double the naive expectation.
This extra noise has the same origin as that discussed in Sec.~\ref{subsec:twopowers}.
In reality, the output modes $X_\tout(t)$ do not all commute; we rendered them commuting by introducing negative-frequency modes.
For each positive-frequency mode, the negative-frequency counterpart plays an equivalent role to the vacuum mode in the heterodyne measurement discussed in Sec.~\ref{subsec:twopowers}, rendering measurements compatible but adding additional quantum noise.
Mechanically, the way the noise term arises in Eq.~\eqref{eq:PSD_penultimate} is precisely analogous to Eq.~\eqref{eq:S_Z_expec}.

Negative-frequency modes do not affect our discussion of number measurements, because we took them to have zero occupancy.
However, the subtlety above implies that our idealized quadrature measurements only describe certain types of real quadrature measurements.
For example, the ADMX experiment uses a nondegenerate parametric amplifier to amplify both quadratures of the signal, which can subsequently be measured classically.
The amplifier only effectively acts on transmission-line modes near $+ \omega_c$, so the negative-frequency modes become irrelevant (removing the extra $+1/2$ of quantum noise from them), but the amplifier itself introduces $+1/2$ of quantum noise, as required by general principles~\cite{PhysRevD.26.1817}.
Thus, our PSD correctly models what is measured in an idealized version of the ADMX experiment, even though the negative-frequency modes are unphysical.

However, the HAYSTAC experiment uses a degenerate parametric amplifier to amplify only a single quadrature.\footnote{For more discussion of the virtues of single-quadrature vs.~two-quadrature measurements, see Refs.~\cite{AlKenany:2016trt,Malnou:2018dxn}.} 
This removes the $+1/2$ from negative-frequency modes without introducing an additional $+1/2$.
To realize this in our idealized model of quadrature measurements, one could remove the negative-frequency contribution by fiat, but a fully rigorous resolution would require abandoning negative-frequency modes and modeling the amplifier.

%%%%%%%%%%%%%%%%%%%%%%%%%%%%%%%%%
\paragraph{Signal Power from Axion DM.}
%%%%%%%%%%%%%%%%%%%%%%%%%%%%%%%%%
%
As an illustrative example, we explicitly compute the signal power of axion DM.
Consider Eq.~\eqref{eq:n_eff}, which enters into the final mean PSD and photon number measurements.
There, $N_\bk$ is the mean axion occupancy per momentum mode $\bk$, which for DM takes the form $N_\bk = (2\pi)^3\, \bar{n}\, p(\bk)$~\cite{Cheong:2024ose}, with $p(\bk)$ the DM momentum distribution and $\bar{n}$ the number density of axions.
The local DM distribution is largely nonrelativistic, and a representative ansatz is the standard halo model,
\be \label{eq:pk_SHM}
p(\bk) = \frac{1}{(\sqrt{\pi} m_a v_0)^3} \exp{-\frac{\qty(\bk + m_a \vb{v}_\odot)^2}{m_a^2 v_0^2}},
\ee
where the DM velocity dispersion is $v_0 \simeq 220\,\text{km/s}$ and the solar speed is $v_\odot \simeq 232\,\text{km/s}$.
Since these quantities are comparable, we define the typical velocity scale $v \sim v_0 \sim v_\odot \sim 10^{-3}$.
For a generic cavity shape, we have $V_c \sim m_a^{-3}$ if the cavity is on the axion resonance, so that the typical axion momentum is $k \sim m_a v \ll V_c^{-1/3}$.
For these momenta, the cavity form factor in Eq.~\eqref{eq:formfactor} satisfies $C(\bk) \simeq C(\vb{0})$.
Equation~\eqref{eq:n_eff} then simplifies to
\be \label{eq:n_eff_NRDM}
\neff[\omega] \simeq \left( \frac{\pi |C(\vb{0})|^2}{2 \Omega} \right) \bar{n} V_c\, p[\omega].
\ee
Using the explicit expressions for the TM$_{010}$ mode of a cylindrical cavity given in Ref.~\cite{suppression_paper}, the expression in parentheses is $\simeq 5.9$.
Above, we have also defined the DM energy distribution
\bea \label{eq:pomega}
p[\omega] &\equiv \int\! d^3 \bk\, p(\bk) \delta(\omega - \omega_\bk) \\
&= \frac{2\omega}{\sqrt{\pi} m_a^2 v_0 v_\odot} \exp{- \frac{v_\odot^2}{v_0^2} -\frac{\omega^2 - m_a^2}{m_a^2 v_0^2}} \sinh(\frac{2v_\odot}{v_0^2} \frac{\sqrt{\omega^2 - m_a^2}}{m_a} ) \Theta(\omega - m_a).
\eea
The second line holds for the standard halo model, and has width $\sim m_a v^2$ as expected.
It is common to express this width in terms of a dimensionless quality factor, $Q_a \sim v^{-2} \sim 10^6$.\footnote{This quality factor should be distinguished from those for the measurement and loss ports, $Q_m$ and $Q_{\ell}$, introduced below Eq.~\eqref{eq:freq_output_noise_DM}.
Those expressions were tied to the cavity loss rate, for instance $Q_m = \omega_c/\kappa_m$.
The axion quality factor solely captures the fractional DM linewidth and it is independent of the coupling to an experiment.
See App.~\ref{app:weakAxionCoupling} for further discussion.}

The above result is sufficient to evaluate the DM contribution to the PSD, $\la S[\omega] \ra$ in Eq.~\eqref{eq:PSD_mean}.
However, to determine the contribution to the expected photon count, $\la N_\tout \ra$ in Eq.~\eqref{eq:mean_avg_count_rate}, we need to evaluate the equivalent expression weighted against the cavity response and integrated over all frequencies.
This gives the axion-induced signal-photon count rate,
\bea \label{eq:Gamma_sig_def}
\Gamma_\sig &\equiv \int \frac{d\omega}{2\pi} \frac{g^2\kappa_m}{(\kappa_\text{tot}/2)^2 + (\omega-\omega_c)^2} \neff[\omega] \\
&\simeq \frac{\pi}{2} \ga^2 B_0^2 |C(\vb{0})|^2\, \bar{n} V_c\, \frac{\kappa_m}{\kappa_\text{tot}} \int \frac{d\omega}{2\pi} \frac{\kappa_\text{tot}}{(\kappa_\text{tot}/2)^2 + (\omega-\omega_c)^2}\, p[\omega]
\eea
where in the second line we used Eq.~\eqref{eq:n_eff_NRDM} and the definition of $g$ from Eq.~\eqref{eq:Hint_simplified}.
The remaining integrand is a product of two narrow distributions: the axion energy distribution, of width $m_a v^2$, and the cavity response, of width $\kappa_\text{tot}$.

In existing axion cavity haloscopes, the cavity has a larger width than DM so that $\kappa_\text{tot} \gg m_a v^2$, or near resonance $Q_a \gg Q_c$.
Then for the standard halo model, we have
\be \label{eq:GammaSig_widecavity}
\Gamma_\sig
\simeq \frac{1}{4} \ga^2 B_0^2 |C(\vb{0})|^2\, \bar{n} V_c \frac{\kappa_m}{(\kappa_\text{tot}/2)^2 + \Delta_a^2} \!\left[ 1 - \frac{m_a \Delta_a}{(\kappa_\text{tot}/2)^2 + \Delta_a^2} \!\left(v_\odot^2 + \frac{3}{2} v_0^2\right)\! + \mathcal{O}(v^4) \right]
\ee
as shown in App.~\ref{app:deltaFunctionExpansion}, where the detuning $\Delta_a = m_a-\omega_c$ is now between the axion and cavity; cf.~Eq.~\eqref{eq:first_bout_eq}.
In that appendix, we also show that in the narrow-cavity regime $\kappa_\text{tot} \ll m_a v^2$ ($Q_c \gg Q_a$ near resonance),
\be \label{eq:GammaSig_narrowcavity}
\Gamma_\sig \simeq \frac{\pi}{2} \ga^2 B_0^2 |C(\vb{0})|^2\, \bar{n} V_c \frac{\kappa_m}{\kappa_\text{tot}} 
\left[ p[\omega_c] - \frac{\kappa_\text{tot}}{2} \frac{d}{d\omega_c} \left( \mathrm{H}(p)[\omega_c] \right) + \mathcal{O}\!\left( \frac{\kappa_\text{tot}^2}{m_a^2 v^4} \right) \right]\!,
\ee
where the Hilbert transform of $p[\omega_c]$ is defined as
\be
\mathrm{H}(p)[\omega_c] = \frac{2}{\pi m_a v_0 v_{\odot}} \left[ \mathrm{F} \!\left( \frac{v_c[\omega_c] - v_{\odot}}{v_0} \right)\! - \mathrm{F} \!\left( \frac{v_c[\omega_c] + v_{\odot}}{v_0} \right) \right]\!,
\ee
with $v_c[\omega_c] = \sqrt{2(\omega_c/m_a-1)}$ the axion velocity at the cavity frequency, and $\mathrm{F}(x)$ the Dawson integral.
When the axion is near resonance with the cavity, the leading-order contribution to both results is
\be \label{eq:Gamma_sig}
\Gamma_\sig
\simeq \ga^2 B_0^2 |C(\vb{0})|^2\, \frac{\bar{n} V_c}{m_a} \frac{\beta}{1+\beta} \min(Q_c, Q_a),
\ee
where we introduced the cavity coupling parameter $\beta$ discussed below Eq.~\eqref{eq:freq_output_noise_DM}.
This is related to the power the axion delivers to the measurement port by $P_\sig \simeq m_a \Gamma_\sig$.
To relate this to the notation in Ref.~\cite{suppression_paper}, note that for $Q_c \lesssim Q_a$ (relevant for existing haloscopes), 
\be \label{eq:Gamma_sig_etaNeff}
\Gamma_\sig \sim \frac{\eta N_\eff}{t_m},
\ee
where we used Eqs.~\eqref{eq:Neff-DM} and \eqref{eq:eta_value} and dropped the order-one ratio involving $\beta$.

%%%%%%%%%%%%%%%%%%%%%%%%%%%%%%%%%
\subsection{Higher-Order Statistics}
\label{subsec:higherOrderIO}
%%%%%%%%%%%%%%%%%%%%%%%%%%%%%%%%%

So far, we have computed the mean of the quadrature PSD $S[\omega]$ and the total photon count $N_\tout$.
Statistical analyses require understanding how the data fluctuates around the expected value, and this is only fixed once we specify the DM $P$-function $P_a$.
Here we show those statistical fluctuations can be computed for general DM states, such that the data can be used to constrain different hypotheses for $P_a$.
The ultimate goal of this subsection is to show that one can distinguish the predictions of the virialized Gaussian hypothesis from the naive coherent-state expectation, both discussed in Sec.~\ref{subsec:DMQstate}.

For concreteness we focus on $N_\tout[\omega]$, the photon count in a single frequency bin; cf.~Fig.~\ref{fig:SvsN_Input_Output} and Eq.~\eqref{eq:psd_discussion}.
In a typical photon-counting experiment, frequency information is discarded, but a frequency-resolved photon-counting measurement can in principle be implemented using frequency filters on the output.
As we discuss, $N_\tout[\omega]$ is related to the quadrature PSD $S[\omega]$, though the latter is continuous-valued and also has contributions from vacuum fluctuations.
Further, $N_\tout[\omega]$ and $S[\omega]$ will in general have different frequency resolutions tracing to the particular way each is measured.
After setting up the calculation using discrete-frequency modes, we show how to compute the variance of the total photon count $\var(N_\tout)$.
We then show how measurements of $\var(N_\tout[\omega])$ can distinguish between different models of DM, and find the integration time required to do so.

%%%%%%%%%%%%%%%%%%%%%%%%%%%%%%%%%
\paragraph{Discrete-Frequency Modes.}
%%%%%%%%%%%%%%%%%%%%%%%%%%%%%%%%%
%
To compute higher moments, it is useful to use discrete-frequency modes.
For instance, to compute the variance of $N_\tout$ using the continuum output mode $b_\tout$, we would need to evaluate
\be \label{eq:Nout_sq}
\la N_\tout^2\ra = \!\int\! \qty[\prod_{n = 1}^4 \frac{d \omega_n}{2\pi}]\! \int_0^T \!\!d t_1 d t_2 e^{i (\omega_1-\omega_2) t_1} e^{i (\omega_3-\omega_4) t_2} \Big\la b^\dag_\tout[\omega_1] b_\tout[\omega_2] b^\dag_\tout[\omega_3] b_\tout[\omega_4] \Big\ra.
\ee
In the limit $T \to \infty$, both the correlator and the time integral yield products of delta functions such as $\delta(\omega_1 - \omega_2) \delta(\omega_3 - \omega_4)$, leading to a singular result.
This is expected; we already saw that $N_\tout$ diverges as $T \to \infty$, because it measures the number of events observed in a time $T$.
Therefore, we expect $\la N_\tout^2 \ra \propto T^2$, with no divergence at finite $T$.
A more serious issue comes from disconnected contributions to the four-point correlator in Eq.~\eqref{eq:Nout_sq}.
Since we have fixed the two-point functions of the input ports (i.e.~in terms of the detector temperature or expected DM occupancy), only the connected four-point function (from $\omega_1 = \omega_2 = \omega_3 = \omega_4$) depends in detail on the DM state.
However, the correlator also includes disconnected pieces, e.g.~proportional to $\delta(\omega_1 - \omega_4) \delta(\omega_2 - \omega_3)$, which also contribute when $\omega_1 = \omega_2 = \omega_3 = \omega_4$.
These pieces are canceled in $\var(N_\tout)$ by contributions from $\la N_\tout \ra^2$ with the cancellation involving subtracting terms proportional to $\delta(0)^2$.
Such subtleties vanish when we move to discrete-frequency modes, where all quantities are manifestly finite.

The discretely sampled input/output operators $\db_{\tin/\tout}(t)$ in Eqs.~\eqref{eq:discrete_b_in} and \eqref{eq:b_out_step} correspond to measurement operators averaged over a finite sampling time $\Delta t$.
Assuming we have $N_s$ samples, the total integration time is $T = N_s \Delta t$.
We can define the discrete-frequency modes $\db[\omega_k]$ using the discrete Fourier transform,
\be \label{eq:DFT}
\db[\omega_k] = \sqrt{\frac{\Delta t}{T}} \sum_{n=0}^{N_s-1} e^{i \omega_k t_n} \db(t_n), \quad
\db(t_n) = \sqrt{\frac{\Delta t}{T}} \sum_{k=0}^{N_s-1} e^{-i \omega_k t_n} \db[\omega_k],
\ee
where $t_n \equiv t_0 + n \Delta t$ and $\omega_k \equiv 2\pi k/T = 2\pi/\Delta t \cdot (k/N_s)$.
The transformed modes also satisfy the canonical commutation relation, as
\be
\comm{\db[\omega_j]}{\db^\dag[\omega_k]}
= \frac{\Delta t}{T} \sum_{m,n} e^{i \omega_j t_m - i \omega_k t_n} \comm{\db(t_m)}{\db^\dag(t_n)}
= \delta_{j,k}.
\ee
Further, as shown in App.~\ref{app:PSDfromDiscreteOp}, the discrete-frequency modes are related to their continuous counterparts in the limit of large $T$ (compared to $\tau_\DM$) and small $\Delta t$ as
\be \label{eq:bfrak_to_bcont}
\db[\omega_k] = \frac{b[\omega_k]}{\sqrt{T}}.
\ee

The above applies immediately to the input and output modes of the measurement port and the input loss port.
The axion signal requires a slightly different treatment since it is regarded as an effective oscillator mode rather than a transmission line.
We define the discrete-frequency modes of the axion as 
\be
\da_\eff(t_n) \equiv a_\eff(t_n), \quad \da_\eff[\omega_k] \equiv \sqrt{\frac{\Delta t}{T}} \sum_{n = 0}^{N_s-1} e^{i \omega_k t_n} \da_\eff(t_n),
\ee
which implies $\da_\eff[\omega_k] = a_\eff[\omega_k]/\sqrt{T \Delta t}$, and $\big\la \da_\eff^\dag[\omega_k] \da_\eff[\omega_k] \big\ra = \neff[\omega_k]/\Delta t$, with $\neff[\omega]$ given in Eq.~\eqref{eq:n_eff}.
Therefore, the discrete-frequency input-output relation is
\be \label{eq:bout_in_relation}
\db_\tout[\omega_k] = \left(1-\frac{\kappa_m}{\kappa_\text{tot}/2-i\Delta_k}\right)\db_\tin[\omega_k] - \frac{\sqrt{\kappa_m\kappa_\ell}}{\kappa_\text{tot}/2 -i\Delta_k} \dl_\tin[\omega_k]- \frac{g\sqrt{\kappa_m \Delta t}}{\kappa_\text{tot}/2 -i\Delta_k} \da_\eff[\omega_k],
\ee
where $\Delta_k = \omega_k - \omega_c$.
This is almost a direct translation of the continuum counterpart in Eq.~\eqref{eq:freq_output_noise_DM}, up to the additional $\Delta t$ dependence in the axion signal.
That term is an artifact of describing the axion using an effective mode and does \textit{not} mean the axion signal goes to zero as $\Delta t \to 0$.
Instead, the $\Delta t$ dependence cancels out in the physical observables below.

%%%%%%%%%%%%%%%%%%%%%%%%%%%%%%%%%
\paragraph{Variance of the Total Photon Count.}
%%%%%%%%%%%%%%%%%%%%%%%%%%%%%%%%%
%
The quadrature PSD and number we considered previously can be expressed in terms of discrete-mode quadrature and number operators:\footnote{The quantity $\dX_\tout[\omega_k]$ is closely analogous to the single-mode quantity $Z$, defined in Eq.~\eqref{eq:defZ_heterodyne}.}
\bea
S[\omega_k] &= \dX^\dag_\tout[\omega_k] \dX_\tout[\omega_k], \quad
\dX_\tout[\omega_k] = \frac{\db_\tout[\omega_k] + \db_\tout^\dag[-\omega_k]}{\sqrt{2}}, \\
N_\tout &= \sum_{k} N_\tout[\omega_k] \equiv \sum_k \db_\tout^\dag[\omega_k] \db_\tout[\omega_k].
\label{eq:Nomegak_def}
\eea
The use of a common set of $\db_\tout[\omega_k]$ for both observables is chosen to highlight their structural similarities.
Nonetheless, there is a key distinction between the two that the above description elides: the frequency resolution, or exact set of discrete $\omega_k$, that each can probe in general will differ.
The difference arises from how the two are measured.
For the quadrature PSD, one repeatedly measures $\dX_\tout(t)$, building up a time series over a total integration time $T$.
Taking the discrete Fourier transform generates an observed $\dX_\tout[\omega_k]$, with frequency measurements spaced by $2\pi/T$; to be explicit, the frequency resolution arises exactly from the phase information in the time domain from the discrete Fourier transform as in Eq.~\eqref{eq:DFT}.
By contrast, $N_\tout[\omega_k]$ \textit{cannot} be constructed from any linear combination of measured $\dX_\tout(t)$ or $n_\tout(t)$ values: there is no phase information in number measurements for the Fourier transform to distill.
Instead, frequency-resolved number measurements require appending a series of frequency filters -- with their own transfer functions and potential noise sources -- to the end of the transmission line.
The resolution of $N_\tout[\omega_k]$ is then set by properties of the filters (although for short integration times the filter resolution degrades).
Note that $\big[N_\tout[\omega_k], N_\tout[\omega_j] \big] = 0$, so that the number in each frequency bin is simultaneously observable.

The variance of the total count can be decomposed into a bin-diagonal variance and an off-diagonal covariance,
\be \label{eq:var-Nout-general}
\var(N_\tout)
= \sum_k \var( N_\tout[\omega_k] )
+ \sum_{k \neq j} \cov( N_\tout[\omega_k], N_\tout[\omega_j]).
\ee
The diagonal contribution can be evaluated as
\be \label{eq:var-diagonal}
\var( N_\tout[\omega_k] )
= \ev{\db^\dag[\omega_k]^2 \db[\omega_k]^2} - \ev{\db^\dag[\omega_k] \db[\omega_k]}^2 + \ev{\db^\dag[\omega_k] \db[\omega_k]},
\ee
where for clarity we have suppressed the subscript on $\db_\tout$.

In general, evaluating the off-diagonal contribution to Eq.~\eqref{eq:var-Nout-general} is more involved.
As we assume the ports are independent, the covariance decomposes into a contribution from DM, the measurement port, and the loss port.
For the final two, we assume a thermal Gaussian state, which is diagonal in the Fock basis and is therefore independent for each frequency mode, implying the covariance vanishes.
The covariance then solely arises from the DM state.
If DM follows the Gaussian model of Eq.~\eqref{eq:VGH}, the argument is identical to that for the Gaussian input ports.
More generally, noting $k \neq j$ and again writing $\db$ for $\db_\tout$, we have
\be
\cov( N_\tout[\omega_k], N_\tout[\omega_j])
= \ev{\db^\dag[\omega_k] \db^\dag[\omega_j] \db[\omega_k] \db[\omega_j]}
- \ev{\db^\dag[\omega_k] \db[\omega_k]} \ev{\db^\dag[\omega_j] \db[\omega_j]}.
\ee
In general this does not vanish.
Nonetheless, for the coherent-state hypothesis in Eq.~\eqref{eq:CSH}, which we apply to the resolved modes $\da_\eff[\omega_k]$, both terms are proportional to $|\beta_k|^2|\beta_j|^2$, implying they cancel against each other and the covariance again vanishes.

%%%%%%%%%%%%%%%%%%%%%%%%%%%%%%%%%
\paragraph{Gaussian-State Number Fluctuations.}
%%%%%%%%%%%%%%%%%%%%%%%%%%%%%%%%%
%
Adopting the virialized Gaussian hypothesis for DM, all input modes are in Gaussian states.
Then, by Wick's theorem and stationarity, the four-point correlator reduces to
\be \label{eq:gaussian_4_point}
\ev{\db^\dag_\tout[\omega_k]^2 \db_\tout[\omega_k]^2} = 2 \ev{\db_\tout^\dag[\omega_k] \db_\tout[\omega_k]}^2\!.
\ee
Combined with Eqs.~\eqref{eq:var-Nout-general} and \eqref{eq:var-diagonal}, this implies
\be
\label{eq:total_count_variance_discrete}
\var(N_\tout)
= \sum_k \qty( \ev{N_\tout[\omega_k]}^2 + \ev{N_\tout[\omega_k]}).
\ee
The first term in the variance reflects classical intensity fluctuations; it is what we would expect in the classical limit where $N_\tout$ is a continuous random variable set by the squared magnitude of a classical amplitude.
The second term is the quantum shot noise; it is also present for the coherent-state DM hypothesis, where the number distribution is Poisson.
More generally, the presence of the first term shows that the number distribution under the virialized Gaussian hypothesis \textit{cannot} be Poisson.
This allows the two states to be distinguished, as we discuss shortly, and we investigate the full number distribution in more detail in Sec.~\ref{subsec:number_measurements}.

%%%%%%%%%%%%%%%%%%%%%%%%%%%%%%%%%
\paragraph{Power from Number vs.~PSD.}
%%%%%%%%%%%%%%%%%%%%%%%%%%%%%%%%%
%
In Sec.~\ref{subsec:twopowers}, we showed that in a toy measurement model, the notions of power one can define from number and quadrature measurements are qualitatively distinct; cf.~Fig.~\ref{fig:SvsN_Input_Output}.
We show this explicitly here, taking the Gaussian DM hypothesis for concreteness.
First, we can relate the number measurement to the power collected at each frequency as
\be \label{eq:discretePOmega}
P[\omega_k] = \frac{\omega_k N_\tout[\omega_k]}{T},
\ee
so that the variance of the total power collected is 
\be \label{eq:var-P-Gauss}
\var(P) = \frac{1}{T^2} \sum_k \omega_k^2 \var(N_\tout[\omega_k]) = \frac{1}{T} \int \frac{d \omega}{2\pi} \qty(\ev{P[\omega]}^2 + \omega \ev{P[\omega]}),
\ee
where in the first step we dropped the off-diagonal contributions as appropriate for a Gaussian state and in the second we converted back to the continuum-mode normalization.\footnote{We can be more explicit regarding the dimensions in the final expression of Eq.~\eqref{eq:var-P-Gauss}.
First, the continuous-mode normalized power $P[\omega] = P^{(c)}[\omega]$ should be interpreted as the power per frequency (with mass dimension 1).
This is to be distinguished from the discrete-mode power $P^{(d)}[\omega_k]$ defined in Eq.~\eqref{eq:discretePOmega}.
In particular, $P^{(c)}[\omega] \sim TP^{(d)}[\omega_k]$ with the identification similar to that in Eq.~\eqref{appeq:continuumVsDiscrete}.
We adopt the same conventions for the continuum versus discrete $P^S$.}
This agrees with the known result for the variance of bolometric measurements in Refs.~\cite{Zmuidzinas:03,Kuo:2024duq}.

We can also construct a notion of frequency-resolved power from quadrature measurements using the PSD, $P^S[\omega_k] = |\omega_k| S[\omega_k]/T$, with the total power resulting from a sum over positive and negative $\omega_k$.
We now show that $P^S \neq P/2$ (recall that $P^S$ is two-sided and $P$ one-sided); in fact, $S[-\omega_k] = S[\omega_k]$.
Explicitly, we have
\be
\ev{S[\omega_k]} = \frac{1}{2} \left[ \ev{\db_\tout^\dag[\omega_k]\db_\tout[\omega_k]} + \ev{\db_\tout[-\omega_k] \db_\tout^\dag[-\omega_k]} \right]\!.
\ee
An analogous computation reveals that for a Gaussian state,
\be \label{eq:psd_var}
\var(S[\omega_k]) = \ev{S[\omega_k]}^2\!.
\ee
Accordingly, moving to a continuum-mode normalization
\be
\var(P^S) = \frac{2}{T} \int \frac{d \omega}{2\pi} \ev{P^S[\omega]}^2,
\ee
which has a distinct structure to Eq.~\eqref{eq:var-P-Gauss} in that the linear power contribution has vanished.

Following the discussion below Eq.~\eqref{eq:total_count_variance_discrete}, the variance here simply reproduces the classical contribution that would arise in a calculation where all commutators are neglected -- for instance, this removes vacuum fluctuations -- and the PSD would be the squared magnitude of a complex Gaussian random number.
It would then be exponentially distributed, as argued for in Refs.~\cite{Cheong:2024ose,Foster:2017hbq}, which implies $\var(S[\omega_k]) = \ev{S[\omega_k]}^2$.
In Sec.~\ref{subsec:quad_meas_stats} we show that in a fully quantum-mechanical calculation where the assumptions of the quantum CLT hold, one finds the full distribution for $S[\omega_k]$ is precisely exponential.

By contrast, $N_\tout[\omega_k]$ takes discrete values and is neither exponentially nor Poisson distributed, given the additional linear term in its variance.
This linear term is relevant in the discussion below, though, of course, it is negligible in the limit of large classical intensity fluctuations, where the two notions of power coincide.
This is the analog of the discussion in Sec.~\ref{subsec:twopowers}, framed in the context of input-output theory.

%%%%%%%%%%%%%%%%%%%%%%%%%%%%%%%%%
\paragraph{Coherent-State Number Fluctuations.}
%%%%%%%%%%%%%%%%%%%%%%%%%%%%%%%%%
%
Above we computed $\var(N_\tout)$ for the Gaussian DM hypothesis.
Here we repeat the calculation for the coherent-state case.
When the DM is not Gaussian, the four-point functions are no longer determined by Wick's theorem from two-point functions.
In particular, for the coherent-state hypothesis in Eq.~\eqref{eq:CSH}, we have
\be \label{eq:aeff_4_point}
\ev{\da^\dag_\eff[\omega_k]^2 \da_\eff[\omega_k]^2} = \ev{\da^\dag_\eff[\omega_k] \da_\eff[\omega_k]}^2
\ee
which differs from the Gaussian-state result in Eq.~\eqref{eq:gaussian_4_point} by a factor of $2$.
For brevity, let
\begin{align}
\la N_\tout^{m+\ell}[\omega] \ra &\equiv \qty(1 - \frac{\kappa_m \kappa_\ell}{(\kappa_\text{tot}/2)^2 + \Delta_{\omega}^2}) n_m[\omega] + \frac{\kappa_\ell \kappa_m}{\qty(\kappa_\text{tot}/2)^2 + \Delta_{\omega}^2} n_\ell[\omega], \\
\chi_a[\omega] &\equiv \frac{g^2 \kappa_m}{(\kappa_\text{tot}/2)^2 + \Delta_{\omega}^2}.
\end{align}
We can then write the number in a frequency bin as 
\be
\ev{N_\tout[\omega]} = \la N_\tout^{m+\ell}[\omega] \ra + \chi_a[\omega] \neff[\omega],
\ee
and, for the coherent-state hypothesis, the variance becomes 
\be \label{eq:countvar_coherent}
\var(N_\tout[\omega]) = \ev{N_\tout[\omega]} \big(\ev{N_\tout[\omega]} + 1 \big) - \chi_a^2[\omega] \neff^2[\omega],
\ee
which is smaller than the Gaussian result that can be inferred from Eq.~\eqref{eq:total_count_variance_discrete} due to the final term; the variance of the DM contribution is exactly that of a Poisson distribution, consistent with our expectation for a coherent state.
Hence, by accumulating sufficient data in spectral measurements, we can distinguish the two hypotheses.

%%%%%%%%%%%%%%%%%%%%%%%%%%%%%%%%%
\paragraph{State Discrimination with Number Measurements.}
%%%%%%%%%%%%%%%%%%%%%%%%%%%%%%%%%
%
Using the results above, we can estimate the time it would take to distinguish between the virialized Gaussian and coherent-state DM hypotheses defined in Sec.~\ref{subsec:DMQstate}.
For simplicity, we take the input measurement and loss ports to be in vacuum.
The mean power, given in Eq.~\eqref{eq:Gamma_sig}, is fixed by the energy in the DM field and basic properties of the axion and detector.
As such, one must use information from higher moments to differentiate the quantum states.
We first show that this can be done in principle by measuring the variance, then consider more powerful observables.

We assume $Q_a > Q_c$, as relevant for currently operating haloscopes, and suppose our frequency filters can resolve the DM signal linewidth into $5$ bins, i.e.~$\Delta \omega \sim m_a/(5Q_a) \sim 1/(5\tau_\DM)$.
A measure of the DM power in the $5$ frequency bins is made in a time $5\tau_\DM$.
So that we have a total of $N_\text{shot}$ DM measurements, we then repeat the procedure $N_\text{shot}/5$ times, implying the total integration time is $T = N_\text{shot} \tau_\DM$.\footnote{The arbitrary choice of $5$ is taken to balance two competing considerations.
On the one hand, we want each measurement to be sufficiently long to ensure the approximation in Eq.~\eqref{eq:bfrak_to_bcont} is accurate.
On the other, resolving the DM signal into a large number of modes would require a complex filtering procedure.
Our results do not change qualitatively if we choose a different resolution scale.}
To estimate the variance of the power, we assume that each of these $N_\text{shot}$ measurements is \textit{independent}.
For the Gaussian hypothesis this is a justified approximation, as studied more carefully in Sec.~\ref{sec:measurement} (see App.~\ref{app:mutual_info} for the technical details).
For the coherent-state hypothesis, this was part of our definition of the model; as discussed in Sec.~\ref{subsec:DMQstate} our goal is to simply demonstrate the Gaussian prediction differs from a naive expectation of how a coherent state behaves.
From here, we can determine how many measurements -- and how long an integration time -- are needed to separate the predictions.

For both the coherent-state and Gaussian models, the variance can be reduced to expressions involving only the mean; see Eqs.~\eqref{eq:gaussian_4_point} and \eqref{eq:aeff_4_point}.
For simplicity, we assume the DM power is distributed equally over the frequency bins (a more accurate treatment does not change the result), and write $N = N_\tout[\omega_k]$.
Then the expected DM signal power within each frequency bin for each shot is
\be \label{eq:mean_inout_distinguish}
\ev{N} = \tau_\DM\, \Gamma_\sig \sim \eta N_\eff (Q_a/Q_c),
\ee
where in the second step we used Eqs.~\eqref{eq:eta_value} and \eqref{eq:Gamma_sig_etaNeff}.
For the parameters currently targeted by ADMX, the mean count is $\mathcal{O}(10^{-2}\text{--}10^{-1})$.
In the calculation below we therefore adopt $\ev{N} \ll 1$; if instead $\ev{N}$ is $\mathcal{O}(1)$ or larger, discriminating the Gaussian and coherent hypotheses becomes far easier and only requires a few measurements.

The Gaussian and coherent-state models have identical means but distinct variances, of $\ev{N}(\ev{N}+1)$ and $\ev{N}$, respectively.
As such, if the mean is known to high accuracy then a measurement of the variance to $\ev{N}^2$ accuracy would allow us to discriminate between the two scenarios.\footnote{As $\ev{N}$ depends on the axion-photon coupling, the local DM density, and specific properties of the detector, it is idealistic to imagine this could be already known exactly.
We show at the end of this subsection one can perform state discrimination whilst estimating both the mean and variance.}
Our experiment provides $N_\text{shot} = T/\tau_\DM$ measurements from which we wish to infer the variance.
Let us label the different measurements, again assumed to be independent, as $N_i$.
With an exactly known mean, $\mu = \la N \ra$, the appropriate estimator for the variance is
\be \label{eq:hat-var}
\hat{\sigma}^2 = \frac{1}{N_\text{shot}} \sum_{i=1}^{N_\text{shot}} (N_i-\mu)^2.
\ee
Note $\la \hat{\sigma}^2 \ra = \var(N)$ so that this is an unbiased estimator.\footnote{The often-used variance estimator that contains a prefactor of $1/(N_\text{shot}-1)$ is only appropriate when the mean is also estimated.
Regardless, for $N_\text{shot} \gg 1$ the distinction is unimportant.}

Our goal is to establish the value of the variance to an accuracy smaller than $\ev{N}^2$, which requires knowing the fluctuations of $\hat{\sigma}^2$.
That is captured by its variance,
\be
\var(\hat{\sigma}^2) = \frac{\mu_4-\var(N)^2}{N_\text{shot}},
\ee
with $\mu_4 = \la (N-\mu)^4 \ra$ the fourth central moment.
The value of $\var(\hat{\sigma}^2)$ we expect to see in our data depends upon the underlying true DM state.
We have not computed the fourth-order moments in input-output theory to determine $\mu_4$.
For the Poisson distribution we would associate with a coherent state, $\mu_4 = \mu(1+3\mu)$.
For the Gaussian hypothesis, the full distribution is derived in Sec.~\ref{sec:measurement}, and has a more complex form for $\mu_4$.
Still, in the $\mu \ll 1$ limit, the two hypotheses have a leading-order result of $\var(\hat{\sigma}^2) \simeq \mu/N_\text{shot}$.
Accordingly, our stated accuracy requires $\sqrt{\mu/N_\text{shot}} < \mu^2$, or in terms of total measurement time,
\be \label{eq:inout_Tdiscrim}
T \simeq \frac{\tau_\DM}{\mu^3}.
\ee

For example, if there exists an axion with parameters such that $\mu = 10^{-2}$, then discovering DM with no background simply requires detecting a single photon, which takes time $T \sim 10^2\,\tau_\DM$, whereas distinguishing the Gaussian prediction from the coherent Poisson hypothesis would take $10^6\,\tau_\DM$.
(In reality both would take longer, since we are completely neglecting noise.)
Given that in the ADMX band $\tau_\DM \simeq 1\,$ms, discovery would take less than a second and state discrimination less than an hour at this particular mass.

That existing experiments could in principle discriminate between two \textit{classical} DM states is consistent with our conclusions in Ref.~\cite{suppression_paper}.
What may appear inconsistent is the scaling in Eq.~\eqref{eq:inout_Tdiscrim}.
In particular, Ref.~\cite{suppression_paper} claimed that the coherent and Gaussian hypotheses can be distinguished with only $1/\mu^2$ measurements, whereas the above experiment required a larger $1/\mu^3$.
There is no contradiction; the difference stems from the use of a different estimator.
In Ref.~\cite{suppression_paper} we used $p_2$, the probability of observing two photons, which is provably a better estimator than the variance for this problem.
In fact, working just with second moments, we can also construct a parametrically optimal estimator.
First, we relax our assumption that the mean is known exactly, and imagine the same dataset is used to estimate the mean via
\be
\hat{\mu} = \frac{1}{N_\text{shot}} \sum_{i=1}^{N_\text{shot}} N_i,
\ee
and the variance using Eq.~\eqref{eq:hat-var} with $\mu \to \hat{\mu}$.
Now the fact that Gaussian and coherent hypotheses differ in the variance is more subtle, as variance is defined in terms of the mean, which is also being estimated.
A more robust quantity to look at is the difference between the variance and the mean, encoded in the Mandel $Q$ parameter,
\be \label{eq:MandelQ}
\hat{Q} = \frac{\hat{\sigma}^2-\hat{\mu}}{\hat{\mu}}.
\ee
This quantity is expected to vanish for a coherent state and instead be equal to the mean count $\mu$ for the Gaussian.
Using a similar analysis to the above, one can show for both hypotheses that $\var(\hat{Q}) \sim 1/N_\text{shot}$ when $\mu \ll 1$.
This recovers the $1/\mu^2$ scaling and would reduce the time to distinguish states to less than a minute for the ADMX example above.
More generally though, to ensure one has an optimal estimator, the full probability distribution and therefore likelihood are required, and we derive these in Sec.~\ref{sec:measurement}.

\section{Measurement in the Schr\"odinger Picture: Measurement Records}
\label{sec:measurement}
%!TEX root = main.tex

In this section, we develop an alternative approach to continuous measurement.
As Sec.~\ref{sec:inout} highlights, input-output theory is powerful.
It can accommodate a quantum description of DM, incorporates dissipation into unitary quantum evolution, and allows one to evaluate moments of observables, which encode properties of the DM quantum state.
Yet the formalism leaves open several questions about the nature of continuously monitoring the cavity-DM system.
For instance, is there a direct procedure for computing the entire probability distribution of observables, rather than just moments?
Can one trace in real time how the state of the cavity-DM system is affected by the process of continuous measurement?
Is there a natural way to trace the temporal correlations in the signal imprinted by DM?
Given the classical description of the system is one of wave mixing, is there a quantum picture that mirrors this simplicity?
Here we show that an affirmative answer to these questions can be provided with an approach that keeps both the DM state and the statistics of the \textit{entire joint measurement record} central.

A core insight in the construction of such a formalism comes from the toy model reviewed in Sec.~\ref{subsec:review}.
In particular, when working in the Schr\"odinger picture and the basis of coherent states, the evolution in Eq.~\eqref{eq:joint_coherent} is trivial and exactly mirrors that of the classical system, with the oscillators periodically exchanging energy.
The $P$-function formalism allowed for the coherent-state simplicity to be lifted to the solution for general cavity and DM states, as given in Eq.~\eqref{eq:approx_cavity_P}.
Further, using the $P$-function we can evaluate the full distribution $p(\mathcal{O})$ of measurement outcomes $\mathcal{O}$, as given in Eq.~\eqref{eq:P_probability}.

In this section, we show that these results lift to a continuous sequence of measurements on the cavity-DM system.
In Sec.~\ref{subsec:single_measurements}, we apply the results from the toy model to the interaction between the cavity and the temporal modes introduced in our discussion of input-output theory, and obtain the probability density $p(\mathcal O_i)$ of a single projective measurement.
We generalize this in Sec.~\ref{subsec:complete_model} to find the \textit{joint} probability distribution $p(\{\mathcal O_i\})$ of all such measurements.
Similar to Eq.~\eqref{eq:P_probability}, it is given by integrating the classical-like probabilities $|\!\braket{\bm{\alpha}}{\{\mathcal O_i\}}\!|^2$ for joint coherent states of the DM field against the quasiprobability density $P_a(\bm{\alpha})$ describing the DM state.
In Secs.~\ref{subsec:number_measurements} and \ref{subsec:quad_meas_stats}, we apply our general result for $p(\{\mathcal O_i\})$ to analyze the statistics of number and quadrature measurements, respectively.
We compute the full probability distribution of both observables for the virialized Gaussian hypothesis.
For quadrature, we confirm the widely used exponential PSD statistics.
For number, we show the full distribution interpolates between a geometric and negative binomial distribution depending on the relative size of how long we integrate to form the measurement bins and the timescale of the signal coherence.
This result, which to our knowledge has not previously been computed, deviates considerably from the Poisson statistics one would associate with a coherent state, and we show that experiments could sharply distinguish the two predictions.

%%%%%%%%%%%%%%%%%%%%%%%%%%%%%%%%%
\paragraph{Assumptions on the DM State.}
%%%%%%%%%%%%%%%%%%%%%%%%%%%%%%%%%
%
The formalism of this section is completely general with respect to the DM $P$-function.
In particular, our results hold for both classical ($P \geq 0$) and nonclassical (allowing $P < 0$) DM states.
In contrast to Sec.~\ref{sec:inout}, we make no overarching assumptions regarding DM stationarity or homogeneity.
When presenting results for specific examples, such as the virialized Gaussian or coherent-state hypotheses of Sec.~\ref{subsec:DMQstate}, the expressions inherit the assumptions underpinning those models.

%%%%%%%%%%%%%%%%%%%%%%%%%%%%%%%%%
\subsection{Warm-Up: Probability Distribution of a Single Measurement}
\label{subsec:single_measurements}
%%%%%%%%%%%%%%%%%%%%%%%%%%%%%%%%%

As an initial step toward the probability of a full measurement record, we first show how to compute $p(\mathcal{O}_i)$, the probability distribution of a single measurement outcome.
In contrast to the projective measurements \textit{on the cavity mode} studied in Ref.~\cite{suppression_paper}, here we consider projective measurements \textit{on the transmission-line modes}, which implement \textit{weak} measurements on the cavity state.
Each of these will usually shift the cavity state slightly toward one that reduces the prior uncertainty on the measured observable.\footnote{The difference between weak and projective measurements can also be understood through the language of POVMs applied to the cavity state instead of pure projection operators; see e.g.~Ref.~\cite{JacobsSteck}.}

We first consider the time evolution of the cavity and transmission line for a single time step from $t_i$ to $t_i + \Delta t$, neglecting the effect of DM.
The relevant time-evolution operator $U_i \equiv U_I(t_i+\Delta t,t_i)$ was already computed in Eq.~\eqref{eq:time_ev_operator}, and here, we focus on how the evolution affects the state of the cavity and the transmission-line temporal mode $\db_\tin(t_i)$.
For simplicity, throughout this section we assume the transmission-line modes are in vacuum.
However, in principle the formalism can accommodate, e.g., thermal input states as well.\footnote{This can be done by transforming the thermal states used in Sec.~\ref{sec:inout} to discrete temporal modes, using Eq.~\eqref{eq:DFT}.
One subtlety is that the thermal discrete-frequency modes are in a product state, but they correspond to discrete temporal modes which are not.
Assuming thermal discrete temporal modes in a product state would instead generically correspond to discrete-frequency modes with nonzero occupancy at negative frequency.}

If the initial joint coherent state of the cavity and transmission line is $\ket{\gamma,0} = \ket{\gamma}_c \otimes \ket{0}_b$, then by the results in Sec.~\ref{subsec:review} this state evolves to 
\be
U_i\ket{\gamma,0} = \big\vt e^{-i\omega_c\Delta t}\gamma\cos\theta,\, -\gamma\sin\theta \big\ra,
\ee
where $\theta \equiv \sqrt{\kappa_m\Delta t}$ parametrizes the oscillation between the cavity and bath.
(Although helpful in the discussions of this subsection, from Sec.~\ref{subsec:complete_model} we revert to writing $\sqrt{\kappa_m\Delta t}$ explicitly.)
Though this angle can be much larger than the analogous $gt$ controlling the DM-cavity evolution in Eq.~\eqref{eq:joint_coherent}, as discussed below Eq.~\eqref{eq:time_ev_operator}, we assume a sampling rate such that $\theta \ll 1$.
The coherent-state evolution generalizes immediately to an arbitrary initial cavity state with $P$-function $P_c(\gamma)$.
After the interaction, the joint density matrix evolves to
\be
U_i\rho U_i^\dag = \int d\gamma \, P_c(\gamma) \,
\big\vt e^{-i\omega_c\Delta t}\gamma\cos\theta,\, -\gamma\sin\theta\big \ra
\big\la e^{-i\omega_c\Delta t}\gamma\cos\theta,\, -\gamma\sin\theta\big\vt.
\ee

Now suppose a projective measurement performed on the temporal mode $\db_\tin(t_i)$ yields outcome $\mathcal{O}_i$.
To find the final cavity density matrix, we apply the projector $\Pi_{\mathcal{O}_i}$ onto the subspace of the temporal mode's Hilbert space consistent with $\mathcal{O}_i$, then trace out the $\db_\tin(t_i)$ mode.
This yields the unnormalized cavity density matrix conditioned on $\mathcal{O}_i$, 
\be \label{eq:rhoc_one_meas}
\rho_c(\mathcal{O}_i) = \tr_i \!\big\{\Pi_{\mathcal{O}_i}U_i \rho U_i^\dag\big\}.
\ee
For nondegenerate measurement outcomes, we can write $\Pi_{\mathcal{O}_i}=\ket{\mathcal{O}_i}\bra{\mathcal{O}_i}$, so that
\be \label{eq:single_meas_cond}
\rho_c(\mathcal{O}_i) = \int d\gamma\, P_c(\gamma)\,
\Big| \big\la \mathcal{O}_i \big\vt -\gamma\sin\theta \big\ra \Big|^2\,
\big\vt e^{-i\omega_c\Delta t}\gamma\cos\theta \big\ra \big\la e^{-i\omega_c\Delta t}\gamma\cos\theta \big\vt.
\ee
The normalization of $\rho_c(\mathcal{O}_i)$ gives the probability of measuring the outcome $\mathcal{O}_i$, 
\be \label{eq:p_onemeasure}
p(\mathcal{O}_i) = \tr{\rho_c(\mathcal{O}_i)} = \int d\gamma\, P_c(\gamma)\, \Big| \big\la \mathcal{O}_i\big\vt {-}\gamma\sin\theta \big\ra \Big|^2.
\ee
This is simply the probability of measuring outcome $\mathcal{O}_i$ in a coherent state $\ket{- \gamma \sin \theta}$, integrated against the ``prior'' quasiprobability distribution $P_c(\gamma)$.

We can also change variables in Eq.~\eqref{eq:single_meas_cond} to express the final cavity state in terms of an (unnormalized) ``posterior'' cavity $P$-function, 
\be
\rho_c(\mathcal{O}_i) = \int d\gamma \, P_c^f(\gamma) \, \ket{\gamma} \bra{\gamma} = \int \frac{d\gamma}{\cos^2\theta}\,P_c\!\left( \frac{\gamma}{\cos\theta} \, e^{i\omega_c \Delta t}\right) \Big| \big\la \mathcal{O}_i\big\vt {-}\gamma\tan\theta e^{i \omega_c \Delta t}\big\ra \Big|^2 \, \ket{\gamma}\bra{\gamma}.
\ee
In practice, $\theta$ will be small.
Compared to $P_c$, the posterior $P_c^f$ differs in three ways: (1) the usual phase rotation from free evolution, (2) a slight $\cos \theta$ compression of the $P$-function's support due to energy loss down the transmission line, and (3) a filter factor that shrinks the $P$-function in areas not consistent with the measurement outcome $\mathcal{O}_i$.

\begin{figure}[t!]
\centering
\includegraphics[width=0.85\linewidth]{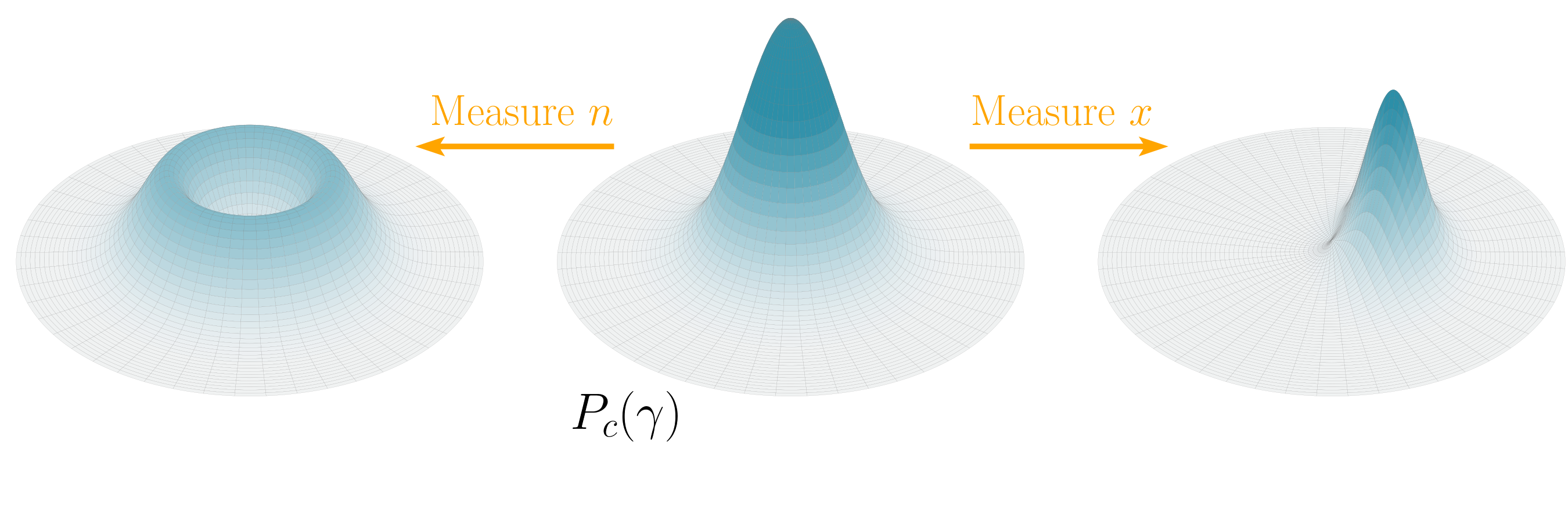}
\vspace{-0.8cm}
\caption{The impact on the cavity $P$-function if we projectively measure number (left) or quadrature (right) at the end of the transmission line, for a cavity initially in a thermal Gaussian state, and definite values for the number and quadrature outcomes.
To emphasize the difference with respect to projective measurements of the cavity itself, we take $\theta = \pi/8$, where the effect is strong, though in practice $\theta \ll 1$ and the posterior cavity state remains far closer to its initial form.}
\label{fig:Pmeasure}
\end{figure}

%%%%%%%%%%%%%%%%%%%%%%%%%%%%%%%%%
\paragraph{Minimal Examples.}
%%%%%%%%%%%%%%%%%%%%%%%%%%%%%%%%%
%
We can see explicitly how the transmission-line measurements impact the cavity $P$-function using simple examples.
For number measurements, the filter factor is a Poisson distribution,
\be
\big\vt \la n \vt \gamma \ra \big\vt^2 
= \text{Pois}\big(n,|\gamma|^2\big) 
\equiv \frac{|\gamma|^{2n} e^{-|\gamma|^2}}{n!} .
\ee
Thus, after a number measurement yielding $n$ quanta, the posterior $P$-function is strongly suppressed away from regions where $|\gamma|^2 \theta^2 \simeq n$.
Similarly, for quadrature measurements, the filter factor is a Gaussian,
\be \label{eq:single_quad_filter}
\big|\la x \vt \gamma\ra \big|^2
= \frac{1}{\sqrt{\pi}}\exp{-(x-x_\gamma)^2 },
\ee
where $x_\gamma = (\gamma+\gamma^*)/\sqrt{2}$.
The posterior cavity $P$-function is then strongly suppressed in regions away from $\theta\, x_\gamma \simeq -x$.

In Fig.~\ref{fig:Pmeasure} we show explicit examples of the posterior $P_c^f$ after a number or quadrature measurement.
Even when $\theta$ is large, the results differ qualitatively from a projective measurement on the cavity itself.
For instance, a number measurement on the cavity collapses it to a Fock state, but a number measurement on the transmission line can never drive the cavity to a nonclassical state, because the filter factor is nonnegative.
(A transmission line can drive the cavity to a nonclassical state, but doing so requires starting the transmission line in a nonclassical state rather than in vacuum as we have assumed here.)

%%%%%%%%%%%%%%%%%%%%%%%%%%%%%%%%%
\subsection{The Joint Distribution of the Record}
\label{subsec:complete_model}
%%%%%%%%%%%%%%%%%%%%%%%%%%%%%%%%%

We now incorporate the effect of the axion, so that the time-evolution operator becomes 
\be
U_i = U_I(t_i+\Delta t,t_i) \simeq \exp{-iH_\text{sys}\Delta t}\exp{\sqrt{\kappa_m \Delta t} \left( c^\dag \db_\tin(t_i) - c \db_\tin^\dag(t_i) \right)}
\ee
where $H_{\text{sys}}$ is the axion-cavity Hamiltonian, 
\be
H_\text{sys} = \omega_c c^\dag c + \int \frac{d^3 \bk}{(2\pi)^3} \, \omega_{\bk} a_{\bk}^\dag a_{\bk} + i\int \frac{d^3 \bk}{(2\pi)^3} \, \Big(g_{\bk}c^\dag a_{\bk} - g^*_{\bk}c a^\dag_{\bk}\Big).
\ee
This is a subset of Eq.~\eqref{eq:fullHamiltonian}, though we have expanded the effective mode $a_\eff$ in its plane-wave components and defined the momentum-dependent couplings $g_{\bk}$ as 
\be
g_{\bk} \equiv \frac{g}{2} \sqrt{\frac{V_c}{\Omega}} \, C(\bk) ,
\ee
with $g=\ga B_0 \sqrt{\Omega}$ as in Eq.~\eqref{eq:Hint_simplified}.
For simplicity, we continue to neglect the loss port $d_\tin(t)$, but it is straightforward to include: since we are neglecting thermal noise, its effect would simply be to replace $\kappa_m \to \kappa_\text{tot} = \kappa_m + \kappa_\ell$ in the cavity damping and response functions, while retaining only $\kappa_m$ in the coupling to the measured output.

%%%%%%%%%%%%%%%%%%%%%%%%%%%%%%%%%
\paragraph{Unitary Time Evolution of the Full System.}
%%%%%%%%%%%%%%%%%%%%%%%%%%%%%%%%%
%
To treat the entire measurement record, we account for the interaction between the cavity and each temporal mode, as well as the subsequent projective measurements of the temporal modes.
Since the input temporal-mode operators $\db_\tin(t_i)$ commute with one another at different times, we can treat the system as undergoing a unitary time evolution for a time $t_N = N\Delta t$, which allows $N$ discrete temporal modes to interact with the cavity.
After these interactions with the cavity, all temporal modes are projectively measured.

Just as in the single-measurement example, our approach is to compute the time-evolved density matrix of the full system.
We then condition this matrix on a set of measurement outcomes $\{\mathcal{O}_i\}$, and take the trace of the conditioned density matrix to find $p(\{\mathcal{O}_i\})$.

Consider an initial joint coherent state $\ket{\gamma_0,\bm{\alpha}_0,\bm{0}}$ for the full system.
This corresponds to $\ket{\gamma_0}$, $\ket{\bm{\alpha}_0} = \bigotimes_{\bk} \ket{\alpha_{\bk}(0)}$, and $\ket{\bm{0}} = \bigotimes_i \ket{0_i}$ for the initial cavity, axion, and transmission-line temporal modes, respectively.
With this choice, the final state is also guaranteed to be a joint coherent state, as the Hamiltonian is quadratic and number-conserving.
We write the time-evolved states as\footnote{This is not the full evolution.
Comparing to Eq.~\eqref{eq:interaction_U_split}, we have neglected the time evolution generated by $H_0$.
We ignore this contribution as its action is trivial: it freely propagates the transmission-line temporal modes, translating their contents from one mode to the next.
For instance, $\vt \ldots, \beta_i, \beta_{i+1}, \beta_{i+2},\ldots \ra$ is translated to $\vt \ldots, \beta_{i-1}, \beta_i, \beta_{i+1},\ldots \ra$.}
\be
\ket{\gamma_i,\bm{\alpha}_i, \bm{\beta}_i} \defeq U_{i-1} \dotsb U_1U_0 \ket{\gamma_0,\bm{\alpha}_0, \bm{0}}.
\ee
This implies that if the cavity and axion start in a general state, described by the joint $P$-function $P_\text{sys}(\gamma_0,\bm{\alpha}_0)$, the full density matrix after a time $t_N$ is 
\be \label{eq:full_evolved_density_matrix}
\rho(t_N) = \int d\gamma_0 d\bm{\alpha}_0\, P_\text{sys}(\gamma_0,\bm{\alpha}_0) \ket{\gamma_N,\bm{\alpha}_N, \bm{\beta}_N} \bra{\gamma_N,\bm{\alpha}_N, \bm{\beta}_N}.
\ee
To compute the values of $\gamma_i$, $\bm{\alpha}_i$, and $\bm{\beta}_i$, we can rewrite their evolution over a single time step as a differential equation in the $\Delta t \to dt$ limit, using 
\be
\ket{\gamma(t+dt),\bm{\alpha}(t+dt), \bm{\beta}(t+dt)} = U_I(t+dt,t)\ket{\gamma(t),\bm{\alpha}(t), \bm{\beta}(t)}.
\ee
The resulting equations arise in an operationally equivalent manner to the Heisenberg equations of motion derived in Sec.~\ref{subsec:io_intro}.
In detail, for the cavity and output modes
\begin{subequations} \label{eq:classical_eoms}
\begin{align} 
\beta(t) &= -\sqrt{\kappa_m \, dt} \, \gamma(t), \\
\dot \gamma(t) &= -\left( i\omega_c + \frac{\kappa_m}{2}\right)\gamma(t) + \int \!\frac{d^3 \bk}{(2\pi)^3}\, g_{\bk}\alpha_{\bk}(t).
\end{align}
\end{subequations}
These can be thought of as the \textit{classical} analogs of Eqs.~\eqref{eq:cont_inout_relations} (with DM added and $b_\tin(t)$ set to zero), arising as the coherent states follow the classical equations of motion.
We reiterate that while Eq.~\eqref{eq:classical_eoms} takes $\Delta t$ differential as a mathematical idealization, realistic individual measurements take some small but finite time $\Delta t$.
However, for the purposes of our discussion, the distinction is irrelevant so long as $\Delta t$ is small enough to resolve the central frequency $m_a$.
The above procedure also produces an equation of motion for the DM modes of
\be \label{eq:classical_eoms_DM}
\dot \alpha_{\bk}(t) = -i\omega_{\bk}\alpha_{\bk}(t) - g^*_{\bk}\gamma(t),
\ee
but as in Sec.~\ref{sec:inout}, we ignore the effects of cavity backaction on these modes, effectively neglecting the second term so that the axion freely evolves.
We elaborate on this choice in what follows.

%%%%%%%%%%%%%%%%%%%%%%%%%%%%%%%%%
\paragraph{Steady-State Regime.}
%%%%%%%%%%%%%%%%%%%%%%%%%%%%%%%%%
%
Since our formulation is in the time domain, we can in principle compute the measurement record for any choice of initial cavity, DM, and transmission-line states.
However, conventionally, a cavity haloscope interrogates each frequency for a time $t \gg 1/\kappa_m$, so for simplicity we assume initial transients in the cavity have decayed away.\footnote{Technically, we also assume the energy density in the DM halo has not been appreciably drained.
By solving Eqs.~\eqref{eq:classical_eoms} and \eqref{eq:classical_eoms_DM} perturbatively in $g_{\bk}$, the solution to $\alpha_\bq(t)$ at second order includes a term $-t\alpha_\bq(0)\int d^3\bk \, |g_{\bk}|^2/\kappa_m$, so one requires $t \ll \kappa_m/\int d^3\bk \, |g_{\bk}|^2$.
An identical assumption was made in Sec.~\ref{subsec:io_intro}; see footnote~\ref{footnote:cavitybackaction}.}
In this regime, we have the steady-state solution
\be \label{eq:gamma_def}
\gamma(t) \simeq \int \frac{d^3 \bk}{(2\pi)^3} \frac{g_{\bk}\alpha_{\bk}(0)}{\frac{\kappa_m}{2}-i\Delta_{\bk}}e^{-i\omega_{\bk} t},
\quad \alpha_{\bk}(t) \simeq \alpha_{\bk}(0) e^{-i\omega_{\bk} t},
\ee
with detuning $\Delta_{\bk} \equiv \omega_{\bk}-\omega_c$.
If we assume our measurements begin once the cavity has already entered this steady-state regime, then the above solutions imply that the initial joint $P$-function of the cavity-DM system in Eq.~\eqref{eq:full_evolved_density_matrix} has the form
\be
P_\text{sys}(\gamma,\bm{\alpha}) = \delta \bigg(\gamma - \int \frac{d^3 \bk}{(2\pi)^3} \chi_{\bk} \alpha_{\bk}\bigg) P_a(\bm{\alpha})
\ee
where we defined a modewise response $\chi_{\bk} \equiv g_{\bk}/(\kappa_m/2 - i\Delta_{\bk})$ and dropped the $0$ subscripts on $\gamma_0, \bm{\alpha}_0$.
Combining the above results, the final density matrix in Eq.~\eqref{eq:full_evolved_density_matrix} can be written in terms of just the DM state, 
\bea
\rho(t_N) = \int d\bm{\alpha}\, P_a(\bm{\alpha}) &\left|\gamma(t_N),\big\{\alpha_{\bk} e^{-i\omega_{\bk} t_N}\big\},\big\{{-}\sqrt{\kappa_m \Delta t} \, \gamma(t_i)\big\} \right\ra \\ 
&\hspace{3.3cm}\otimes \left\la\gamma(t_N),\big\{\alpha_{\bk} e^{-i\omega_{\bk} t_N}\big\},\big\{{-}\sqrt{\kappa_m \Delta t} \, \gamma(t_i)\big\} \right|\!.
\eea

%%%%%%%%%%%%%%%%%%%%%%%%%%%%%%%%%
\paragraph{The Joint Probability of the Measurement Record.}
%%%%%%%%%%%%%%%%%%%%%%%%%%%%%%%%%
%
Applying the logic of Eq.~\eqref{eq:rhoc_one_meas}, after a projective measurement of all $N$ temporal modes with outcomes $\mathcal{O}_i$ and corresponding projectors $\Pi_{\mathcal{O}_i}$, the conditioned density matrix of the cavity-axion system is 
\be \label{eq:conditioned_rho}
\rho_\text{sys}(\{\mathcal{O}_i\}) = \tr_{\db}\left\{\Pi_{\mathcal{O}_1}\dotsb\Pi_{\mathcal{O}_N}\rho(t_N)\right\} ,
\ee
where $\tr_{\db}$ denotes a trace over all temporal modes.
The normalization of $\rho_\text{sys}(\{\mathcal{O}_i\})$ gives the joint probability of the complete measurement record $\{\mathcal{O}_i\}$, 
\be \label{eq:joint_record_prob}
p(\{\mathcal{O}_i\}) = \tr{\rho_\text{sys}(\{\mathcal{O}_i\})} = \int d\bm{\alpha}\, P_a(\bm{\alpha})\prod_i\Bigl|\braket{\mathcal{O}_i}{-\sqrt{\kappa_m \Delta t}\, \gamma(t_i)}\Bigr|^2 .
\ee
Compared to the single-measurement analog in Eq.~\eqref{eq:p_onemeasure}, here we are tracking the DM evolution and have assumed a cavity steady state, so that the expression can be written in terms of $P_a$.
The same intuition holds: we compute the overlap of the observable with a specific transmission-line state evolved classically through the coherent-state dynamics, now weighted against the axion prior $P_a$.

We now revisit the minimal examples we considered for a single measurement.
For number measurements, the probability of observing a given record $\{n_i\}$ is
\be \label{eq:number_meas_result}
p(\{ n_i\}) = \int d\bm{\alpha}\, P_a(\bm{\alpha})\prod_i \text{Pois}\big(n_i,|\gamma(t_i)|^2\kappa_m \Delta t\big).
\ee
For quadrature measurements, we have
\be \label{eq:quadrature_measurement_record}
p(\{r_i\}) = \int d\bm{\alpha}\, P_a(\bm{\alpha})\left(\frac{\Delta t}{\pi}\right)^{\!N/2} \exp{-\sum_{i=1}^N \Delta t\left[r_i + \sqrt{\kappa_m}\, X_\gamma(t_i) \right]^2}
\ee
where $r_i$ denotes the measurement values associated with the reweighted discrete temporal mode $\dX_\tout(t_i) / \sqrt{\Delta t}$, in contrast to the value $x$ used in Eq.~\eqref{eq:single_quad_filter}.
We have also defined
\be \label{eq:X_gamma_t}
X_\gamma(t) = \frac{1}{\sqrt 2}\left(\gamma(t) + \gamma^*(t) \right) = \frac{1}{\sqrt 2}\int \frac{d^3 \bk}{(2\pi)^3}\left(\chi_{\bk}\alpha_{\bk}e^{-i\omega_{\bk}t} + \text{c.c.} \right)\!.
\ee
The joint probability $p(\{r_i\})$ is similar in form to a path integral, in the sense that each value of $\bm{\alpha}$ leads to a time-dependent DM-driven trajectory $X_\gamma(t)$.
Heuristically, the probability of a trajectory $\{r_i\}$ depends on its similarity to a DM trajectory $- \sqrt{\kappa_m}\, X_\gamma(t_i)$, integrated over DM states weighted by $P_a(\bm{\alpha})$.

%%%%%%%%%%%%%%%%%%%%%%%%%%%%%%%%%
\paragraph{Example: Number Measurement for a Single Axion Mode.}
%%%%%%%%%%%%%%%%%%%%%%%%%%%%%%%%%
%
As an explicit example, suppose that only one DM mode is occupied, and it is coupled to the cavity on resonance.
Then $\gamma(t)$ in Eq.~\eqref{eq:gamma_def} simplifies to
\be \label{eq:single_mode_amp}
\gamma(t) = \sqrt{\eta}\, \alpha \, e^{-i\omega_c t}, \qquad \eta= 4 g^2/\kappa_m^2
\ee
where $\sqrt{\eta}$ is the fraction of the DM field amplitude that the cavity absorbs in the steady state and is consistent with the definition in Eq.~\eqref{eq:eta_value}.
Applying this to Eq.~\eqref{eq:number_meas_result} yields
\bea \label{eq:number_joint_record}
p(\{n_i\}) &= \int d\alpha\, P_a(\alpha) \prod_i\frac{\big(\eta\kappa_m\Delta t|\alpha|^2\big)^{n_i}}{n_i!}\exp{-\eta\kappa_m\Delta t |\alpha|^2} \\
&= \left(\frac{n_\text{tot}!}{N^{n_\text{tot}}\prod_i n_i!} \right)\int d\alpha\, P_a(\alpha)\, \text{Pois}\big(n_\text{tot},\eta\kappa_m T|\alpha|^2\big)
\eea
where we defined the total number of counts $n_\text{tot} = \sum_i n_i$ in the second equality, and $T = N\Delta t$.

The result shows that the DM state impacts the probability $p(n_\text{tot})$ of detecting $n_\text{tot}$ total counts, given by the integral in Eq.~\eqref{eq:number_joint_record}.
This integral takes the same form as a single-shot projective number measurement of the axion mode, with an effective efficiency
\be
\bar{\eta}(T) \defeq \eta \kappa_m T.
\ee
We can thus apply the results of Ref.~\cite{suppression_paper} to conclude that
\be \label{eq:num_meas_cases}
p(n_{\text{tot}}) = \begin{cases} \text{Pois}\big(n_\text{tot}, \bar{\eta}(T) N_\eff\big) & \text{coherent DM} \\
(\bar{\eta}(T) N_\eff)^{n_\text{tot}}/(1+ \bar{\eta}(T) N_\eff)^{n_\text{tot}+1} & \text{Gaussian DM} \\
\binom{N_\eff}{n_\text{tot}}\bar{\eta}(T)^{n_\text{tot}}(1-\bar{\eta}(T))^{N_\eff-n_\text{tot}} & \text{Fock DM}
\end{cases}
\ee
where each DM state has the same mean occupancy $N_\eff$ as defined in Eq.~\eqref{eq:Neff-DM}.

In the second line of Eq.~\eqref{eq:number_joint_record}, the prefactor is $m(\{n_i\})$, the multinomial probability for independently distributing $n_\text{tot}$ counts into $N$ bins and obtaining counts $n_i$.
Strikingly, how the counts are distributed amongst the time bins is \textit{completely independent of the DM state}.
Mechanically, this occurs because the joint probability depends only on the constant time-independent intensity $|\gamma(t_i)|^2 = \eta|\alpha|^2$ at each time step.
Physically, that the distribution of events in time contains no information on the DM state originates from the fact a single-mode field has an infinite coherence time.
For the remainder of this section we exclusively consider DM to be in a multimode state, so that it has a finite coherence time, except for a brief discussion of implications of a single-mode state for the waiting-time distribution.
We return to discuss the single-mode scenario and its implications in greater detail in Sec.~\ref{subsec:nonclassical_singlemode}.

%%%%%%%%%%%%%%%%%%%%%%%%%%%%%%%%%
\subsection{Binned Number Statistics and the Waiting-Time Distribution}
\label{subsec:number_measurements}
%%%%%%%%%%%%%%%%%%%%%%%%%%%%%%%%%

Here, we study in detail the probability of an axion haloscope measuring a certain distribution of clicks from DM.
The general result for a measurement record $\{ n_i \}$ collected over time steps of duration $\Delta t$ is given in Eq.~\eqref{eq:number_meas_result} and our goal, therefore, is to unpack that result.
First, we reorganize the bins over which the photon counts are collected, moving from $\Delta t$ to bins of arbitrary length $\tau$.
From here, we show how the statistics of the number distribution depend on the state of the DM.
We compute the exact count distribution for the virialized Gaussian hypothesis and show that it can deviate significantly from the Poisson statistics characteristic of a coherent state.
We use this to calculate the time required to distinguish coherent and Gaussian DM states, as we did in Sec.~\ref{subsec:higherOrderIO}.
Finally, we compare our results to what one can obtain from the distribution of the waiting time between detected events.

%%%%%%%%%%%%%%%%%%%%%%%%%%%%%%%%%
\paragraph{Binned Number Measurements.}
%%%%%%%%%%%%%%%%%%%%%%%%%%%%%%%%%
%
To express the time evolution of the coherent states as a differential equation, we took $\Delta t$ to be infinitesimal in Eq.~\eqref{eq:classical_eoms}.
To convert this to an actual experimental record, where each measurement is taken over finite time, we bin the photon measurements into windows of length $\tau$.
We index the smaller $\Delta t$ and larger $\tau$ bins by $i$ and $j$.
Our task is to derive the photon-number measurement record $\{\tilde{n}_j\}$ in the larger bins.
This is given by weighting all cases where the record in the small bins, $\{n_i\}$, occurs such that within each of the larger bins we have $\sum_{i \in j} n_i \equiv n^\text{tot}_j = \tilde{n}_j$, 
\bea
p(\{\tilde{n}_j\}) &= \sum_{\{n_i\}} p(\{n_i\}) \prod_j \delta_{\tilde{n}_j, n^\text{tot}_j} \\
&= \int d\bm{\alpha}\, P_a(\bm{\alpha}) \sum_{\{n_i\}}\prod_{i,j} \text{Pois}\big(n_i,\,|\gamma(t_i)|^2\kappa_m \Delta t\big)\, \delta_{\tilde{n}_j, n^\text{tot}_j}.
\eea
Note that the sum over $\{n_i\}$ is independent of the DM state: we are simply performing discrete convolutions of Poisson distributions within the integrand.
Since the sum of independent Poisson-distributed variables is also Poisson distributed, the expression simplifies to
\be \label{eq:full_number_dist}
p(\{\tilde{n}_j\}) = \int d\bm{\alpha}\, P_a(\bm{\alpha})\prod_j \text{Pois}\bigg(\tilde{n}_j,\,\kappa_m\int_{j\tau}^{(j+1)\tau} \!\!dt\, |\gamma(t)|^2\bigg).
\ee
In the final step we again took the small time bins to be infinitesimal, $\Delta t \to dt$, thereby converting the sum to an integral.
This result describes the physically relevant measurement record in bins of finite size; we now drop the tildes and simply call the set of outcomes $\{n_j\}$.

For a single count $n_j$, the marginal probability distribution is given by
\be\label{eq:single_bin_prob}
p(n_j) = \int d\bm{\alpha}\, P_a(\bm{\alpha}) \, \text{Pois}\bigg(n_j,\,\kappa_m\int_{j\tau}^{(j+1)\tau} \!\!dt\, |\gamma(t)|^2\bigg).
\ee
This expression is considerably more complex than the single-mode example given in Eq.~\eqref{eq:number_joint_record}.
In particular, from Eq.~\eqref{eq:gamma_def}, the factor of $|\gamma(t)|^2$ is sensitive to all of the DM modes and how they interfere as time evolves, the same interference that generates the DM coherence time.
Of course, for sufficiently short measurement increments, $\tau \to 0$, the interference effects cannot develop, and the result reduces to the single-mode expressions in Eq.~\eqref{eq:num_meas_cases}.

We can now use the general $\tau$ result to understand the number measurement statistics we expect DM to imprint upon a detector.
We first study the mean and variance for Gaussian DM, showing explicitly how the results can diverge from the Poisson distribution one might naively expect for a counting experiment.
We then move beyond the first two moments to compute the entire distribution of number fluctuations, which we can use to optimally distinguish the Gaussian and coherent hypotheses.

\begin{figure}[t!]
\centering
\includegraphics[width=0.48\linewidth]{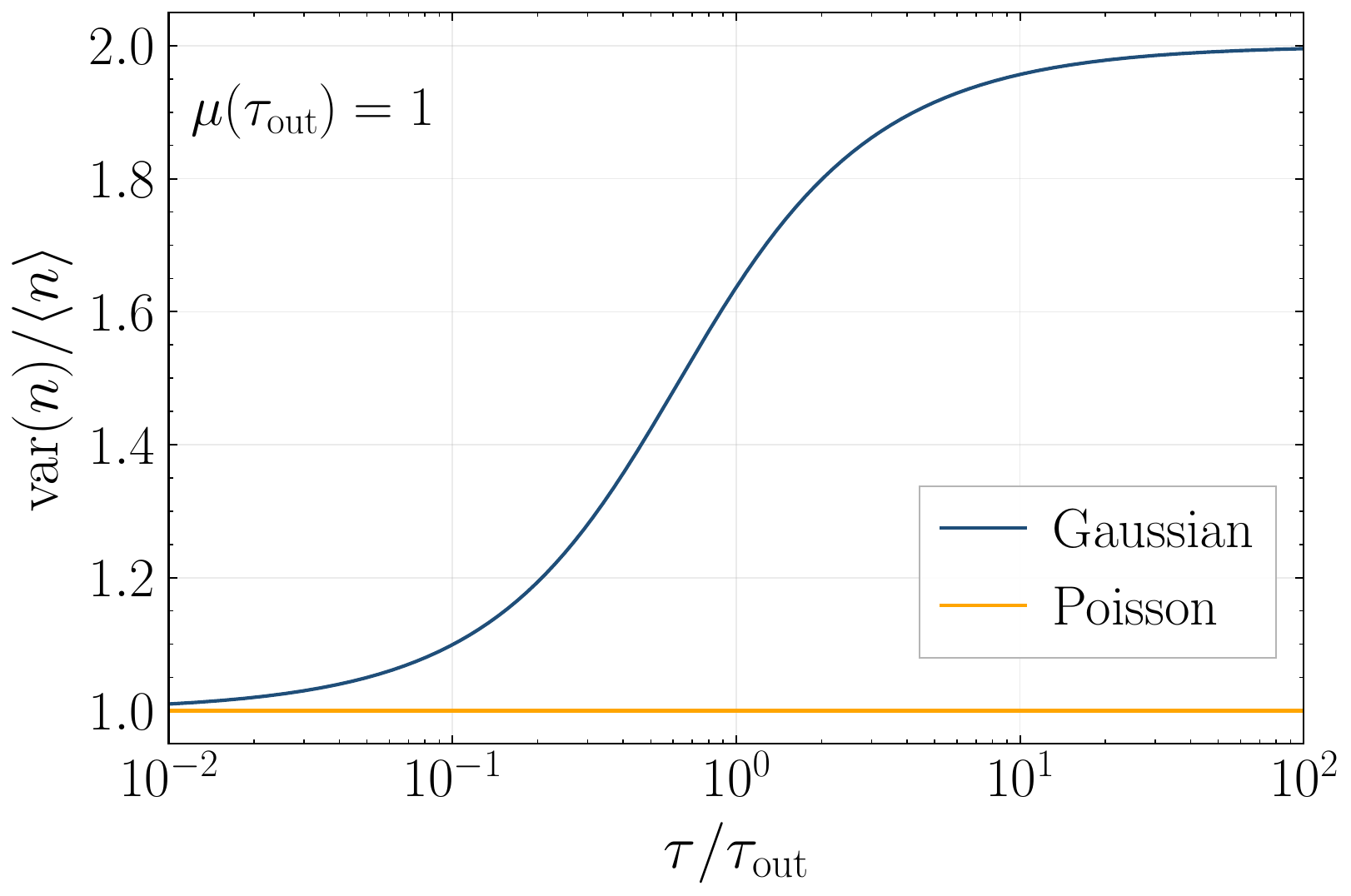}
\hspace{0.2cm}
\includegraphics[width=0.48\linewidth]{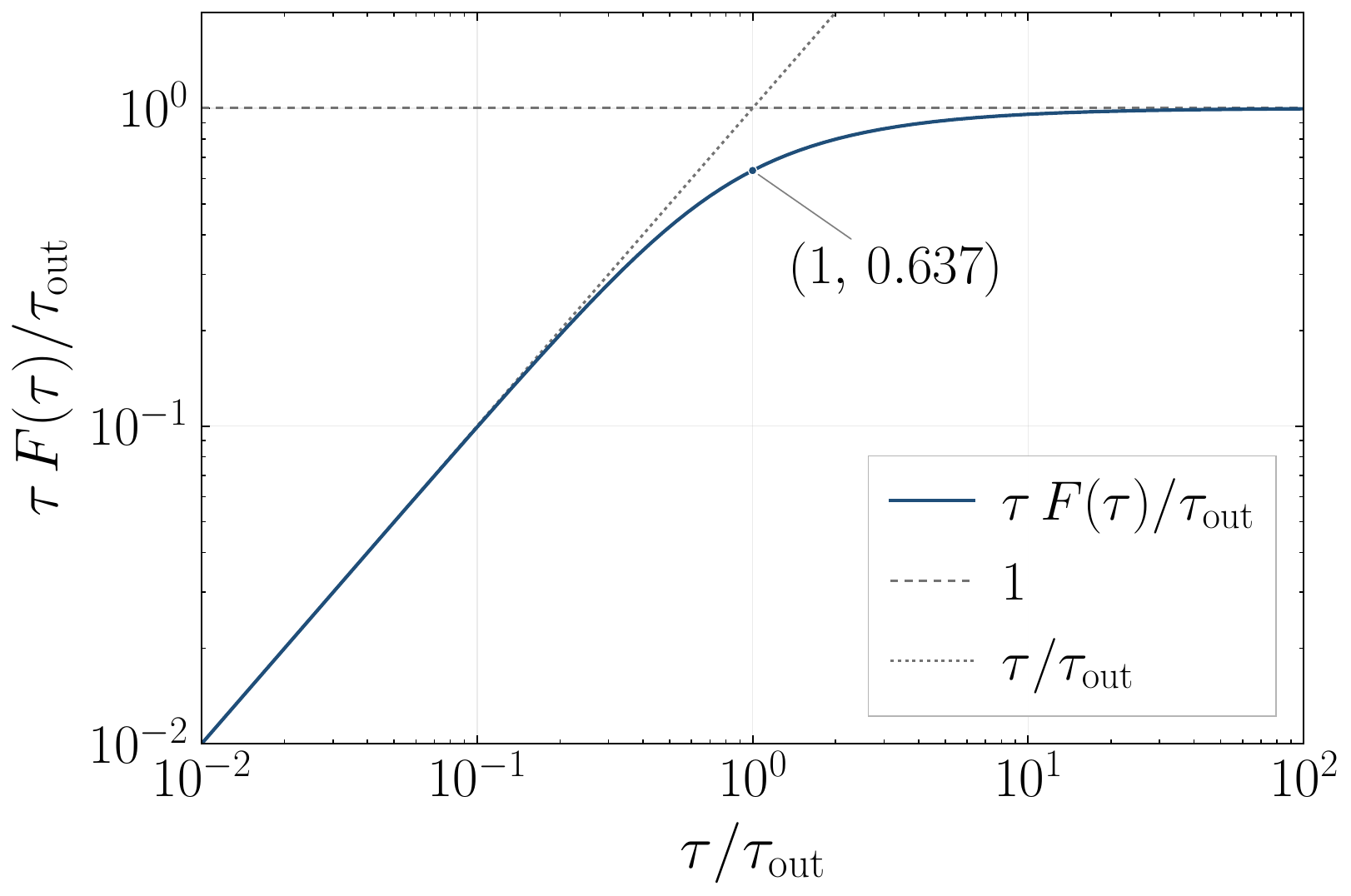}
\vspace{-0.4cm}
\caption{(Left) Variance over mean of the number distribution for the virialized Gaussian hypothesis as we vary the size of the bins within which the counts are collected, as compared to the Poisson prediction of $\var(n)=\la n \ra = \mu(\tau)$.
There is a sharp transition in behavior around $\tau \sim \tau_\tout$ where the bin size matches the coherence time of the DM signal and in general the variance can deviate significantly from the Poisson contribution.
For small $\tau$, the Gaussian result is $\simeq 1+\mu(\tau)$, which asymptotically matches the Poisson prediction as $\mu(\tau) \propto \tau$.
For large $\tau$, the Gaussian result plateaus to a constant $1+\mu(\tau_\tout)$.
(Right) The deviation from Poisson is controlled by $F(\tau)$ defined in Eq.~\eqref{eq:F_def}, with the result as displayed manifestly consistent with the limits in Eq.~\eqref{eq:F_limits}.
For both figures, we adopt the explicit standard halo model result in Eq.~\eqref{eq:Ftau_SHM}, with $v_0 \simeq 220\,\text{km/s}$ and $v_\odot \simeq 232\,\text{km/s}$.}
\label{fig:Poisson_deviation}
\end{figure}

%%%%%%%%%%%%%%%%%%%%%%%%%%%%%%%%%
\paragraph{Variance of the Photon Count.}
%%%%%%%%%%%%%%%%%%%%%%%%%%%%%%%%%
%
For Gaussian DM, from Eq.~\eqref{eq:single_bin_prob} we can compute the mean and variance of the counts in a single bin as
\be \label{eq:meanvar}
\la n \ra = \mu(\tau), \quad
\var(n) = \mu(\tau) + F(\tau)\mu^2(\tau),
\ee
where we have introduced the following functions of the bin length,\footnote{The Poisson deviation function $F(\tau)$ plays a nearly identical role to $\theta(V)$ defined in Ref.~\cite{Cheong:2024ose}.
Both control how fluctuations scale as bin size is varied with respect to the natural coherence scale of the signal; $F(\tau)$ does this for time and $\theta(V)$ for space.
One difference is that $F(\tau)$ includes effects of the detector, whereas $\theta(V)$ was defined purely for DM.
Nevertheless, in the wide-cavity limit, only properties of DM are relevant, and the two become very similar; cf.~Fig.~\ref{fig:Poisson_deviation} and the equivalent result for $\theta(V)$.}
\begin{align}
\label{eq:mu_def}
\mu(\tau) &= \kappa_m \tau \la |\gamma(0)|^2\ra_P, \\
F(\tau) &= \frac{1}{|\la |\gamma(0)|^2\ra_P|^2}\frac{1}{\tau} \int_{-\tau}^\tau \!dt \, \left(1-\frac{|t|}{\tau} \right)|\la \gamma^*(0)\gamma(t)\ra_P|^2.
\label{eq:F_def}
\end{align}
Here, $\la \cdot \ra_P$ corresponds to an expectation over the DM $P$-function, which we take to follow the virialized Gaussian state of Eq.~\eqref{eq:VGH}.
For example, we can explicitly evaluate the mean via
\bea
\kappa_m \tau \la |\gamma(0)|^2\ra_P
&= \kappa_m \tau \int d\bm{\alpha}\, P_a(\bm{\alpha})\,\left|\int \frac{d^3 \bk}{(2\pi)^3} \frac{g_{\bk}\alpha_{\bk}}{\frac{\kappa_m}{2}-i\Delta_{\bk}} \right|^2 \\
&= \tau \frac{V_c}{4\Omega} \int \frac{d^3 \bk}{(2\pi)^3} \frac{g^2 \kappa_m\,|C(\bk)|^2 N_{\bk}}{(\kappa_m/2)^2+(\omega_{\bk}-\omega_c)^2} = \tau \Gamma_\sig,
\eea
where the final step used the definitions in Eqs.~\eqref{eq:n_eff} and \eqref{eq:Gamma_sig_def}.
This is exactly consistent with the mean we computed for input-output; see Eq.~\eqref{eq:mean_inout_distinguish}.
Using Eq.~\eqref{eq:Gamma_sig_etaNeff}, in the regime haloscopes currently operate, we have $\mu(\tau) \sim \bar{\eta}(\tau) N_\eff$.
The variance in Eq.~\eqref{eq:meanvar} is also consistent with the input-output results in Eq.~\eqref{eq:total_count_variance_discrete} for $T = \tau \gg \tau_\tout$; this can be seen by noting $\mu(T) = \sum_k \ev{N_\tout[\omega_k]}$ and $F(T) \mu^2(T) = \sum_k \ev{N_\tout[\omega_k]}^2$.
(That the input-output expression is only valid for large $T$ is a result of the approximation inherent in Eq.~\eqref{eq:bfrak_to_bcont}.)

The Poisson deviation function $F(\tau)$ quantifies the amount by which $\var(n)$ exceeds the prediction of Poisson statistics, which would simply have $\var(n) = \mu(\tau)$.
We can understand the asymptotic behavior of the function as follows.
For $\tau \to 0$, the autocorrelation $|\la \gamma^*(0)\gamma(t)\ra_P|^2$ only probes short times and tends to a constant; we can therefore directly evaluate $F(\tau) \to 1$.
For $\tau \to \infty$, we now probe the autocorrelation at long times, where we eventually expect it to decay rapidly due to interference between different modes.
If we take the timescale for its decay as $\tau_\tout$, which can differ from $\tau_\DM$ as $\gamma(t)$ depends on DM and the instrument response, then the integrand vanishes for $t \gg \tau_\tout$, and we obtain the limiting cases
\be \label{eq:F_limits}
F(\tau) = 
\begin{cases} 
1 & \tau \ll \tau_\tout, \\
\tau_\tout/\tau & \tau \gg \tau_\tout,
\end{cases}
\ee
as shown in App.~\ref{app:coherence_time}.
Thus, at fixed $\tau$, in the limit $\tau_\tout \to 0$, the variance approaches that of a Poisson distribution.\footnote{This limit is relevant for particle-like DM as the coherence time scales with $1/m_\DM$.} 
However, at fixed $\tau_\tout$, in the limit $\tau \to \infty$ we have $F(\tau) \mu^2(\tau) \propto \tau$, so the variance never approaches Poisson.
For this reason, the number statistics of Gaussian DM can generally differ from the coherent-state prediction.
This is highlighted in Fig.~\ref{fig:Poisson_deviation} where we show both the Gaussian variance and the Poisson deviation function.

If we have an ansatz for the DM phase-space distribution, we can explicitly evaluate $F(\tau)$.
As an example, we adopt the standard halo model of Eq.~\eqref{eq:pk_SHM} so that DM is nonrelativistic and $g_\bk$ is approximately constant.
In the wide-cavity regime, $Q_a \gg Q_c$, we find
\bea \label{eq:Ftau_SHM}
F(\tau) &\simeq \frac{\sqrt{2\pi}}{m_a v_0 v_\odot \tau}\Biggl\{\text{Erf}\!\left(\frac{v_0 v_\odot m_a\tau}{\sqrt{2+v_0^4 m_a^2\tau^2/2}} \right) \\
&\qquad \qquad \qquad +\frac{2}{m_av_0^2 \tau}e^{-2v_\odot^2/v_0^2}\left[\text{Erfi}\!\left(\frac{\sqrt{2} \, v_\odot/v_0}{\sqrt{1+m_a^2 v_0^4 \tau^2/4}}\right)-\text{Erfi}\!\left(\frac{\sqrt2 \, v_\odot}{v_0}\right) \right] \Biggr\},
\eea
where $\text{Erfi}(x) = -i \, \text{Erf}(ix)$.
Using this form, we show $\tau F(\tau)/\tau_\tout$ in Fig.~\ref{fig:Poisson_deviation}, from which we can see explicit consistency with the limits in Eq.~\eqref{eq:F_limits}.
In the wide-cavity limit, any dependence upon the detector drops out, and the Poisson deviation is purely controlled by DM properties.
Indeed, in this limit, $\tau_\tout$ becomes exactly $\tau_\DM$, as we can confirm explicitly,
\be
\tau_\tout = \lim_{\tau \to\infty}\tau F(\tau) \simeq \frac{\sqrt{2\pi} \, \text{Erf}\big(\sqrt{2} v_\odot/v_0\big)}{m_a v_0 v_\odot} = 2.80\,\text{ms} \left(\frac{1\,\mu \text{eV}}{m_a}\right)\!,
\ee
where the final result holds for $v_0 \simeq 220\,\text{km/s}$ and $v_\odot \simeq 232\,\text{km/s}$.
This exactly matches the form of $\tau_\DM$ given in Ref.~\cite{Cheong:2024ose}.

%%%%%%%%%%%%%%%%%%%%%%%%%%%%%%%%%
\paragraph{Full Photon Count Distribution.}
%%%%%%%%%%%%%%%%%%%%%%%%%%%%%%%%%
%
We now compute the full photon count distribution $p(n)$ for the virialized Gaussian hypothesis of Eq.~\eqref{eq:VGH}.
The derivation is involved and deferred to App.~\ref{app:single_bin_PGF}.
The result in two asymptotic limits of $\tau/\tau_\tout$ can be derived more straightforwardly, however, similar to the asymptotic expressions for $F(\tau)$.
We emphasize that the variation of the number statistics as a function of $\tau/\tau_\tout$ is unsurprising: for $\tau \ll \tau_\tout$ we probe a single effective DM mode and for $\tau \gg \tau_\tout$ we probe many.

In the $\tau \ll \tau_\tout$ limit, the integrand $|\gamma(t)|^2$ is approximately constant---the relative phases between DM modes are roughly constant such that there is no meaningful interference.
Consequently, the expression for $p(n)$ in Eq.~\eqref{eq:single_bin_prob} reduces to the same integral that appears in Eq.~\eqref{eq:number_joint_record}.
The Gaussian hypothesis statistics are therefore geometric:
\be \label{eq:Geom_dist}
p(n) = \frac{1}{1+\mu(\tau)}\left(\frac{\mu(\tau)}{1+\mu(\tau)}\right)^n
\quad \text{for}\ \tau \ll \tau_\tout.
\ee

In the opposite limit, we can approximate the count $n$ as being the sum of $M \equiv \tau/\tau_\tout \gg 1$ independent geometric variables.
Then the count follows a negative binomial distribution,
\be \label{eq:NB_dist}
p(n) = \binom{n+M-1}{M-1}\,\frac{1}{(1+\mu(\tau_\tout))^M}\left(\frac{\mu(\tau_\tout)}{1+\mu(\tau_\tout)}\right)^n 
\quad \text{for}\ M \equiv \tau/\tau_\tout \gg 1.
\ee
Going forward, we refer to this marginal distribution using the shorthand $p(n) = \text{NB}(n, \mu(\tau), M)$.
For $M=1$, the negative binomial distribution reduces to the geometric distribution.

\begin{figure}[t!]
\centering
\includegraphics[width=1\linewidth]{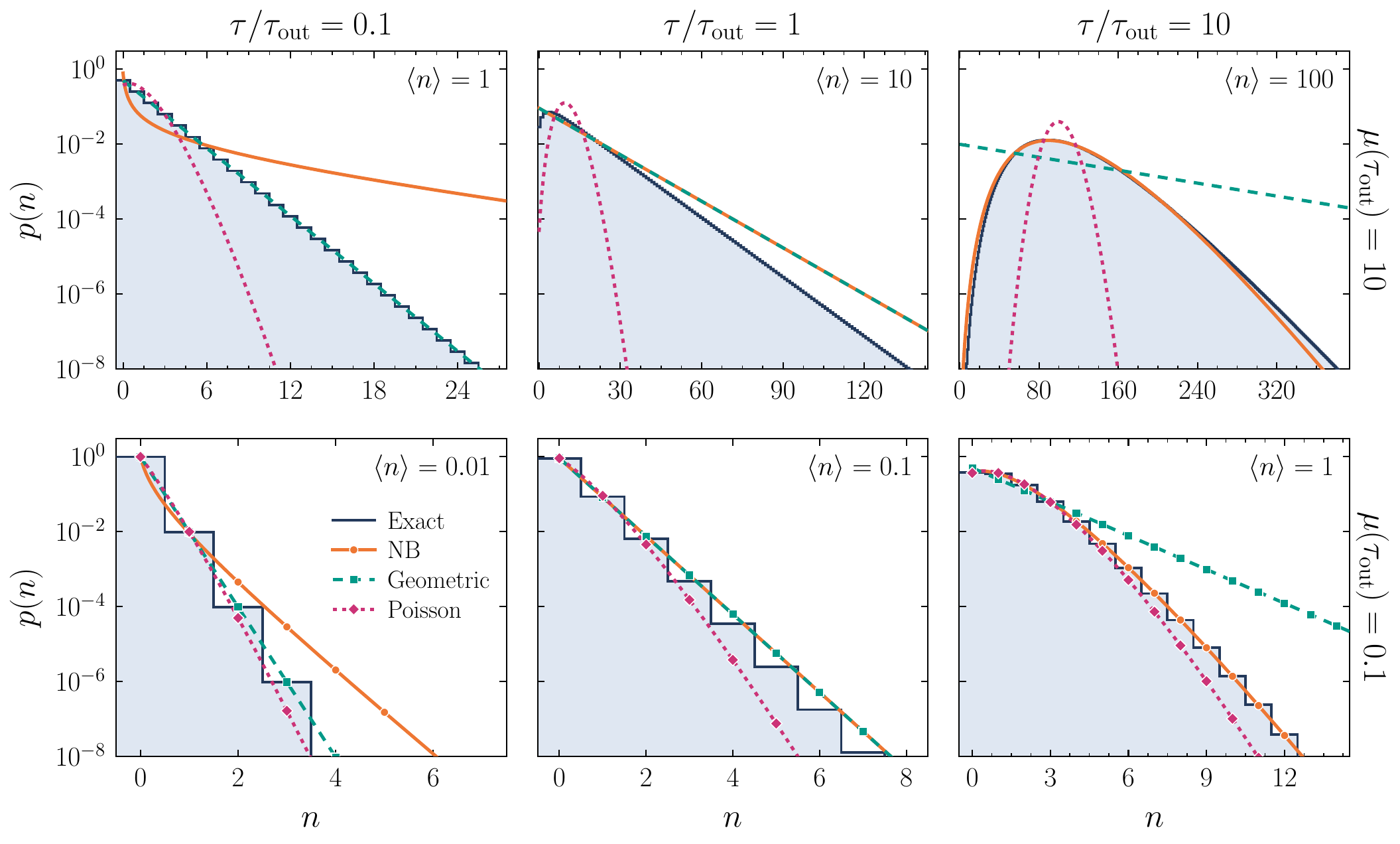}
\vspace{-1.cm}
\caption{The exact number distribution for the virialized Gaussian hypothesis for bin sizes $\tau/\tau_\tout \in \{0.1,\,1,\,10\}$, and mean counts per coherence time $\mu(\tau_\tout) \in \{10,\,0.1\}$.
In each case, we show the mean counts per bin $\la n \ra = \mu(\tau) = \mu(\tau_\tout) \tau / \tau_\tout$, the exact result (blue), and also the prediction from a negative binomial (orange), geometric (green), and Poisson (red) distribution.
For large and small $\tau/\tau_\tout$, the exact distribution is well approximated by the negative binomial and geometric results, respectively.
The Poisson distribution never provides a good description of the full distribution.
All results are computed assuming the standard halo model in Eq.~\eqref{eq:pk_SHM}.}
\label{fig:p_n_comparison}
\end{figure}

We show the exact result for different $M$ and $\mu(\tau_\tout)$ in Fig.~\ref{fig:p_n_comparison}.
(To show the negative binomial distribution for arbitrary $M$, we analytically continue Eq.~\eqref{eq:NB_dist}, including to $M < 1$.)
The exact result is compared against three other distributions: the negative binomial distribution (accurate for $M \gg 1$), the geometric distribution (accurate for $M \ll 1$), and the Poisson distribution (accurate for the full $p(n)$ nowhere).
In the limit $\la n \ra \ll 1$, where the axion would at best be at the edge of discovery, only $p(0) \simeq 1$ and $p(1) \simeq \la n \ra$ are non-negligible, and all distributions -- having the same mean -- agree.
Yet as soon as the signal starts to be resolved, the distinction becomes apparent.
Although this deviation from Poisson is an ancient result in quantum optics~\cite{Mandel:1959}, axion haloscopes performing number measurements often treat the distribution as Poisson; see e.g.~Refs.~\cite{Lamoreaux:2013koa,Braggio:2024xed}.

%%%%%%%%%%%%%%%%%%%%%%%%%%%%%%%%%
\paragraph{Distinguishing Gaussian and Coherent States.}
%%%%%%%%%%%%%%%%%%%%%%%%%%%%%%%%%
%
To further emphasize that the exact virialized Gaussian hypothesis prediction differs from the coherent-state prediction, we can compute the time it would take to distinguish these hypotheses even when the mean is small.
This is the same calculation we performed in Sec.~\ref{subsec:higherOrderIO} to distinguish the Gaussian and coherent hypotheses of Sec.~\ref{subsec:DMQstate}, although now we can perform the computation using the full probability distribution rather than just the first two moments.

Our strategy is to divide the full measurement into bins of length $\tau$.
If we measure the system for a time $T$, that then provides a measurement record $\{n_j\}$ with $N_\text{shot}=T/\tau$ values.
For $M$ sufficiently large, the bins are separated by far more than a coherence time and become independent.
Indeed, for the Gaussian hypothesis, Fig.~\ref{fig:p_n_comparison} shows that, for $M = 10$, the negative binomial -- a sum of independent geometric distributions -- is an excellent approximation to the full distribution; it is a good approximation even when $M$ is roughly a few.
We justify the approximate independence of these measurements in App.~\ref{app:mutual_info}.
We thus take $M$ to be at least a few and then treat each bin as an independent, identically distributed negative binomial variable.
This choice is solely so that we can obtain an analytic result; smaller bins are optimal as a larger $M$ can only wash information out from the measurement record.
The case of $M \ll 1$ can be studied numerically, although it does not lead to a parametric enhancement in the distinguishability.
For the coherent-state alternative hypothesis, we take the bins to be independent and identically Poisson distributed.
We fix both hypotheses to have identical means, so that they are only distinguishable through their fluctuations.

In summary, the two joint photon count distributions are given by
\be \label{eq:GvsC_number_dists}
p_G(\{n_j\}) = \prod_j \text{NB}(n_j, \mu(\tau), M),
\quad p_c(\{n_j\}) = \prod_j \text{Pois}(n_j, \mu(\tau)).
\ee
To distinguish the hypotheses, we define a test statistic $\Theta$ as twice the log-likelihood ratio of the probabilities, given a dataset $\{n_j\}$,
\be \label{eq:Theta-def}
\Theta \equiv 2\log(\frac{p_G(\{n_j\})}{p_c(\{n_j\})}).
\ee
Positive and negative $\Theta$ favor the Gaussian- and coherent-state hypotheses, respectively.

We now compute the integration time required for $\Theta$ to exceed a fixed threshold $\Theta_0$.
Note that the test statistic is itself a random variable, since it is a function of the joint measurement record $\{n_j\}$.
To calculate its scaling with time, we need to find its expectation value with respect to a specific $p(\{n_j\})$.
For concreteness, we fix the virialized Gaussian state as the true model; that is, we compute how long it would take to rule out the coherent-state model given that DM is in a Gaussian state.
Then the expected test statistic is
\be \label{eq:KL_G_vs_C}
\la \Theta\ra_G = 2\sum_{\{n_j\}}p_G(\{n_j\})\log(\frac{p_G(\{n_j\})}{p_c(\{n_j\})}) = 2 D(G||c).
\ee
Here, the sum is over all possible number measurement records $\{n_j\}$, and $D(G||c)$ is the Kullback--Leibler (KL) divergence between the Gaussian- and coherent-state hypotheses.
Because each of the distributions factorizes over the bins, the KL divergence is a sum over the divergences between the marginal distributions of each bin, $D(G||c) = N_\text{shot} D_j(G||c)$.

For large values of $\mu(\tau_\tout)$, distinguishing the hypotheses is trivial; see Fig.~\ref{fig:p_n_comparison}.
We therefore assume we are in the limit where $\mu(\tau_\tout) \ll 1$, in which case we can compute the leading contribution to the single-shot KL divergence,
\be
D_j(G||c)\simeq \frac{1}{2}M^2\mu^2(\tau_\tout)\left[\left(1+\frac{1}{M}\right)\log(1+\frac{1}{M}) - \frac{1}{M} \right]\!.
\ee
Substituting this into Eq.~\eqref{eq:KL_G_vs_C}, taking $M = 5$ for concreteness (matching our choice in Sec.~\ref{subsec:higherOrderIO}), and demanding $N_\text{shot}$ large enough that $\la \Theta\ra_G > \Theta_0$, we find the following measurement time required to distinguish the hypotheses,\footnote{For $M \gg 1$, the required time scales as $T \propto M$.
Intuitively, once the bins are large enough to be independent, coarse-graining further washes out information and leads to a worse test for distinguishing the Gaussian and coherent states.
If we instead take $M \ll 1$, the bins become correlated and do not each carry independent information.
In that case, the scaling in Eq.~\eqref{eq:gc_dist_time} remains, although the numerical prefactor can be slightly reduced.}
\be \label{eq:gc_dist_time}
T \simeq 11 \Theta_0 \frac{\tau_\tout}{\mu^2(\tau_\tout)}.
\ee
The $1/\mu^2$ scaling of the timescale to distinguish the Gaussian and coherent hypotheses is consistent with that derived in Ref.~\cite{suppression_paper} for discrete projective measurements, albeit with $\tau_\tout$ used in place of the single-shot measurement time $t_m \sim \kappa_m^{-1}$.
The result is also consistent with the $1/\mu^2$ scaling we claim can be achieved with input-output; see the discussion around Eq.~\eqref{eq:MandelQ}.
Here, however, we can obtain the scaling and exact numerical coefficients.

%%%%%%%%%%%%%%%%%%%%%%%%%%%%%%%%%
\paragraph{The Waiting-Time Distribution.}
%%%%%%%%%%%%%%%%%%%%%%%%%%%%%%%%%
%
The statistics of counting measurements can also be encoded in the probability that two clicks in the detector are separated by a time $\tau$ (distinct from the bin size discussed above), given a starting click.
The resulting waiting-time distribution $w(\tau)$ is a common tool for studying counting statistics and here we show it can be derived straightforwardly from the measurement-record distribution, $p(\{n_i\})$.
To connect the two, we observe that
\be \label{eq:pn_wt}
p\big(n_1=1, n_2 = 0, n_3 = 0, \ldots, n_{\tau - 1} = 0, n_\tau=1 \big) = p(n_1=1)w(\tau)\Delta t,
\ee
since both sides represent the joint probability of detecting a photon at time $t_1$ and subsequently waiting a time $\tau$ before detecting another.
We can then use Eq.~\eqref{eq:full_number_dist} to compute the waiting-time distribution for an arbitrary DM $P$-function,
\be \label{eq:waiting_time}
w(\tau) = \frac{\kappa_m}{\la |\gamma(0)|^2\ra_P}\left \la |\gamma(0)|^2|\gamma(\tau)|^2\exp\left(-\kappa_m \int_0^\tau \!\!dt\, |\gamma(t)|^2\right)\right\ra_P\!.
\ee
This expression has been derived previously in Refs.~\cite{PhysRevA.39.1200, PhysRevA.38.2423} using operator-algebra techniques; here, we see that it is an immediate consequence of the measurement record.
Note that the waiting-time distribution also encodes the Mandel $Q$ parameter,
\be
w(0) \Gamma_{\sig}^{-1} = \frac{Q}{\la |\gamma(0)|^2\ra_{P}} + 1.
\ee
If DM is homogeneous, the result simplifies to $w(0) = Q\,\kappa_m + \Gamma_{\sig}$.

\begin{figure}[t!]
\centering
\includegraphics[width=0.48\linewidth]{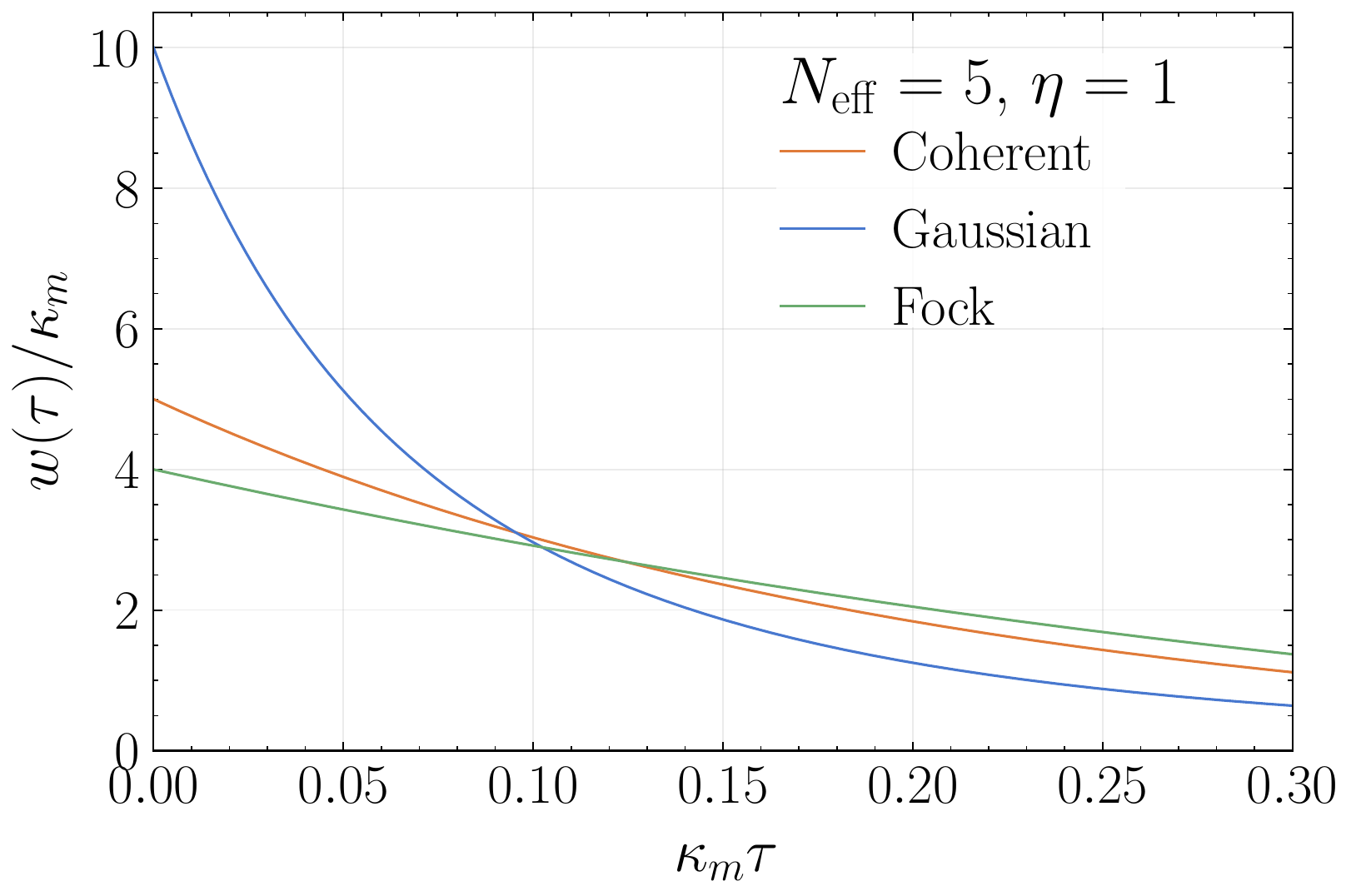}
\hspace{0.2cm}
\includegraphics[width=0.48\linewidth]{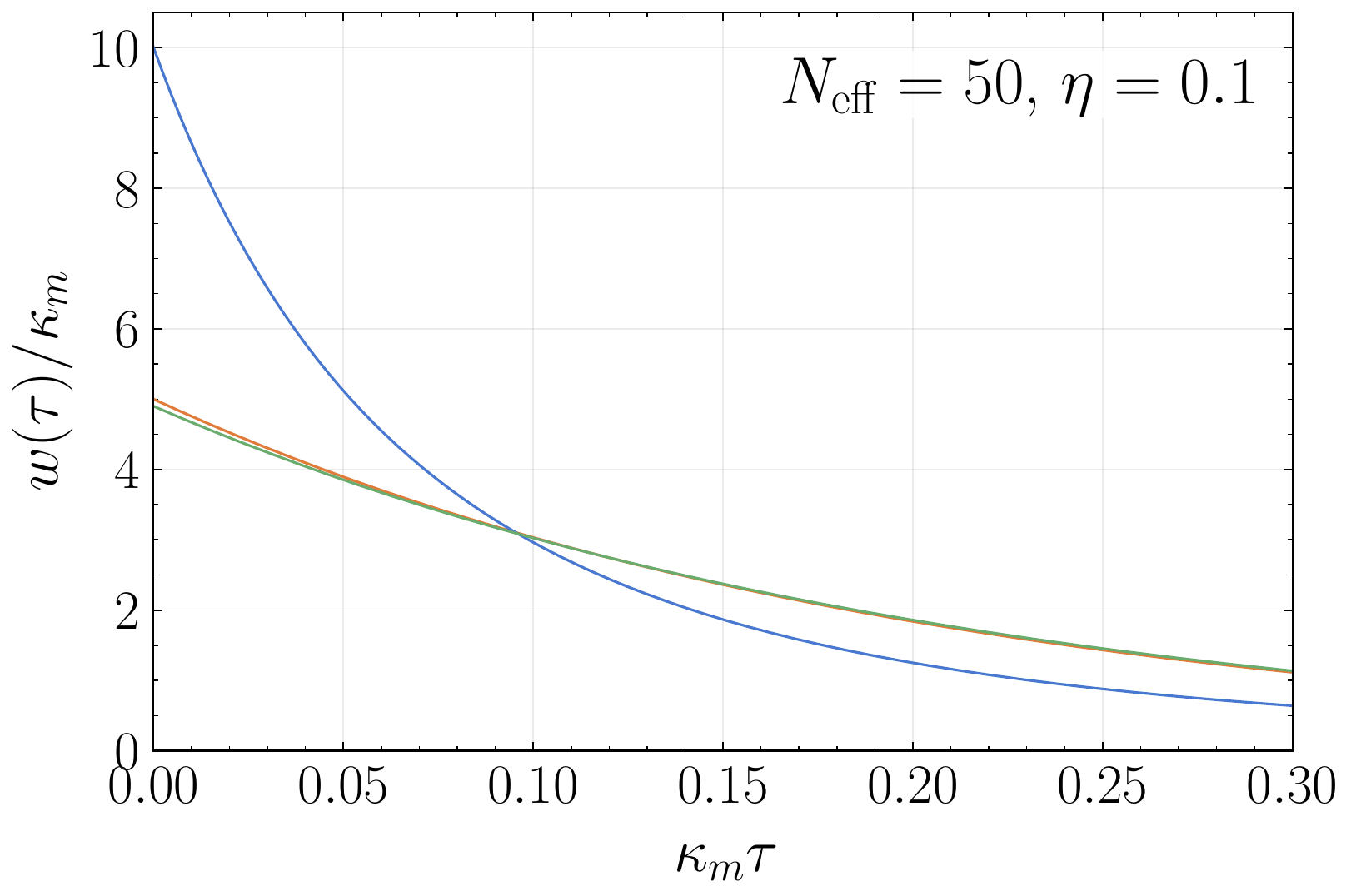}
\vspace{-0.4cm}
\caption{The waiting-time distribution for DM in a single-mode coherent, Gaussian, and Fock state.
We fix $\eta N_\eff = 5$ and show two values of $\eta$, both unphysically large.
Recall $\bar{\eta}(T) = \eta \kappa_m T$, so the result is for a fixed mean within a time $\kappa_m^{-1}$.
For small efficiencies, the coherent and Fock states become indistinguishable.
}
\label{fig:WTD}
\end{figure}

Computing the waiting-time distribution for an arbitrary $P_a$ is in general involved.
To provide some intuition, we can take the simplifying scenario where DM is described by a single resonant mode, as in Eq.~\eqref{eq:num_meas_cases}.
In this case, for different choices of the DM state, we obtain
\be \label{eq:wtd_cases}
w(\tau) = \frac{\bar\eta(\tau)}{\tau} \times \begin{cases} N_\eff \exp(- \bar\eta(\tau) N_\eff) & \text{coherent DM} \\
2N_\eff/(1 + \bar\eta(\tau) N_\eff)^{3} & \text{Gaussian DM} \\
(N_\eff-1) (1 - \bar\eta(\tau) )^{N_\eff-2} & \text{Fock DM}
\end{cases}
\ee
Each of these cases is shown in Fig.~\ref{fig:WTD} for two values of the efficiency.
As expected from our results in Refs.~\cite{Bao:2025nsd,suppression_paper}, the Fock and coherent states become extremely difficult to distinguish in the limit of small efficiency, while the coherent and Gaussian states do not. 
One can formalize this by computing the KL divergence between waiting-time distributions, to find the number of independent draws required to distinguish pairs of states.

In closing, we note that the subtlety for single-mode DM discussed below Eq.~\eqref{eq:num_meas_cases} also applies to the waiting-time distribution. 
Draws from the waiting-time distribution are only approximately independent if they are separated by much more than the DM coherence time. 
For single-mode DM, the coherence time is infinite, and all draws are correlated. 
As we discuss in Sec.~\ref{subsec:nonclassical_singlemode}, this only makes it more difficult to distinguish states.

%%%%%%%%%%%%%%%%%%%%%%%%%%%%%%%%%
\subsection{Quadrature Measurements and the PSD}
\label{subsec:quad_meas_stats}
%%%%%%%%%%%%%%%%%%%%%%%%%%%%%%%%%

We now consider a record of quadrature measurements.
Our starting point is the probability $p(\{r_i\})$ of making a specific series of quadrature measurements, given in Eq.~\eqref{eq:quadrature_measurement_record}.
Rather than the raw quadrature values, experiments tend to analyze the PSD and we show in what follows how the statistics of the PSD can be computed from the quadrature distribution.

We first obtain the joint distribution of the magnitudes of the discrete Fourier transform (DFT) of this measurement record through a marginalization of the full distribution $p(\{r_i\})$.
From here, computing the PSD is straightforward and we confirm the mean matches our input-output theory result from Sec.~\ref{subsec:OutputObservables}.
We then show that if DM satisfies the virialized Gaussian hypothesis, then the PSD in each frequency bin is exponentially distributed and \textit{independent}, justifying the common treatment in the literature.
Finally, we return to the raw data distribution $p(\{r_i\})$ and find the time required to distinguish certain axion states.

%%%%%%%%%%%%%%%%%%%%%%%%%%%%%%%%%
\paragraph{The Joint PSD Distribution.}
%%%%%%%%%%%%%%%%%%%%%%%%%%%%%%%%%
%
The complex DFT of a dataset $\{r_i\}$ collected in a time $T$ is defined as
\be \label{eq:DFT_def}
\tilde r_k = \sum_{i=1}^N \Delta t \, r_i \, e^{i\omega_k t_i} \defeq \rho_k\, e^{i \phi_k}
\ee
where the magnitude $\rho_k \geq 0$, $k \in \{1,2,\ldots,(N-1)/2\}$, and $\omega_k = 2\pi k /T$.\footnote{To understand the counting of $k$ modes, we have $N$ \textit{real} quadrature measurements, which then translate to $(N-1)/2$ \textit{complex} DFT components, plus one real DC component.
Further discussion is provided in App.~\ref{app:DFT_PDF}.}
The discrete PSD in each bin is then defined as
\be
S_k = \rho_k^2/T.
\ee
The joint distribution for $\{\rho_k\}$ is found by a change of variables, which involves a marginalization over each quadrature $r_i$, weighted by $p(\{r_i\})$, 
\be \label{eq:dft_mag_marginalization}
p(\{\rho_k\}) = \int \prod_i dr_i \, p(\{r_i\}) \prod_{k=1}^{(N-1)/2} \delta\!\left(\rho_k - \left|\sum_{i=1}^N \Delta t \, r_i \, e^{i\omega_k t_i} \right| \right)\!.
\ee
Using the form for $p(\{r_i\})$ in Eq.~\eqref{eq:quadrature_measurement_record}, we show in App.~\ref{app:DFT_PDF} that this result can be reduced to an integral of the DM $P$-function,
\be \label{eq:joint_dft_mag_dist}
p(\{\rho_k\}) = \int d\bm{\alpha}\, P_a(\bm{\alpha})\prod_{k=1}^{(N-1)/2} R\Bigl(\rho_k, \sqrt{\kappa_m}|\tilde X_\gamma(k)|,\tfrac{1}{4} T\Bigr).
\ee
Here $R(x,\nu,\sigma^2)$ is the Rician distribution\footnote{The Rician describes the distance $x$ from the origin of a random variable drawn from a 2D circular Gaussian distribution with variance $\sigma^2$ centered a distance $\nu$ away from the origin.}
\be \label{eq:rician_def}
R(x,\nu,\sigma^2) = \frac{x}{\sigma^2}\exp{-\frac{x^2+\nu^2}{2\sigma^2}}\, I_0\left(\frac{x\nu}{\sigma^2}\right)\!.
\ee
In the argument of the Rician we also define the DFT of the classical amplitude $X_\gamma(t)$, given in Eq.~\eqref{eq:X_gamma_t}, as
\be \label{eq:X_gamma_FT}
\tilde X_\gamma(k) = \sum_{i=1}^N \Delta t\, X_\gamma(t_i)\,e^{i \omega_k t_i}.
\ee
The distribution in Eq.~\eqref{eq:joint_dft_mag_dist} determines the PSD distribution, but also contains more information: it determines \textit{all} moments of $\rho_k$, while the PSD determines only even moments.

As a first example, we recover the axion contribution to the mean of a PSD bin, which we previously calculated in input-output theory.
Using the fact that the second moment of the Rician $R(x,\nu,\sigma^2)$ is $2\sigma^2 +\nu^2$, and substituting in the explicit expression for $\tilde X_\gamma(k)$,
\be
\la S_k\ra = \frac{1}{T}\la \rho_k^2\ra = \frac{1}{2} + \frac{\kappa_m}{T}\left\la \left| \sum_{i=1}^N \frac{\Delta t}{\sqrt 2} \int\! \frac{d^3 \bp}{(2\pi)^3} \left(\chi_{\bp}\alpha_{\bp}e^{-i\omega_{\bp}t_i} + \text{c.c.} \right)e^{i\omega_k t_i}\right|^2\right\ra_P\!,
\ee
where to avoid confusion with the DFT index $k$, we use $\bp$ rather than $\bk$ for the axion momenta.
We then assume both stationarity and homogeneity as in Sec.~\ref{sec:inout}, so that $\la \alpha_{\bp}^*\alpha_{\bq}\ra_P =N_{\bp}(2\pi)^3\delta^{(3)}(\bp-\bq)$ and all other second moments vanish.
We further take $\Delta t \to dt$ and $T$ longer than the signal coherence time in order to convert the double sum over time steps into a Fourier transform, yielding
\bea
\la S_k\ra &= \frac{1}{2} + \frac{\kappa_m}{2}\int \frac{d^3 \bp}{(2\pi)^3} \Bigl[|\chi_{\bp}|^2 N_{\bp}\, 2\pi \delta(\omega_k - \omega_{\bp})+ (\omega_k \leftrightarrow -\omega_k)\Bigr] \\
&= \frac{1}{2} + \frac{1}{2} \left[\frac{g^2 \kappa_m}{(\kappa_m/2)^2 + (\omega_k-\omega_c)^2} \neff[\omega_k] + (\omega_k \leftrightarrow -\omega_k) \right]\!.
\eea
This exactly matches the PSD in Eq.~\eqref{eq:PSD_mean}, up to the fact that here we are neglecting noise from the measurement and loss ports.
In particular, we still have the extra quantum noise due to negative-frequency modes, discussed in Sec.~\ref{subsec:OutputObservables}.

%%%%%%%%%%%%%%%%%%%%%%%%%%%%%%%%%
\paragraph{DM Factorization Implies PSD Factorization.}
%%%%%%%%%%%%%%%%%%%%%%%%%%%%%%%%%
%
The same axion mode $\alpha_\bp$ can contribute to many different $\rho_k$ bins, given its appearance in $X_\gamma(t)$.
Thus, certain axion $P$-functions enforce strong correlations between the frequency bins.
Nevertheless, we now show that when $P_a$ factorizes in momentum space, the PSD bins are approximately independent.

The simplest instance of the quantum CLT is to assume that each DM plane-wave mode is independent, in the sense that their joint $P$-function factorizes,
\be \label{eq:DM_P_factorize}
P_a(\bm{\alpha}) = \prod_{\bp}P_{\bp}(\alpha_{\bp}).
\ee
However, as discussed in Sec.~\ref{subsec:DMQstate}, this is more than is needed for the quantum CLT to hold and Gaussian DM emerge.
Instead, we only need $P_a$ to factorize at a scale below that resolved by the detector.
In reality there must be some physical scale below which the DM degrees of freedom begin to be correlated, as it is hard to envision dynamics that can independently populate the DM with infinite precision in momentum space.
As in Sec.~\ref{subsec:DMQstate}, we call the momentum-correlation scale $\Delta p_\text{corr}$.\footnote{We emphasize that $\Delta p_\text{corr}$ is not the inverse of the position-space coherence length discussed in Ref.~\cite{Cheong:2024ose}.}
It may be astronomically small -- of order the inverse size of the DM halo in the Milky Way -- but it still should be finite.
All that matters for the purposes of realistically implementing the assumption in Eq.~\eqref{eq:DM_P_factorize}, however, is that we cannot observe long enough to resolve the correlated modes.
Quantitatively, we require $1/T \sim \Delta p_\text{det} \gg \Delta p_\text{corr}$ or $T \Delta p_\text{corr} \ll 1$.

We can formally implement a finite correlation scale by coarse-graining the plane-wave modes over a correlation volume $V_p = (\Delta p_\text{corr})^3$.
To do this in practice, we rewrite integrals as
\be
\int \frac{d^3\bp}{(2\pi)^3}f(\bp) \simeq V_p \sum_{\bp} \bar f(\bp),
\ee
where the integrands become correlation-volume-averaged versions of themselves, as in 
\be
\bar f(\bp) = \frac{1}{V_p}\int_{V_p} \frac{d^3\bq}{(2\pi)^3}\,f(\bp + \bq).
\ee
We can imagine that $V_p$ is so small that in many cases it probes scales far smaller than those over which the integrand varies, implying we can take $\bar f(\bp) \simeq f(\bp)$.
After this change, the degrees of freedom of the DM we care about have a well-defined, factorizable $P$-function in the sense of Eq.~\eqref{eq:DM_P_factorize}, where the product is now over discrete correlation volumes.

We now wish to show that, under this assumption, the distribution of the $\rho_k$, given by Eq.~\eqref{eq:joint_dft_mag_dist}, also factorizes, 
\be
p(\{\rho_k\}) \simeq \prod_k p_k(\rho_k).
\ee
This does not follow immediately from factorization of the $P$-function because the same DM mode $\alpha_{\bp}$ can contribute to many different $\rho_k$.
However, for a sufficiently long integration time $T$, the DFT resolves finer-grained frequencies, and the dependence on the $\alpha_{\bp}$ in each $\rho_k$ bin becomes approximately disjoint.

To see this in more detail, note that we can write $\tilde X_\gamma(k)$ as
\be
\tilde X_\gamma(k) = \frac{V_p}{\sqrt 2}\sum_{\bp} \Bigl[ \chi_{\bp}\alpha_{\bp} f_-(k,{\bp}) + \chi_{\bp}^*\alpha^*_{\bp} f_+(k,{\bp})\Bigr],
\ee
with
\be
f_{\pm}(k,{\bp}) = \sum_{i=1}^N\Delta t\, \exp{ it_i(\omega_k \pm \omega_{\bp})}.
\ee
For $\omega \Delta t\ll 1$, this geometric series is well approximated by
\be
f_{\pm}(k,{\bp}) \simeq T\exp\!\big\{i(\omega_k \pm \omega_{\bp})T/2\big\}\,\text{sinc} \big((\omega_k \pm \omega_{\bp})T/2\big).
\ee
We then assume the following hierarchy
\be \label{eq:T_hierarchy}
\omega_k^{-1}, \omega_{\bp}^{-1} \ll T \ll \Delta p_\text{corr}^{-1}.
\ee
In this limit, the momentum-correlation scale still cannot be resolved, but we can approximate the coefficients $f_{\pm}(k,\bp)$ as kernels that isolate approximately disjoint sets of $\alpha_{\bp}$.
These sets correspond to spherical shells in momentum space with radius $\sim (\omega_k^2-m_a^2)^{1/2}$ and thickness $\sim \omega_{k}/(({\omega_{k}^2 - m_{a}^2 })^{1/2}T)$.
Each DFT component $\tilde X_\gamma(k)$ is then a sum of an approximately disjoint set of $\alpha_{\bp}$-mode contributions with frequencies very close to $\omega_k$.
It then follows that the joint distribution of the $\rho_k$ factorizes, so that the PSD bins are approximately independent.

%%%%%%%%%%%%%%%%%%%%%%%%%%%%%%%%%
\paragraph{Statistics of a PSD Bin.}
%%%%%%%%%%%%%%%%%%%%%%%%%%%%%%%%%
%
We now show that when the $P$-function factorizes, by the quantum CLT each PSD bin $S_k = |\tilde{r}_k|^2 / T$ is exponentially distributed.
As justified above, we can think of $\tilde{r}_k$ as approximately constructed only from DM modes with frequency near $\omega_k$.
The number of independently contributing modes in this spherical shell is
\be
N_\text{modes} \simeq \frac{\omega_k\sqrt{\omega_k^2 - m_a^2}}{T V_p} \gg 1
\ee
due to the hierarchy in Eq.~\eqref{eq:T_hierarchy}.
Since many independent modes contribute, the quantum CLT applies, so that each $\tilde r_k$ is a (circular) 2D Gaussian random variable.
If we further take the DM $P$-function to be stationary, then $\la \tilde r_k\ra = 0$.
Consequently $S_k = |\tilde r_k|^2/T$ is $\chi^2$-distributed with two degrees of freedom, which is exactly the exponential distribution.
The above results formally justify that the virialized Gaussian hypothesis, which holds exactly where the quantum CLT can be applied, has independent exponential statistics in each frequency bin, exactly as widely assumed in the literature; see e.g.~Ref.~\cite{Foster:2017hbq}.

We conclude that strong correlations between momentum modes of the DM $P$-function (large $\Delta p_\text{corr}$) must be present to have any chance of distinguishing the DM state from the uncorrelated case through a PSD measurement.
Otherwise, if the scale of factorization is small enough in momentum space, independent exponential statistics for each bin are inevitable.

%%%%%%%%%%%%%%%%%%%%%%%%%%%%%%%%%
\paragraph{Likelihoods for Generalized Gaussian States.}
%%%%%%%%%%%%%%%%%%%%%%%%%%%%%%%%%
%
For Gaussian DM, we have established the full behavior of the PSD statistics.
If we have an alternative DM hypothesis we wish to test with a quadrature measurement record, the PSD is not optimal, as it discards valuable phase information contained in the quadrature measurement record.
Here we demonstrate that we can apply the likelihood approach of Sec.~\ref{subsec:number_measurements} to the full record of quadrature measurements.

As we have seen, this approach requires evaluating integrals over the DM $P$-function.
These integrals are straightforward for generalized Gaussian states, defined in App.~\ref{app:QuantumStates}, which include the coherent, thermal Gaussian, and squeezed states.
For this class we can then explicitly evaluate Eq.~\eqref{eq:quadrature_measurement_record}, finding a Gaussian distribution for the measurement record,\footnote{That this result holds even for nonclassical generalized Gaussian states (such as squeezed states) is nontrivial, and is established in Ref.~\cite{Weedbrook2012GaussianInfo}.}
\be \label{eq:gen_gauss_record}
p(\br) = \frac{1}{\sqrt{\det(2\pi C_r)}}\exp{-\frac{1}{2}(\br-\bar{\br})^TC_r^{-1}(\br-\bar{\br})} ,
\ee
where we have written the quadrature record as a vector with $\br_i = r_i$.
The mean and covariance are specified in terms of $P$-function moments through\footnote{As the $P$-function is a distribution, it can be easier to evaluate expectation values using the optical equivalence theorem, $\la \mathcal O(\alpha)\ra_P = \la :\! \mathcal O(a)\!:\ra$.}
\be \label{eq:gen_gauss_mean_cov}
\bar{\br}_i = -\sqrt{\kappa_m} \big\la X_\gamma(t_i) \big\ra_P,
\quad (C_r)_{ij} = \frac{\delta_{ij}}{2\Delta t} + \big\la (\sqrt{\kappa_m}\, X_\gamma(t_i)+ \bar{\br}_i)(\sqrt{\kappa_m}\, X_\gamma(t_j)+ \bar{\br}_j)\big\ra_P.
\ee
In the continuum limit $\Delta t \to dt$, the first term in the covariance matrix of the measurement record is to be interpreted as a delta function, $\frac{1}{2}\delta(t-t')$.
Also note that for stationary DM states, the second term of the covariance matrix is precisely the real part of the Glauber correlation function $\text{Re} ( G^{(1)}(t_i-t_j))$ of the effective mode of the DM inside the cavity.

We can now compare two generalized Gaussian hypotheses for the DM state by computing the test statistic $\Theta$, as we did below Eq.~\eqref{eq:Theta-def}.
Taking the expectation value of $\Theta = 2\log(p_1/p_2)$ with respect to the true model (taken as $p_1$ without loss of generality) yields
\begin{align}
\la \Theta \ra_1 &= \left \la 2\log{\frac{p_1(\br)}{p_2(\br)}}\right\ra_1\\
&=\tr{(C_r^{(2)})^{-1}C_r^{(1)} - \log((C_r^{(2)})^{-1}C_r^{(1)}) - \mathds{1}} + (\bar{\br}^{(1)}- \bar{\br}^{(2)})^T (C_r^{(2)})^{-1}(\bar{\br}^{(1)}- \bar{\br}^{(2)}), \nonumber
\end{align}
where the superscript $(i)$ indicates which model the mean or covariance matrix belongs to.

This can be used to compute the time to distinguish any two generalized Gaussian states.
As an example, we take hypothesis $2$ to be a virialized Gaussian state, and hypothesis $1$ to be a joint squeezed-vacuum state with the same average occupation number in each mode.
For concreteness, suppose the directions of the mode squeezings all align at some time during the measurement window, of length $T$.
Then, in App.~\ref{app:quad_likelihood_test}, we show that the test statistic obeys
\be \label{eq:squeezed_test_stat}
\la \Theta\ra_S \simeq \frac{2\mu(\tau_\tout)}{1+2\mu(\tau_\tout)} - \log(1+\frac{2\mu(\tau_\tout)}{1+2\mu(\tau_\tout)}),
\ee
which \textit{does not} grow with the total time $T$.

Intuitively, this is because outside the window when the squeezing axes align, the misalignment of the squeezing axes drives the effective mode to look indistinguishable from a Gaussian.
More generally, this is a severe obstruction to observing nonclassicality from squeezed axion states.
The alignment of the squeezing axes only persists for a time $\sim 1 / \Delta \omega_a$, so observable squeezing must either be produced locally or have a very narrow bandwidth.
A related obstruction involving spatial structure was recently discussed in Ref.~\cite{Fang:2026qrn}.

\section{Discussion}
\label{sec:discussion}
%!TEX root = main.tex

The extreme weakness of the axion-photon coupling pushes experiments to the limits of measurement precision dictated by quantum mechanics.
Yet the growing interest in quantum-enhanced sensing has also led to confusion around the applicability of the classical-field description of the axion.
The goal of this work and Refs.~\cite{suppression_paper,Bao:2025nsd} is to dispel that confusion.
This work shows how to incorporate a quantum description of the axion, in the realistic case of a continuously monitored haloscope.
This allows one to justify the classical field approximation or, arguably better yet, work with the fundamental quantum description throughout.

One benefit of the quantum perspective is that it allows one to think more carefully about the state of the axion field, as discussed in Sec.~\ref{subsec:DMQstate}.
For instance, axion DM is widely described as being in a ``coherent state,'' yet the virialized Gaussian hypothesis is a far more likely description of the local DM state.
These two states lead to distinct observable signatures.
For instance, coherent states yield Poisson number fluctuations, while the virialized Gaussian hypothesis yields the broader distribution derived in Sec.~\ref{subsec:number_measurements}.
Moreover, our formalism can be applied to DM in an arbitrary quantum state.

We end with several points of discussion.
In Sec.~\ref{subsec:nonclassical_singlemode}, we consider how our results impact the observability of nonclassical DM effects.
In Sec.~\ref{subsec:comparison}, we briefly review our two continuous-measurement formalisms and compare them to other formalisms used in particle-physics contexts.
Finally, in Sec.~\ref{subsec:future} we look to future directions.

%%%%%%%%%%%%%%%%%%%%%%%%%%%%%%%%%
\subsection{Nonclassical Effects and $P$-Function Correlations}
\label{subsec:nonclassical_singlemode}
%%%%%%%%%%%%%%%%%%%%%%%%%%%%%%%%%

We now discuss the detectability of intrinsically quantum effects of DM.
This problem was a core focus of Refs.~\cite{Bao:2025nsd,suppression_paper}, where two obstructions were identified.
First, as discussed in detail in Sec.~\ref{subsec:DMQstate}, if the axion $P_a$ factorizes in momentum space at a scale below what our detector can resolve and is stationary at that scale, then by the quantum CLT we instead see a positive, classical Gaussian.
Second, even if $P_a$ has sufficient momentum-space correlations to avoid this, intrinsically quantum effects are imprinted on detectors with an additional suppression of $\eta$ that is not enhanced by $N_\eff$.
Both challenges remain present with the addition of a transmission line.
Our study of continuous measurement here suggests a further obstruction: there can be a tension between avoiding the quantum CLT and obtaining enough information to establish a nonclassical $P_a$.
In particular, the single-mode example discussed for number measurements in Sec.~\ref{sec:measurement} evades the CLT, but its infinite coherence time implies measurements only provide a single effective draw from $p(n_\text{tot})$ that can be used to scrutinize $P_a$.

We emphasize that detecting nonclassical effects of axion DM is a different and far harder task than discovering the axion.
DM discovery only requires measuring a non-zero mean signal power.
Confirming nonclassicality requires establishing the axion $P$-function achieves negative values; this can be done, for example, by demonstrating DM has a smaller than classically achievable variance in number or quadrature measurements, as occur for a Fock or squeezed state, respectively.
Obtaining independent measurements from the underlying DM $P_a$ can only help determine whether the axion is in the class of nonclassical states.
However, it is not essential: as discussed in Ref.~\cite{Bao:2025nsd}, if we could prepare our detector in a nonclassical state with occupancy $\sim1/\eta$, in principle we could construct a one-shot measurement protocol to test DM nonclassicality.
Needless to say, preparing such a state is well beyond current technology.

%%%%%%%%%%%%%%%%%%%%%%%%%%%%%%%%%
\paragraph{Independent Measurements of the DM State.}
%%%%%%%%%%%%%%%%%%%%%%%%%%%%%%%%%
%
Consider the joint probability of the measurement record $\{\mathcal{O}_i\}$, which can be written as a generalization of Eq.~\eqref{eq:joint_record_prob},
\be
p(\{\mathcal{O}_i\}) = \int d\bm{\alpha} \, P_a(\bm{\alpha}) \prod_i p(\mathcal{O}_i |\bm{\alpha}),
\ee
where $p(\mathcal{O}_i |\bm{\alpha})$ is the probability of measuring outcome $\mathcal{O}_i$ if the DM state were in a coherent state indexed by $\bm{\alpha}$.
This joint probability is an integral over the joint $P$-function, implying that the set of measurement outcomes will in general have nontrivial correlations.
Equivalently, new measurements need not provide independent information that can be used to further interrogate $P_\eff$.
In our study of projective measurements in Refs.~\cite{Bao:2025nsd,suppression_paper}, we made the simplifying assumption that the measurements are perfectly independent,
\be
p(\{\mathcal{O}_i\}) = \prod_i p_i(\mathcal{O}_i),
\ee
and identically distributed, $p_i(\mathcal{O}_i) = p(\mathcal{O}_i)$.
Even then, detecting nonclassical states was shown to be strictly out of reach for near-term experiments; the only viable path involved detectors initially prepared in macroscopic quantum states.

However, the requirement of independent, identically distributed observables is a strong condition and generically enforces a stringent constraint on $P_a$: it must factorize into many independent effective mode $P$-functions that couple to our cavity one at a time, one for each measurement.
Recalling that the marginal probability for a single measurement $p(\mathcal{O}_i)$ can be expressed in terms of an effective mode,
\be\label{eq:single_O_prob}
p(\mathcal{O}_i)=\int d\bm{\alpha} \, P_a(\bm{\alpha})\, p(\mathcal{O}_i |\bm{\alpha}) = \int d\alpha_\eff \, P_\eff(\alpha_\eff)\,  p(\mathcal{O}_i|\alpha_\eff)
\ee
in the sense of Eq.~\eqref{eq:Peff}, we may then write the complete joint probability as
\be\label{eq:factorized_eff_prob}
p(\{\mathcal{O}_i\}) = \int d\alpha_\eff^{(1)}\, P_\eff(\alpha_\eff^{(1)}) p(\mathcal{O}_1|\alpha_\eff^{(1)})\int d\alpha_\eff^{(2)}\, P_\eff(\alpha_\eff^{(2)}) p(\mathcal{O}_2|\alpha_\eff^{(2)})\dotsb.
\ee
As shown in App.~\ref{app:mutual_info}, the virialized Gaussian state approximately satisfies this condition for measurements that are separated by at least the coherence time $\tau_\tout$.
However, for more general axion states which contain nonclassical effective mode states, we are unaware of any that satisfy the above factorization even approximately.\footnote{A multimode coherent state defined by $P_a(\bm{\alpha}) = \delta(\bm{\alpha}-\bm{\beta})$ does not satisfy Eq.~\eqref{eq:factorized_eff_prob}, as the $P_\eff$ appearing in each factor would not be identical, implying $p(\mathcal O_i) \neq p(\mathcal O_j)$ due to the varying interference between the DM modes at different times.
As discussed in Sec.~\ref{subsec:DMQstate}, if we average over these amplitude fluctuations, we recover the virialized Gaussian state.}

We have not attempted in this work to resolve the dynamical origin of intrinsically quantum DM states which, e.g.,~requires tracking the state through virialization~\cite{Eberhardt:2022rcp,Eberhardt:2023axk} as the growth of quantum corrections competes against decoherence~\cite{Allali:2020shm,Allali:2021puy}.
However, we have argued that it is not straightforward to construct such multimode nonclassical states; as discussed in Sec.~\ref{subsec:DMQstate}, even a multimode coherent state often behaves similarly to a Gaussian.
Regardless, if we assume that such a state does exist, and that $P_\eff$ is also that of a highly nonclassical state, then such a model would assist with the inference of density matrix properties in providing multiple independent measurements.
Other cases, where the underlying momentum modes are further correlated, can reduce the number of independent measurements available.
This suggests that the scenario studied in Refs.~\cite{suppression_paper,Bao:2025nsd} -- where we concluded observing nonclassical effects was well out of reach for any near-term experiment -- rested on a potentially optimistic assumption about how much information we can extract about the DM state.
This only further increases the difficulty of observing nonclassicality in axion DM.

%%%%%%%%%%%%%%%%%%%%%%%%%%%%%%%%%
\paragraph{Single-Mode DM.}
%%%%%%%%%%%%%%%%%%%%%%%%%%%%%%%%%
%
An extreme example of the failure of independence is shown in Eq.~\eqref{eq:number_joint_record}.\footnote{In principle, one could also study this from the perspective of the waiting-time distribution, although the full joint waiting-time distribution contains the same information as the full number record.}
There, for single-mode DM, the entire measurement record only contains information of a single draw from $p(n_\text{tot})$.
How the counts are distributed in time is \textit{completely} independent of the DM state.
Even if this single draw were sampled from a nonclassical distribution, we would have to determine that using just one probe of the $P$-function.
In this case it can become intractable to even distinguish between classical DM states.
The above discussion reveals that this is the general behavior of single-mode DM states: their effectively infinite coherence time forces all measurement results to reflect a single draw of $\alpha$ from the $P$-function.
The effect is analogous to cosmic variance in modes of the cosmic microwave background, in that although there is a probability distribution determining the amplitude of each mode, one mode allows for one measurement, preventing us from interrogating the underlying distribution from which it is drawn.

Given this, we can understand a further sense in which observing nonclassical effects could be even harder than argued in Refs.~\cite{Bao:2025nsd,suppression_paper}.
Although weak coupling was the ultimate obstruction to observing nonclassical effects in those works, there was a first barrier one had to overcome for that to be the relevant challenge: the quantum CLT, which as discussed in Sec.~\ref{subsec:DMQstate} applies when the signal's $P$-function contains many independent modes the detector cannot resolve.
We argued that to avoid the effective mode's $P$-function being driven to a Gaussian, one needs correlations between the DM modes, with the extreme example being a density matrix with only a single mode populated.
A physical motivation for the case where DM has both a longer coherence time and populates a smaller number of modes than the virialized Gaussian is an axion Bose--Einstein condensate~\cite{Semikoz:1995rd,Sikivie:2009qn,Guth:2014hsa,Budker:2023sex,Paranjape:2026hkr}.
However, we can now see this introduces a new challenge: unless the detector can resolve independent modes, the entire joint probability is an integral of $\prod_i p(\mathcal{O}_i|\alpha)$ over a \textit{single} $P$-function $P_a(\alpha)$, so measurement outcomes do not provide independent measurements, reducing the information available to study the DM state.

As a final comment, we emphasize that these correlations do not prevent us from distinguishing single-mode DM from the virialized Gaussian hypothesis, even when both have the same fixed $\Gamma_\sig$.
For quadrature measurements, this is intuitively clear: the PSD for the multimode Gaussian can reveal the DM lineshape after a measurement time $T > \tau_\DM$, whereas the width of the single-mode distribution remains unresolved.
However, using the formalism of this paper we can also do the same for number measurements, which carry no phase information.
The number distribution for the Gaussian scenario is discussed in Sec.~\ref{subsec:number_measurements}.
For the single-mode hypothesis, we take $P_a$ to be a coherent state for specificity, in which case the joint distribution (for bins of arbitrary length $\tau$) can be written exactly from Eq.~\eqref{eq:number_meas_result} as
\be
p(\{n_j\}) = \prod_j \text{Pois}(n_j, \Gamma_\sig \tau).
\ee
Coincidentally, this is the same model studied and compared to the Gaussian case in the discussion surrounding Eq.~\eqref{eq:GvsC_number_dists}, so we see that Poisson statistics from DM are realized exactly by a coherent state occupying a single mode.
Thus, the time it would take to distinguish these hypotheses with number measurements is already given in Eq.~\eqref{eq:gc_dist_time}, and could be achieved in a time $T \sim \tau_\tout/\mu^2(\tau_\tout)$.

%%%%%%%%%%%%%%%%%%%%%%%%%%%%%%%%%
\subsection{Comparison to Other Formalisms}
\label{subsec:comparison}
%%%%%%%%%%%%%%%%%%%%%%%%%%%%%%%%%

The primary content of this paper is two fully quantum-mechanical descriptions of the continuous measurement of axion DM: input-output theory and the measurement-record formalism.
Here we summarize the key differences between our two approaches and compare both to alternative formalisms, such as master equations and influence functionals, which have also been used in a particle-physics context.

The goal of any formalism is to outline how DM imprints itself on a detector and its observables, as a function of the DM quantum state.
Of course, once the initial state of the system, the dynamics, and the measurement are specified, every complete quantum description must yield the same probability for the measurement record.
In general, if $\mathcal{I}_{\{\mathcal{O}_i\}}$ denotes the (possibly non-unitary) quantum operation describing the evolution conditioned on an observable record $\{\mathcal{O}_i\}$, then as in Eqs.~\eqref{eq:conditioned_rho} and~\eqref{eq:joint_record_prob},
\be \label{eq:instrument}
\rho_{\text{sys}}(\{\mathcal{O}_i\}) = \mathcal{I}_{\{\mathcal{O}_i\}}\big[\rho_{\text{sys}}(0)\big], \qquad
p(\{\mathcal{O}_i\}) = \tr\{\rho_{\text{sys}}(\{\mathcal{O}_i\})\},
\ee
where $\rho_{\text{sys}}(0)$ is the initial density matrix of the system.
Different formalisms correspond to differing choices in how Eq.~\eqref{eq:instrument} is evaluated, or to computing only marginal statistics contained in the full distribution $p(\{\mathcal O_i\})$.
They differ in the picture in which they evolve, in which degrees of freedom they trace out, and in whether a concrete measurement is specified.

Input-output theory, outlined in Sec.~\ref{sec:inout}, operates in the Heisenberg picture and bypasses directly evaluating the trace in favor of correlation functions of the observables.
It relates the inputs that drive the cavity, including the axion field, to the measurable output fields.
It is conceptually similar to scattering theory, framed in terms of asymptotic in and out states.
Input-output theory is naturally suited to the frequency domain, where the Heisenberg equations of motion can be used to remove dependence on the cavity, as in Eqs.~\eqref{eq:cont_inout_relations} and \eqref{eq:first_bout_eq}.
It is also a particularly elegant framework for computing the mean value of detector observables, such as the signal power.
Under the assumptions of homogeneity and stationarity, DM states with identical mean occupation spectrum produce the same mean photon flux and PSD, as controlled by $\neff[\omega]$ given in Eq.~\eqref{eq:n_eff}.
DM states can be differentiated via higher-order statistics; as shown in Sec.~\ref{subsec:higherOrderIO}, higher-order correlation functions can be computed within input-output theory, albeit with more computational effort.

The measurement-record approach of Sec.~\ref{sec:measurement}, which is novel, instead evaluates the trace directly, reducing $p(\{\mathcal{O}_i\})$ to an integral over the DM $P$-function as in Eq.~\eqref{eq:joint_record_prob}.
This provides the full probability distribution of the observables, which can be used to construct optimal statistical analyses.
The calculation is straightforward because the underlying dynamics, governed by the effective Hamiltonian of Eq.~\eqref{eq:Hint_simplified}, are simple.
Coherent states remain coherent states, and their amplitudes evolve according to the classical equations of motion in Eq.~\eqref{eq:classical_eoms}, so that the quantum dynamics are no harder to solve than the equivalent classical problem.
As experimental measurements are recorded sequentially in time, the measurement-record approach operates naturally in the time domain, as we exemplified by computing the full number distribution for arbitrary bin sizes, shown in Fig.~\ref{fig:p_n_comparison}.

Several alternative approaches involve treating DM as an environment to be traced out, or more generally tracing out every degree of freedom except those measured.
The effect of DM on the reduced density matrix of the detector is then encoded in correlation functions or influence functionals.
Such methods have been widely used in the axion literature, from Rydberg-atom detectors~\cite{Kitagawa:1999am,Bradley:2003kg} and matter-wave/atom interferometers~\cite{Badurina:2026owr,Murgui:2025unt}, to the evolution of axion DM in the galaxy~\cite{Eberhardt:2023axk} and in cosmology~\cite{Cao:2022bua,Cao:2022kjn} (see also Refs.~\cite{Burgess:2022nwu,Salcedo:2024smn,Salcedo:2024nex}).
These works either take the Feynman--Vernon influence functional, obtained by integrating out the environment on the Schwinger--Keldysh contour~\cite{Feynman:1963fq,Su:1987pi,BenTov:2021jsf}, or the master equation that evolves the reduced density operator directly, which in a phase-space representation yields a Fokker--Planck equation.

These methods tend to focus on the probabilities of a specific set of correlated measurement outcomes; equivalently, they can be framed as projective or single-shot measurements.
A reduced state naturally delivers one-time quantities and low-order correlation functions, such as the mean excited-atom numbers~\cite{Kitagawa:1999am} or detector correlation functions~\cite{Bernal:2024hcc}.
These are often reflective of the applications; for the example of atom interferometers, once prepared the atoms are left to evolve under the interaction with DM before a single measurement of their phase difference is performed at the end.
As such, these alternative methods are more relevant for the projective measurement context considered in our earlier works~\cite{Bao:2025nsd,suppression_paper}, rather than the continuous measurements considered here.
In that context, the main benefit of our earlier approach is that it kept dependence on the DM's $P$-function explicit throughout, allowing a more transparent separation of intrinsically quantum effects.

%%%%%%%%%%%%%%%%%%%%%%%%%%%%%%%%%
\subsection{Further Directions}
\label{subsec:future}
%%%%%%%%%%%%%%%%%%%%%%%%%%%%%%%%%

This paper represents a step toward a theoretical description of axion DM that approaches the sophistication of modern axion experiments.
For concreteness, we focused on a standard cavity haloscope in a static magnetic field and read out through a semi-infinite transmission line, which we take to have a thermal or vacuum input.
These are all choices that could be generalized and extended, as we now discuss.

An immediate next step would be a detailed study of DM analyses performed with number measurements.
We performed the first computation of the single-bin number statistics for virialized Gaussian DM, shown in Fig.~\ref{fig:p_n_comparison}.
This is the analog of the known exponential distribution of PSD measurements.
However, much more is known about PSD measurements, or quadrature measurements more generally.
As exemplified by Eqs.~\eqref{eq:PSD_mean} and \eqref{eq:pomega}, the PSD depends on the DM energy distribution, so that PSD measurements allow access to the DM speed distribution~\cite{Foster:2017hbq,OHare:2017yze,Knirck:2018knd,OHare:2025jpr}.
Additional information can be extracted with phase-sensitive quadrature measurements at spatially separated detectors using DM interferometry, which accesses the full three-dimensional velocity distribution of DM~\cite{Foster:2020fln,Fukuda:2025zcf}.
Number measurements are phase-insensitive, but as shown in Sec.~\ref{subsec:number_measurements}, the number distribution depends on the autocorrelation function of the axion field, which is the Fourier transform of the PSD.
Thus, number measurements also carry information about the DM speed distribution, and it would be interesting to understand to what extent it can be extracted.
It would also be important to understand any distinctive fingerprints DM leaves in the photon count distribution that can be used to veto unexpected backgrounds, in much the same way one uses the expected DM PSD lineshape for quadrature analyses; Fig.~\ref{fig:Poisson_deviation} already gives a hint in this direction, with DM having a characteristic change in its variance when the binning matches $\tau_\tout$.
Further, the existence of intensity interferometry~\cite{Brown:1956zza,HanburyBrown:1956bqd} demonstrates that in principle there can be correlated information between instruments performing number measurements.

One could also develop a fully quantum description of other types of axion detectors, e.g.~involving nuclear spins~\cite{Graham:2013gfa,Budker:2013hfa,Garcon:2019inh,Aybas:2021nvn,Dror:2022xpi,Beadle:2025dgy}, electron spins~\cite{Barbieri:2016vwg,QUAX:2020adt,Berlin:2023ubt}, ``heterodyne'' cavities loaded with an AC field~\cite{Berlin:2019ahk,Berlin:2020vrk}, or lumped-element circuits~\cite{Sikivie:2013laa,Chaudhuri:2014dla,Kahn:2016aff,DMRadio:2022jfv}.
The last of these has also been operated as a broadband detector under the moniker of ABRACADABRA~\cite{Kahn:2016aff,Ouellet:2018beu,Ouellet:2019tlz,Salemi:2021gck}; in that case there is no narrow resonance, so our approach cannot be applied directly, since the RWA is no longer appropriate.
Another general consideration is that many techniques for transducing the signal from the detector do not transfer the information to asymptotic states that move far from the detector, as done in input-output theory.
Instead, for many readout techniques backaction is a critical noise source that must be included and is by construction absent given our use of semi-infinite transmission lines.

Beyond specific detectors, it will also be important to establish quantum limits on DM detectability.
We have largely avoided discussion of beyond-the-standard-quantum-limit scaling.
Understanding how various instruments scale as next-generation quantum technologies become available will prove important as these devices push to where we most expect to discover the axion.
A variety of schemes have appeared~\cite{Lamoreaux:2013koa,Wurtz:2021cnm,Jiang:2022vpm,Shi:2022wpf,Zheng:2025qgv,He:2025ovo}, some of which are claimed to be optimal, while the information-theoretic perspective is only beginning to take shape~\cite{Gardner:2024frf,Allen:2025hdx,Bogorad:2026ggm}.
Here we simply note that the quantum state of DM could have an impact on the optimal method for its detection.
This brings back the issues discussed in Sec.~\ref{subsec:DMQstate} of resolving the exact quantum state of DM, although as argued there, it is difficult to envision a virialized DM state that is not Gaussian.

Lastly, as argued in Ref.~\cite{suppression_paper}, we expect many of our general results -- suppressed quantum effects, utility of effective modes, simplicity of the complete quantum description -- to extend to other signals such as quadratically coupled DM (see, e.g., Refs.~\cite{Hees:2018fpg,Banerjee:2022sqg,Kim:2022ype,Kim:2023pvt,Beadle:2023flm}), neutrinos, or gravitational waves.
Showing this in detail would validate the generality of these lessons and provide those experiments with additional tools to optimize their searches.

\begin{acknowledgments}
We thank Daniel Carney, Sebastian Ellis, and Giacomo Marocco for discussions.
YB and LTW are supported by the Department of Energy grant DE-SC0009924.
DYC is supported by the Enrico Fermi and KICP fellowship from the Enrico Fermi Institute and the Kavli Institute for Cosmological Physics at the University of Chicago and the Kavli Foundation.
The research of NLR, JT, and KZ was supported by the Office of High Energy Physics of the U.S.~Department of Energy under contract DE-AC02-05CH11231.
The work of JT was further supported by the NSF Graduate Research Fellowship Program under Grant DGE2146752.
We acknowledge the use of Fable 5 and GPT-6 Astra for refining derivations and proofreading draft versions of the manuscript.
\end{acknowledgments}

% \newpage
\appendix
\addtocontents{toc}{\protect\setcounter{tocdepth}{1}}
\section*{Appendix}
%!TEX root = main.tex

%%%%%%%%%%%%%%%%%%%%%%%%%%%%%%%%%
\section{Common Technical Results}
%%%%%%%%%%%%%%%%%%%%%%%%%%%%%%%%%

%%%%%%%%%%%%%%%%%%%%%%%%%%%%%%%%%
\subsection{Continuous and Discrete Mode Normalization}
\label{app:ContinuousVsDiscrete}
%%%%%%%%%%%%%%%%%%%%%%%%%%%%%%%%%

The plane-wave mode expansion of a scalar field can be written either in infinite volume or in a fiducial quantization volume $\mathcal{V}$ where the modes are discretely spaced in momentum, 
\be
\phi(\bx) = \begin{dcases} 
\int\! \frac{d^3 \bk}{(2\pi)^3}\, \frac{1}{\sqrt{2 \omega_{\bk}}} \qty(a_{\bk} e^{i \bk \cdot \bx} + a^\dag_{\bk} e^{- i \bk \cdot \bx}) & \text{continuous} \\ 
\frac{1}{\sqrt{\mathcal{V}}} \sum_{\bk} \frac{1}{\sqrt{2\omega_{\bk}}} \qty(a_{\bk} e^{i \bk \cdot \bx} + a^\dag_{\bk} e^{- i \bk \cdot \bx}) & \text{discrete}.
\end{dcases}
\ee
The normalization of the mode operators for each case leads to the commutation relations
\be \label{eq:acomm_condis}
\big[ a_{\bk}, a_{\bq}^\dag\big] = \begin{dcases} 
(2\pi)^3 \delta^{(3)}(\bk - \bq) & \text{continuous} \\ 
\delta_{\bk, \bq} & \text{discrete}.
\end{dcases}
\ee
with the mass dimensions of the mode operators being $-3/2$ and $0$, respectively.
Comparing these expressions, one can convert between continuous ($a^{(c)}$) and discrete modes ($a^{(d)}$) via
\be \label{appeq:continuumVsDiscrete}
\int\! \frac{d^3 \bk}{(2\pi)^3} \longleftrightarrow \frac{1}{\mathcal{V}} \sum_{\bk}, \qquad (2\pi)^3 \delta^{(3)}(\bk - \bq) \longleftrightarrow \mathcal{V} \, \delta_{\bk, \bq}, \qquad a^{(c)}_{\bk} \longleftrightarrow \sqrt{\mathcal{V}} \, a^{(d)}_{\bk}
\ee
An analogous construction applies for discrete-frequency modes with the replacement $\mathcal{V} \to T$, $d^3 \bk/(2\pi)^3 \to d\omega/(2\pi)$, and $(2\pi)^3 \delta^{(3)}(\bk - \bq) \to 2\pi \delta(\omega - \omega')$.

An often useful result is that for a homogeneous axion state, 
\be
\la a_{\bk}^\dagger a_{\bq} \ra = \begin{cases} N_{\bk} (2\pi)^3 \delta^{(3)}(\bk - \bq) & \text{continuous} \\ 
N_{\bk} \delta_{\bk, \bq} & \text{discrete} \end{cases} 
\ee
where the dimensionless quantity $N_{\bk}$ is the occupancy per unit phase-space volume in the continuous picture, and the mode occupancy in the discrete picture.
Analogously, as we discuss in App.~\ref{app:PSDfromDiscreteOp}, the frequency modes for stationary states of the transmission line can be evaluated as 
\be \label{appeq:frequencyModes}
\la b_{\omega}^\dagger b_{\omega'} \ra = \begin{cases} N_{\omega} (2\pi) \delta(\omega - \omega') & \text{continuous} \\ 
N_{\omega} \delta_{\omega, \omega'} & \text{discrete} \end{cases} 
\ee
in which $N_\omega$ can be interpreted as the number flux per frequency in the continuous picture and as the occupancy of the discrete-frequency mode in the discrete picture.

%%%%%%%%%%%%%%%%%%%%%%%%%%%%%%%%%
\subsection{Quantum States and their $P$-Functions}
\label{app:QuantumStates}
%%%%%%%%%%%%%%%%%%%%%%%%%%%%%%%%%

In Sec.~\ref{subsec:review}, we introduced the phase-averaged coherent state in Eq.~\eqref{eq:phaseMixtureP} and the (thermal) Gaussian state in Eq.~\eqref{eq:gaussian_state_def}.
Another useful example is the Fock state $\ket{N}$, which has 
\be
P(\alpha) = L_N\qty(-\pdv{\alpha}\pdv{\alpha^*}) \delta(\alpha)
\ee
where $L_N$ is the $N$\textsuperscript{th}-order Laguerre polynomial.
Besides the coherent state, all of these states are stationary, implying $\ev{(a^\dag)^m a^n}$ vanishes for $m \neq n$.
Some low-lying moments are:
\begin{equation} \begin{array}{c|ccc}
\hline\hline
\text{state} & \ev{a} & \ev{a^\dag a} & \ev{(a^\dag)^2 a^2} \\
\hline
\text{coherent} & \alpha_0 & \abs{\alpha_0}^2 & \abs{\alpha_0}^4 \\
\text{Fock} & 0 & N & N(N-1) \\
\text{phase-averaged coherent} & 0 & N & N^2 \\
\text{(thermal) Gaussian} & 0 & N & 2N^2 \\
\hline\hline
\end{array} \end{equation}

In Sec.~\ref{sec:measurement}, we also refer to generalized Gaussian states.
This is a broad class of states, including coherent, squeezed, and thermal states, characterized by Gaussian Wigner functions, 
\be
W(\alpha) = \frac{1}{\sqrt{\det(2\pi C_0)}}\exp{-\frac{1}{2}(\boldsymbol{\alpha}-\bar{\boldsymbol{\alpha}})^TC_0^{-1}(\boldsymbol{\alpha}-\bar{\boldsymbol{\alpha}})},
\ee
where $\boldsymbol{\alpha} = (\text{Re}(\alpha), \text{Im}(\alpha))$ and $C_0$ is a positive-definite covariance matrix.
With these conventions, the vacuum state has $C_0 = \mathds1/4$.
One can then show that all associated $s$-ordered characteristic functions -- see Eq.~\eqref{eq:s_ordered} -- are Gaussian as well, and in particular, 
\be \label{eq:gen_gaussian_char}
\tilde P^{(1)}(\lambda) = \exp{-2\boldsymbol{\lambda}^T J^T\left(C_0-\frac{\mathds{1}}{4}\right)J\boldsymbol{\lambda} + 2i\bar{\boldsymbol{\alpha}}^T J\boldsymbol{\lambda}}
\ee
where $\boldsymbol{\lambda} = (\text{Re}(\lambda), \text{Im}(\lambda))$, and $J$ is the $2 \times 2$ symplectic form matrix chosen so that $2i\boldsymbol{\alpha}^TJ\boldsymbol{\lambda} = \lambda\alpha^*-\lambda^*\alpha$.
When $C_0 - \mathds{1}/4$ is positive definite, the $P$-function is
\be \label{eq:gen_gaussian_P}
P(\alpha) = \frac{1}{\sqrt{\det(2\pi (C_0 - \mathds{1}/4))}}\exp{-\frac{1}{2}(\boldsymbol{\alpha}-\bar{\boldsymbol{\alpha}})^T \left(C_0-\frac{\mathds{1}}{4}\right)^{-1}(\boldsymbol{\alpha}-\bar{\boldsymbol{\alpha}})},
\ee
which takes a multivariate Gaussian form.

In the pure coherent state case, where $C_0 = \mathds 1/4$, the above should be interpreted as $\delta(\boldsymbol{\alpha}-\bar{\boldsymbol{\alpha}})$.
More generally, $C_0 - \mathds{1}/4$ may have a negative eigenvalue, in which case the integral that defines $P(\alpha)$ (the inverse transform of $\tilde P^{(1)}(\lambda)$) does not converge; in fact, this is always the case for squeezed vacuum states.
In these cases, we can still obtain a formal $P(\alpha)$ using singular distributions, as has been discussed for Fock states above, or even by naively taking Eq.~\eqref{eq:gen_gaussian_P} at face value.
Both of these stand in for the fact that calculations (e.g.~normally ordered moments) rely only on the Gaussian form of the characteristic function Eq.~\eqref{eq:gen_gaussian_char}, which is always well-defined.
For instance, it ensures that all normally ordered central moments of these states (i.e.~moments of the form $\la (a^\dag - \la a^\dag\ra)^m (a - \la a\ra)^n\ra$) obey Wick's theorem.

%%%%%%%%%%%%%%%%%%%%%%%%%%%%%%%%%
\subsection{Properties of the Coherence Time}
\label{app:coherence_time}
%%%%%%%%%%%%%%%%%%%%%%%%%%%%%%%%%

The coherence time of any stationary quantity $\xi(t)$ is defined as (see e.g.~Ref.~\cite{Cheong:2024ose})\footnote{Although $\tau_\text{coh}$ is defined for arbitrary states of a quantum field, its interpretation as the sole characteristic timescale controlling the decay of correlations is only strictly true for generalized Gaussian states, where the complete statistics are encoded in the second moments alone.}
\be
\tau_\text{coh} \equiv \int_{-\infty}^{\infty} dt \,|g_{\xi}^{(1)}(t)|^2 = \frac{1}{|\la \xi^{*}(0)\xi(0) \ra |^2}\int_{-\infty}^{\infty} dt \, |\la \xi^{*}(0)\xi(t) \ra |^2.
\ee
Using Parseval's theorem, we can also rewrite this expression in the frequency domain with a PSD $\la S_{\xi}[\omega]\ra $ (cf.~Eq.~\eqref{eq:PSDasFourier}), as 
\be
\tau_{\text{coh}} = \frac{1}{\left|\int \frac{d\omega}{2\pi}\la S_{\xi}[\omega]\ra\right|^2} \int \frac{d\omega}{2\pi}\, |\la S_{\xi} [\omega]\ra|^2.
\ee
If $\xi$ is taken to be the output port $\gamma$ of the transmission line, the result is the coherence time $\tau_{\text{out}}$ of the output field.
In the limit where the PSD is signal dominated such that $n_{m}$ and $n_{\ell}$ are negligible, $\tau_{\text{out}}$ can be expressed in terms of
\be
\tau_{\text{out}} = \frac{2\pi}{\left[\int d\omega \,|\chi(\omega)|^2 \neff[\omega] \right]^2} \int d\omega \,|\chi(\omega)|^4 \neff[\omega]^2
\ee
with susceptibility
\be
|\chi(\omega)|^2 = \frac{g^2 \kappa_{m}}{(\kappa_{\text{tot}}/2)^2 + (\omega - \omega_{c})^2}.
\ee
Consider the narrow-cavity regime, $\kappa_\text{tot} \ll m_a v^2$, where $|\chi(\omega)|^2$ is narrower than that of $\neff[\omega]$.
In this case, $\neff[\omega]$ is constant within the $\omega$ ranges where the $\chi(\omega)$ response is appreciable.
Thus, $\tau_{\text{out}}$ is determined by $\kappa_{\text{tot}}$ as
\be
\tau_{\text{out}} \simeq 2\pi \frac{\int d\omega |\chi(\omega)|^4}{| \int d\omega |\chi(\omega)|^2|^2} = \frac{2}{\kappa_{\text{tot}}}.
\ee
In the opposite wide-cavity limit, $\kappa_\text{tot} \gg m_a v^2$, 
\be
\left.\tau_{\text{out}}\right|_{\kappa_{\text{tot}} \gg m v^2} \simeq 2\pi \frac{\int d\omega \,\neff[\omega]^2}{| \int d\omega \,\neff[\omega]|^2} = \tau_\DM,
\ee
where $\tau_\DM$ is the DM coherence time introduced in Sec.~\ref{subsec:review}.
Writing $t_m = 1/\kappa_\text{tot}$, we can combine the results as
\be
\tau_{\text{out}} \sim \max( \tau_{\DM}, t_m).
\ee

Finally, we can demonstrate the asymptotic behavior of the Poisson deviation function $F(\tau)$ in Eq.~\eqref{eq:F_def}.
(Note there we evaluate the expectation value assuming a Gaussian DM state, although this definition can also be applied to the $P$-function of a general stationary state.)
We take a representative example with a coherence time $\tau_{\text{out}}$,
\be \label{eq:app_representative_gamma}
\frac{|\la \gamma^{*}(0) \gamma (\tau)\ra_{P}|^2}{|\la |\gamma(0)|^2\ra_{P} |^2} = e^{-\pi \tau^2/ \tau_{\text{out}}^2}.
\ee
This yields the exact result and limiting cases discussed in Sec.~\ref{subsec:number_measurements},
\be
F(\tau) = \frac{\tau_{\text{out}}}{\tau} \,\text{Erf} \left(\frac{\sqrt{\pi}\tau}{\tau_{\text{out}}} \right) + \frac{\tau_{\text{out}}^2}{\pi \tau^2} \left(e^{- \pi \tau^2 / \tau_{\text{out}}^2} -1 \right) \simeq \begin{cases} 1 & \tau \ll \tau_{\text{out}} \\ \tau_{\text{out}} / \tau & \tau \gg \tau_{\text{out}} \end{cases}
\ee
%!TEX root = main.tex

%%%%%%%%%%%%%%%%%%%%%%%%%%%%%%%%%
\section{Input-Output Calculations}
%%%%%%%%%%%%%%%%%%%%%%%%%%%%%%%%%

%%%%%%%%%%%%%%%%%%%%%%%%%%%%%%%%%
\subsection{Details of String-Oscillator Model} 
\label{app:string-Oscillator}
%%%%%%%%%%%%%%%%%%%%%%%%%%%%%%%%%

In this appendix, we show the details of the string-oscillator model discussed in Sec.~\ref{sec:string-OscillatorSystem}.
Consider a system consisting of a semi-infinite string coupled to an oscillator of mass $m$ and spring constant $k$, as depicted in Fig.~\ref{fig:string_oscillator_sketch}.
In particular, we consider attaching the mass to the string's endpoint at $x = 0$.
Let $z(t)$ denote the displacement of the mass from the equilibrium position and $\psi(x, t)$ that of the string.
The system is subject to the constraint $\psi(x=0,t) = z(t)$ which can be enforced through a Lagrange multiplier $\lambda(t)$
\be
\Lag = \frac{m \dot{z}^2}{2} - \frac{k z^2}{2} + \int_{-\infty}^{0} d x \, \qty( \frac{\mu \dot{\psi}^2}{2} - \frac{{\cal T} \qty(\partial_x \psi)^2}{2} ) + \lambda(t)\qty[\psi(0, t) - z(t)],
\ee
where ${\cal T}$ and $\mu$ are the string tension and mass density.
The equation of motion for $\psi(x,t)$ is 
\be
\mu \ddot{\psi} - {\cal T} \partial_x^2 \psi = \lambda(t) \delta(x) 
\implies \lambda(t) = {\cal T} \eval{\partial_x \psi}_{x = 0^-}
\ee
in which we integrated near $x = 0$.
The Lagrange multiplier contributes to the evolution of $z$ through $\Lag \supset -\lambda(t) z(t)$.
Hence, using the equation above, we can conclude that the following Lagrangian gives the same evolution of $z$
\be
\Lag = \frac{m \dot{z}^2}{2} - \frac{k z^2}{2} + \int_{-\infty}^{0} d x \, \qty( \frac{\mu \dot{\psi}^2}{2} - \frac{{\cal T} \qty(\partial_x \psi)^2}{2} ) 
- {\cal T} z \eval{\partial_x \psi}_{x = 0}.
\ee
Now, the equation of motion for $z$ can be obtained by fixing $\psi$ while varying $z$ as
\be
\ddot{z} = -\omega_c^2 z - \kappa c \eval{\partial_x \psi}_{x = 0}
\ee
in which we defined $\omega_c \defeq \sqrt{k / m}$, $\kappa \defeq \sqrt{{\cal T} \mu}/m$, and $c \defeq \sqrt{{\cal T}/\mu}$.

For a real scalar, one may consider the mode expansion
\begin{gather}
\psi(x, t) = \psi_{\tin}(x,t) + \psi_\tout(x,t) + \hc, \label{appeq:IOWaveformDefine} \\
\psi_\tin(x, t) \defeq \int_0^\infty \frac{d k}{2\pi} \, A_k e^{i (k x - \omega_k t)}, \quad 
\psi_\tout(x, t) \defeq \int_0^\infty \frac{d k}{2\pi} \, B_k e^{-i (k x + \omega_k t)},
\end{gather}
in which $\omega_k \defeq c\abs{k}$.
Then, for each $\omega = \omega_k$, the equation of motion of $z$ becomes 
\be
(-\omega_k^2 + \omega_c^2) z[\omega_k] = - i \kappa \omega_k \qty(A_k - B_k)
\ee
in which $z[\omega_k]$ denotes the Fourier transform of $z(t)$.
However, the boundary condition $\psi(0, t) = z(t)$ fixes $z[\omega_k] = A_k + B_k$ so that the output $B_k$ is linearly related to the input $A_k$ as 
\be
B_k = -\qty( 1 - \frac{\kappa \omega_k}{\kappa \omega_k/2 - i (\omega_k^2 - \omega_c^2)/2} ) A_k 
\simeq - \qty( 1 - \frac{\kappa}{\kappa/2 - i (\omega_k - \omega_c)} ) A_k,
\ee
in which we approximated $\omega_k \simeq \omega_c$ in the last step.
For a dispersionless string, we can map the input field to a time series by observing the waveform at some fixed position $x$ so that 
\be
\beta_\tin(t) = \psi_{\tin}\qty(x = c t_0, t), \quad 
\beta_\tout(t) = \psi_{\tout}\qty(x = -c t_f, t)
\ee
where we assume that the initial data is read at $t_0 < 0$ and final data is read at $t_f > 0$.

To generalize the result, we consider coupling the mass to another string $\phi(y,t)$ propagating along the $y$ direction with the interaction $\Lag \supset - \bar{\cal T} z \eval{\partial_y \phi}_{y = 0}$.
Then, the input-output relation becomes $D_{\bar{k}} = z[\omega_{\bar{k}}] - C_{\bar{k}}$ in which 
\be
\phi_{\tin}(y, t) = \int_0^\infty \frac{d \bar{k}}{2\pi} \, C_{\bar{k}} e^{i (\bar{k} y - \omega_{\bar{k}} t)}, \quad 
\phi_{\tout}(y, t) = \int_0^\infty \frac{d \bar{k}}{2\pi} \, D_{\bar{k}} e^{-i (\bar{k} y + \omega_{\bar{k}} t)},
\ee
where $\bar{c} = \sqrt{\bar{\cal T}/\bar{\mu}}$ and $\omega_{\bar{k}} = \bar{c}\bar{k}$.
For simplicity, we assume that $\bar{c} = c$.
The Fourier-transformed equation of motion for the oscillator at frequency $\omega = \omega_k$ is
\be
\qty(-\omega_k^2 + \omega_c^2) z[\omega_k] = - i \kappa \omega_k \qty(A_k - B_k) - i \bar{\kappa} \omega_{k}(C_{k} - D_{k}), 
\ee
where $\bar{\kappa} \defeq \sqrt{\bar{\cal T} \bar{\mu}}/m$, leading to 
\bea \label{appeq:io_ABC}
B_k &= -\qty( 1 - \frac{\kappa \omega}{(\kappa + \bar{\kappa}) \omega / 2 - i (\omega^2 - \omega_c^2)/2} ) A_k + \qty( \frac{\bar{\kappa} \omega}{(\kappa + \bar{\kappa})\omega/2 - i (\omega^2 - \omega_c^2)/2} ) C_{k} \\
&\simeq -\qty( 1 - \frac{\kappa}{(\kappa + \bar{\kappa}) / 2 - i (\omega - \omega_c)} ) A_k + \qty( \frac{\bar{\kappa}}{(\kappa + \bar{\kappa})/2 - i (\omega - \omega_c)} ) C_{k}
\eea
where $A_k + B_k = z[\omega] = C_{k} + D_{k}$ was used to eliminate the $z[\omega]$ and $D_{k}$ dependence.
Here, we interpret $A_k$ and $C_{k}$ as the input Fourier amplitudes from both $x$ and $y$ directions.

The presence of a second string changes the output $B_k$ in two ways, as we discuss in Sec.~\ref{sec:string-OscillatorSystem}.
First, the total decay width increases due to the replacement of $\kappa \to \kappa + \bar{\kappa}$ in the denominator.
Second, there is an additional susceptibility that couples the output waveform of $\psi$ to the input waveform of $\phi$.

%%%%%%%%%%%%%%%%%%%%%%%%%%%%%%%%%
\paragraph{Canonical Formalism of the String-Oscillator System.}
%%%%%%%%%%%%%%%%%%%%%%%%%%%%%%%%%
%
To make contact with the DM haloscope in the main text, we discuss the canonical formalism of the string-oscillator system.
After applying the Legendre transform, the Hamiltonian for the system becomes 
\bea
H = \frac{p_z^2}{2m} + \frac{m \omega_c^2 z^2}{2} &+ \int_{-\infty}^0 d x\, \qty[\frac{\pi_\psi^2}{2\mu} + \frac{{\cal T} \qty(\partial_x \psi)^2}{2}] + {\cal T} z \eval{\partial_x \psi}_{x = 0} \\
&+ \int_{-\infty}^0 d y\, \qty[\frac{\pi_\phi^2}{2\bar{\mu}} + \frac{\bar{\cal T} \qty(\partial_y \phi)^2}{2}] + \bar{\cal T} z \eval{\partial_y \phi}_{y = 0} .
\eea
With a mode expansion 
\begin{align}\label{eq:app_canonical_string_mode_expansion}
z = \sqrt{\frac{1}{2m \omega_c}} \qty(\gamma^* + \gamma), \quad &p_z = m \dot{z} = i \sqrt{\frac{m \omega_c}{2}} \qty(\gamma^* - \gamma), \\
\psi(x) = \int \frac{d k}{2\pi} \, \sqrt{\frac{1}{2\mu \omega_k}} \qty( \beta^*_k e^{-i k x} + \beta_k e^{i kx} ), \quad 
&\pi_\psi(x) = \mu \dot{\psi} = i \int \frac{d k}{2\pi} \, \sqrt{\frac{\mu \omega_k}{2}} \qty( \beta^*_k e^{-i k x} - \beta_k e^{i k x} ), \nonumber\\
\phi(y) = \int \frac{d k}{2\pi} \, \sqrt{\frac{1}{2\bar{\mu} \omega_{k}}} \qty( \alpha^*_{k} e^{-i k y} + \alpha_{k} e^{i k y} ), \quad 
&\pi_\phi(y) = \bar{\mu} \dot{\phi} = i \int \frac{d k}{2\pi} \, \sqrt{\frac{\bar{\mu} \omega_{k}}{2}} \qty( \alpha^*_{k} e^{-i k y} - \alpha_{k} e^{i k y} ), \nonumber 
\end{align}
we find that the Hamiltonian becomes 
\bea
H = \omega_c \abs{\gamma}^2 + \int \frac{d k}{2\pi} \, \omega_k \abs{\beta_{k}}^2 + \int \frac{d k}{2\pi} \, \omega_{k} \abs{\alpha_{k}}^2
& + i \qty(\gamma^* + \gamma) \int \frac{d k}{2\pi} \, g_k \qty(\beta_k - \beta_k^*) \\
& + i \qty(\gamma^* + \gamma) \int \frac{d k}{2\pi} \, \bar{g}_{k} \qty(\alpha_{k} - \alpha_{k}^*),
\eea 
in which $g_k = k {\cal T}/(2\sqrt{m \mu \omega_c \omega_k}) = \sqrt{c \kappa \omega_k}/(2\sqrt{\omega_c})$.
Making the RWA allows us to write the Hamiltonian as 
\be
H = \omega_c \abs{\gamma}^2 + \int \frac{d k}{2\pi} \, \omega_k \abs{\beta_k}^2 + \int \frac{d k}{2\pi} \, \omega_{k} \abs{\alpha_{k}}^2 + \qty[\int \frac{d k}{2\pi} \, i g_k \gamma^* \beta_k + \int \frac{d k}{2\pi} \, i \bar{g}_{k} \gamma^* \alpha_{k} + \hc],
\ee
which closely resembles the Hamiltonian in Eq.~\eqref{eq:fullHamiltonian} after identifying $\gamma \leftrightarrow c$, $\beta_k \leftrightarrow b_\omega$, and $\alpha_k \leftrightarrow a_k$.
The mapping 
\be
A_k = \frac{\beta_k}{\sqrt{2 \mu \omega_k}}, \quad 
B_k = \frac{\beta_{-k}}{\sqrt{2 \mu \omega_k}}, \quad
C_{k} = \frac{\alpha_{k}}{\sqrt{2\bar{\mu} \omega_{k}}},
\ee
recovers the input-output relation derived in Eq.~\eqref{appeq:io_ABC}.
One can explicitly write $\beta_k$ in terms of an inverse Fourier transform
\be
\beta_k \simeq \int_{-\infty}^{0} d x \, \left[ \sqrt{\frac{\mu \omega_k}{2}} \psi(x) e^{-i k x} + \frac{i}{\sqrt{2\mu \omega_k}} \pi_\psi(x) e^{-i k x} \right]\!,
\ee
and the canonical Poisson bracket $\comm{\psi(x)}{\pi_{\psi}(y)}_{\psi,\pi} = \delta(x - y)$ implies that 
\be
\comm{\beta_k}{\beta_q^*}_{\psi,\pi} 
= -i \int_{-\infty}^0 d x\, e^{-i (k-q) x}
\simeq -\pi i \delta(k - q),
\ee
up to principal-value terms, where $\comm{u}{v}_{\psi,\pi} = (\fdv*{u}{\psi(x)}) (\fdv*{v}{\pi_\psi(x)}) - (\fdv*{u}{\pi_\psi(x)}) (\fdv*{v}{\psi(x)})$.
Despite their non-vanishing Poisson bracket, $\beta_k$ and $\beta_k^*$ still commute under multiplication, while the quantum operators $b_k$ and $b_k^\dagger$ do not.
Nonetheless, the string-oscillator model provides a simple classical analog of the axion measurement problem.

%%%%%%%%%%%%%%%%%%%%%%%%%%%%%%%%%
\paragraph{Relation to Transmission Lines.}
%%%%%%%%%%%%%%%%%%%%%%%%%%%%%%%%%
%
Our toy model is an adequate model for a transmission line, which is commonly modeled as a one-dimensional infinite LC circuit~\cite{Clerk:2008tlb}.
The one-dimensional field $\psi(x,t)$ here is the flux variable defined as 
\be
\psi(x,t) = \int_{-\infty}^{t} d \tau\, V(x,\tau),
\ee
obeying the Lagrangian
\be
\Lag = \int d x\, \left[ \frac{c}{2} \dot{\psi}^2 - \frac{1}{2\ell} (\partial_x \psi)^2 \right]\!,
\ee
in which $c$ denotes the linear capacitance density and $\ell$ denotes the linear inductance density.
By identifying $c = \mu$ and $\ell = 1/\mathcal{T}$, we see that the LC-coupled transmission line behaves as a one-dimensional string.
The voltage one measures is proportional to the $Y$ quadrature of the field $\psi$, i.e.~$\pi_\psi = c \dot{\psi}$.

%%%%%%%%%%%%%%%%%%%%%%%%%%%%%%%%%
\subsection{Axion-Induced Cavity Decay}
\label{app:weakAxionCoupling}
%%%%%%%%%%%%%%%%%%%%%%%%%%%%%%%%%
%
Here, we justify the weak axion-cavity coupling assumption invoked in the main text, e.g.~Sec.~\ref{subsec:io_intro}.
First, the DM-cavity coupling is much smaller than the typical measurement coupling, 
\be
g = \ga B_0 \sqrt{\Omega} \simeq 5\times 10^{-17} m_a \qty(\frac{B_0}{10\,\text{T}}),
\ee
assuming for a QCD axion that $\ga \simeq \alpha_e m_a / (4\pi \Lambda_\text{QCD}^2)$ with $\alpha_e$ the electromagnetic fine-structure constant.
On resonance, the decay rate of cavity photons to axions is $\kappa_a \sim g^2/\omega_c \sim 2\times 10^{-33} \omega_c$, which is much smaller than $\kappa_\text{tot} \sim 10^{-5} \omega_c$.
We thus expect $g \ll \kappa_\text{tot}$ and $g^2 t \ll \kappa_\text{tot}$ for any relevant measurement time considered.
However, one might worry that this estimate of the decay rate is missing an enhancement by the axion occupation number.
We now show explicitly that this is not the case.

The full axion-cavity-transmission-line Hamiltonian after the RWA is given by Eq.~\eqref{eq:fullHamiltonian}, with the relevant parts reproduced below
\bea
H &\supset \omega_c c^\dag c + \int \frac{d^3 \bk}{(2\pi)^3} \, \omega_{\bk} a_{\bk}^\dag a_{\bk} + \left[ i g c^\dag \int \frac{d^3 \bk}{(2\pi)^3} \, \sqrt{\frac{V_c}{4\Omega}} C(\bk) a_{\bk} + \hc \right]\!,
\eea
where $\omega_{\bk} = (m_a^2 + \abs{\bk}^2)^{1/2}$ is the axion angular frequency.
The Heisenberg equation for the axion mode $a_{\bk}$ is
\be
\dot{a}_{\bk} = i \comm{H}{a_{\bk}} = -i \omega_{\bk} a_{\bk} - g \sqrt{\frac{V_c}{4\Omega}} C^*(\bk) c
\ee
which has the solution
\be
a_{\bk}(t) = a_{\bk}(t_0) e^{-i \omega_{\bk}(t - t_0)} - \int_{t_0}^t d t'\, e^{-i \omega_{\bk}(t - t')} g \sqrt{\frac{V_c}{4\Omega}} C^*(\bk) c(t')
\ee
given that $c(t)$ is known.
Substituting this back into the equation of motion for $c$ gives
\be
\dot{c} = i \comm{H}{c} \supset g \int \frac{d^3 \bk}{(2\pi)^3} \, \sqrt{\frac{V_c}{4\Omega}} C(\bk) a_{\bk}(t) = g a_\eff(t).
\ee
This term is responsible for the decay of the cavity mode into the axion bath because once we substitute $a_{\bk}(t)$ with its solution, a damping term $\dot{c} \supset -\kappa c$ is introduced.
In particular, if we define the freely evolved effective mode in Eq.~\eqref{eq:free_effective_mode} as $a^{(0)}_\eff(t)$, we have
\be \label{appeq:cavityDecay}
\dot{c}(t) \supset 
g a^{(0)}_\eff(t) - \int_{t_0}^t d t'\, \kappa_a(t-t') c(t'),
\ee
with a response function defined as 
\be
\kappa_a(t) \defeq \frac{g^2 V_c}{4\Omega} \int \frac{d^3 \bk}{(2\pi)^3} \, \abs{C(\bk)}^2 e^{-i \omega_{\bk} t}.
\ee
The first term in Eq.~\eqref{appeq:cavityDecay} is how the cavity is driven by the axion and retained in our calculations in the main text.
The second term describes the axion-induced cavity decay.
The response function $\kappa_a(t)$ is intentionally defined such that its Fourier transform
\be
\kappa_a[\omega] = \frac{g^2 V_c}{4\Omega} \int \frac{d^3 \bk}{(2\pi)^3} \, 2\pi \delta(\omega - \omega_{\bk}) \abs{C(\bk)}^2,
\ee
can be interpreted as the decay rate of the cavity mode with frequency $\omega$ into the axion bath.
However, causality dictates that the time-domain response is given by $\kappa_a(t) \Theta(t)$ rather than $\kappa_a(t)$.
In the frequency domain, this translates to 
\be
\int_0^\infty d t\, e^{i \omega t} \kappa_a(t) 
= \frac{\kappa_a[\omega]}{2} - i \, \PV\! \int \frac{d\omega'}{2\pi} \, \frac{\kappa_a[\omega']}{\omega' - \omega},
\ee
where $\PV$ is the Cauchy principal value.
The first term dictates that the cavity has a decay term of the form $\kappa_a[\omega]c[\omega]/2$, whereas the second term is an axion-induced frequency shift.
That both are controlled by the same $\kappa_a[\omega]$ is just the Kramers--Kronig relation and a consequence of causality.
The factor of $1/2$ in the decay term exactly matches that in Eq.~\eqref{eq:cont_inout_relations_c}.
Indeed, in the Markovian approximation, $\kappa(t) \propto \kappa \delta(t)$ so that $\kappa[\omega]$ is independent of frequency and the frequency-shift term exactly vanishes.
The frequency-dependent part of the shift also vanishes for $\omega>m_a$ in the nonrelativistic approximation, with $\omega_{\bk} \approx m_a + \abs{\bk}^2/(2m_a)$ and $\abs{C(\bk)}^2 \approx \abs{C(0)}^2 \propto 1/m_a$.
Nonetheless, we present both terms here for completeness; both effects are $\order{g^2}$ and as claimed not enhanced by the axion occupation number, hence they are negligible.

We emphasize that the above decay rate is not used to define the axion quality factor $Q_a$; cf.~$Q_m$ and $Q_\ell$ as defined below Eq.~\eqref{eq:freq_output_noise_DM}.
In particular, the inverse fractional linewidth of the DM spectrum, $Q_a \sim 10^{6}$, is unrelated to the DM-cavity coupling.
Though $Q_a$ is called a quality factor, the cavity in fact loses energy into DM at a rate $g^2/\omega_c \ll m_a/Q_a$.

%%%%%%%%%%%%%%%%%%%%%%%%%%%%%%%%%
\subsection{Properties of the Output Operator}
\label{app:equivalentTemporalInputOutput}
%%%%%%%%%%%%%%%%%%%%%%%%%%%%%%%%%

Here we provide further details of the output operator defined in Eq.~\eqref{eq:b_out_def} within Sec.~\ref{subsec:io_intro}.
As noted in footnote~\ref{footnote:output-GC}, our definition of $b_\tout$ is not immediately consistent with the canonical input-output formulation.
Here we demonstrate our input and output operators can be matched exactly to the widely used Gardiner--Collett (GC) conventions~\cite{Gardiner:1985uof}.
Having established that connection, we show that the output commutator result in Eq.~\eqref{eq:b_out_commutator} follows immediately.

%%%%%%%%%%%%%%%%%%%%%%%%%%%%%%%%%
\paragraph{Two Definitions.}
%%%%%%%%%%%%%%%%%%%%%%%%%%%%%%%%%
%
The conventional GC input-output operators are defined as\footnote{The GC definitions in Eqs.~(2.10) and (2.18) of Ref.~\cite{Gardiner:1985uof} have prefactors of $1/\sqrt{2\pi}$ rather than $1/(2\pi)$.
We have suppressed this minor point of difference that originates solely from the fact we take the frequency modes to have commutator $[b_\omega,b^\dag_{\omega'}] = 2\pi\delta(\omega-\omega')$, whereas the equivalent expression used by GC has no $2\pi$.}
\bea \label{eq:binout_GC}
b^\GC_\tin(t) &\defeq \int \frac{d\omega}{2\pi}\, e^{-i \omega(t-t_i)} b_\omega(t_i), \\
b^\GC_\tout(t) &\defeq \int \frac{d\omega}{2\pi} \, e^{-i \omega(t-t_f)} b_\omega(t_f),
\eea
where the initial and final times $t_i$ and $t_f$ are taken to be asymptotically far in the past and future, respectively.
These expressions differ from what we used in our discussion in Sec.~\ref{sec:inout}, where in Eqs.~\eqref{eq:b_tin_def} and \eqref{eq:b_out_def} we defined
\bea
b_\tin(t) &\defeq \int \frac{d\omega}{2\pi}\, e^{-i \omega t} b_\omega(0), \\
b_\tout(t) &\defeq \int \frac{d\omega}{2\pi}\, e^{i \omega t_d} b_\omega(t).
\eea
The difference between the input modes is trivial: choosing $t_i = 0$, we have $b_\tin(t) = b^\GC_\tin(t)$.
That $b_\tout(t)$ and $b^\GC_\tout(t)$ are related is less obvious.
Note not only that our input and output modes are less symmetric, but further that in $b^\GC_\tout(t)$ the time argument does \textit{not} label the time evolution of the underlying $b_{\omega}$ operator; in particular, the operators appearing in the definition of $b^\GC_\tout(t)$ are all evaluated at the same time $t_f$, which is an important point we return to.
Nonetheless, we next show the two output modes are equivalent.
In doing so, we take $t_i = 0$ throughout for notational conciseness; it is straightforward to restore.

As an intermediate step, we derive the relation between GC input and output modes.
Recall we already have from Eq.~\eqref{eq:bout_to_bin} that
\be \label{eq:our_inout}
b_\tout(t) = U_I^\dag (t, 0) b_\tin(t-t_d) U_I(t, 0).
\ee
We can derive an equivalent expression for the modes in Eq.~\eqref{eq:binout_GC} as follows,
\bea \label{eq:GC_inout}
b^\GC_\tout(t) &= \int \frac{d\omega}{2\pi} \, e^{-i \omega t} \, U_\text{full}^\dagger(t_f, 0) \qty[e^{i \omega t_f} b_\omega(0)] U_\text{full}(t_f, 0) \\
&= U_\text{full}^\dagger(t_f, 0) U_0(t_f) \left( \int \frac{d\omega}{2\pi} \, e^{-i \omega t}\, b_\omega(0) \right) U_0^\dagger(t_f) U_\text{full}(t_f, 0) \\
&= U_I^\dagger(t_f, 0) b^\GC_\tin(t) U_I(t_f, 0),
\eea
where we used the decomposition of the full time-evolution operator in Eq.~\eqref{eq:interaction_U_split}, and the fact that $U_0$ simply evolves the phase.

We next use Eqs.~\eqref{eq:our_inout} and \eqref{eq:GC_inout} to show that $b_\tout(t) = b^\text{GC}_\tout(t-t_d)$.
Recall $t_f$ is taken to be a time in the far future, so we can always consider $t_f > t$ and decompose $U_I(t_f,0) = U_I(t_f,t) U_I(t,0)$.
This gives,
\bea \label{eq:out_link_midpoint}
b^\GC_\tout(t-t_d) 
&=\,U_I^\dagger(t_f, 0) b_\tin(t-t_d) U_I(t_f, 0) \\
&=\,U_I^\dag (t,0) \qty(U_I^\dagger(t_f, t) b_\tin(t-t_d) U_I(t_f, t)) U_I(t, 0)
\eea
using $b^\GC_\tin(t) = b_\tin(t)$.
Comparing this to Eq.~\eqref{eq:our_inout}, the key is the impact of the time evolution in parentheses.

To understand the important time evolution, the input mode commutator of Eq.~\eqref{eq:b_tin_commutator} is crucial.
Recall that result depended explicitly on extending the modes to negative frequencies, as discussed at the beginning of Sec.~\ref{subsec:io_intro}.
To be explicit,
\bea
\![b_\tin(t), b^\dag_\tin(t')] 
= \int \frac{d\omega}{2\pi} \int \frac{d\omega'}{2\pi}\, e^{-i(\omega t-\omega' t')} \, [b_\omega(0), b^\dag_{\omega'}(0)]
= \int \frac{d\omega}{2\pi}\, e^{-i\omega (t-t')}
= \delta\!\,(t-t'),
\eea
with the final step depending critically on the presence of negative frequencies.
Combining this with the explicit form of $H_I(t)$ in Eq.~\eqref{eq:HI(t)}, we conclude that
\be
\comm{H_I(s)}{b_\tin(t-t_d)} = 0 \qq{if} s \neq t - t_d.
\ee
Accordingly, with $t_d > 0$, $U_I(t_f, t)$ commutes with $b_\tin(t-t_d)$, so that Eq.~\eqref{eq:out_link_midpoint} simplifies to
\be
b^\GC_\tout(t-t_d) = U_I^\dag (t,0) b_\tin(t-t_d) U_I(t, 0) = b_\tout(t),
\ee
demonstrating the claimed equivalence.
Intuitively, $b_\tout(t)$ is dependent upon $b_\tin(t-t_d)$ as the mode that interacted with the cavity at $t - t_d$ is the one that hits our detector at a time $t$.
Further, the subsequent interaction-picture evolution of the input mode $b_\tin(t-t_d)$ from a time $t$ to $t_f$ is trivial, as it never interacts with the cavity again.
Note that retaining $t_d > 0$ was crucial, so that even though we send $t_d \to 0$ in our calculations, this is solely for notational clarity, and in detailed calculations it should be retained.

%%%%%%%%%%%%%%%%%%%%%%%%%%%%%%%%%
\paragraph{The Local Output Commutator.}
%%%%%%%%%%%%%%%%%%%%%%%%%%%%%%%%%
%
The fact that our output modes are defined in terms of modes at different times, $b_{\omega}(t)$, makes it less convenient to confirm that they satisfy the local commutation relation in Eq.~\eqref{eq:b_out_commutator}.
The GC operators, $b^\GC_\tout(t)$, do not have this property, depending instead on $b_{\omega}(t_f)$, and we can use the above connection between the definitions to establish the local commutator.
In detail,
\bea
\big[ b_\tout(t), b^\dag_\tout(t') \big]
&= \big[ b^\GC_\tout(t-t_d), b^{\GC\,\dag}_\tout(t'-t_d) \big] \\
&= \int \frac{d\omega}{2\pi}\int \frac{d\omega'}{2\pi} \, e^{-i \omega(t-t_d-t_f)} e^{i \omega'(t'-t_d-t_f)} \big[b_\omega(t_f), b^\dag_{\omega'}(t_f) \big] \\
&= \int \frac{d\omega}{2\pi} \, e^{-i \omega(t-t')} \\
&=\delta\!\,(t-t').
\eea

%%%%%%%%%%%%%%%%%%%%%%%%%%%%%%%%%
\subsection{Approximate Expressions for Axion Signal Power}
\label{app:deltaFunctionExpansion}
%%%%%%%%%%%%%%%%%%%%%%%%%%%%%%%%%

Here we provide the details of how $\Gamma_\sig$ in Eq.~\eqref{eq:Gamma_sig_def} can be evaluated in various limits.

%%%%%%%%%%%%%%%%%%%%%%%%%%%%%%%%%
\paragraph{General Strategy.}
%%%%%%%%%%%%%%%%%%%%%%%%%%%%%%%%%
%
One may formally expand a delta function near a point using
\be
\delta(x - z) = \sum_{n=0}^\infty \frac{(-z)^n}{n!} \frac{d^{n}}{dx^{n}} \delta(x),
\ee
and because for any distribution $f(z)$, 
\be
f(x) = \ev{\delta(x - z)} = \sum_{n=0}^{\infty} \frac{\ev{(-z)^n}}{n!} \frac{d^{n}}{dx^{n}} \delta(x),
\ee
in which the expectation values are taken with respect to $z$ rather than $x$, we can perform a formal expansion of peaked distributions in terms of moments.
Thus, 
\be
f(x) = \sum_{n} \frac{\qty(-1)^n \ev{z^n}}{n!} \frac{d^{n}}{dx^{n}} \delta(x).
\ee

%%%%%%%%%%%%%%%%%%%%%%%%%%%%%%%%%
\paragraph{Expansion of the DM Energy Distribution.}
%%%%%%%%%%%%%%%%%%%%%%%%%%%%%%%%%
%
For the frequency distribution, note that the underlying standard halo model distribution of DM is a Maxwell--Boltzmann distribution with width $v_0/\sqrt{2}$ centered at $-\vb{v}_\odot$,
\be
f(\vb{v}) = \frac{1}{\pi^{3/2} v_0^{3}} \exp{-\frac{\qty(\vb{v} + \vb{v}_\odot)^2}{v_0^2}}.
\ee
It is easier to consider the expansion of $p[\omega]$ around $m_a$ by noticing that 
\be
\omega_{\bk} \simeq m_a \underbrace{+ \frac{m_a v^2}{2} - \frac{m_a v^4}{8} + \order{v^6}}_{\defeq z(v)},
\ee
and
\be
p[\omega] = \int d^3 \vb{v} \, f(\vb{v}) \delta(\omega - m_a - z)
= \sum_{n} \frac{\qty(-1)^n \ev{z^n}}{n!} \frac{d^{n}}{d\omega^{n}} \delta(\omega - m_a).
\ee
We can compute the first two moments for the standard halo model,
\begin{gather}
\ev{z} = \frac{m_a}{2} \ev{v^2} - \frac{m_a}{8} \ev{v^4} + \order{v^6} 
\simeq \frac{m_a}{2} \qty[ v_\odot^2 + \frac{3}{2} v_0^2 - \frac{1}{4} \qty( v_\odot^4 + 5 v_0^2 v_\odot^2 + \frac{15}{4} v_0^4 ) ],\\
\ev{z^2} = \frac{m_a^2}{4} \ev{v^4} + \order{v^6} 
\simeq \frac{m_a^2}{4} \qty( v_\odot^4 + 5 v_0^2 v_\odot^2 + \frac{15}{4} v_0^4 ),
\end{gather}
so that 
\be
p[\omega] \simeq \delta(\omega - m_a) - \ev{z} \frac{d}{d\omega} \delta(\omega - m_a) + \frac{\ev{z^2}}{2} \frac{d^{2}}{d\omega^{2}} \delta(\omega - m_a) + \order{v^6}.
\ee
Combined, these results give Eq.~\eqref{eq:GammaSig_widecavity}.

%%%%%%%%%%%%%%%%%%%%%%%%%%%%%%%%%
\paragraph{Expansion of the Cauchy--Lorentz Distribution.} 
%%%%%%%%%%%%%%%%%%%%%%%%%%%%%%%%%
%
The cavity-response function, which has the form of a Cauchy--Lorentz distribution, introduces an obstruction to the above analysis.
In detail, a naive delta-function expansion fails as the variance of a Cauchy--Lorentz distribution is not well defined.
Nonetheless, we note that
\be
\chi[\omega] \defeq \frac{1}{\pi} \frac{\kappa/2}{(\omega - \omega_c)^2 + (\kappa/2)^2}
= \sum_{n=0}^{\infty} \frac{\qty(-1)^n \qty(\kappa/2)^n}{n!} \int \frac{d t}{2\pi} \, \abs{t}^n e^{i (\omega - \omega_c)t},
\ee
where we integrate $t$ over the full real line.
Now, consider a sufficiently well-behaved function $g[\omega]$ that decays sufficiently quickly at large $\abs{\omega}$ which we integrate against the Cauchy--Lorentz distribution,
\be
\int d\omega \, \chi[\omega] g[\omega]
= \sum_{n} \frac{\qty(-1)^n \qty(\kappa/2)^n}{n!} \int \frac{d\omega d t}{2\pi} \, g[\omega] \abs{t}^n e^{i(\omega - \omega_c) t}.
\ee
The even-$n$ terms can be handled by taking
\be
\int \frac{d\omega d t}{2\pi} \, g[\omega] \abs{t}^{2n} e^{i (\omega - \omega_c)t} 
= \int \frac{d\omega d t}{2\pi} \, g[\omega] \qty(-1)^n \pdv[2n]{\omega} e^{i (\omega - \omega_c)t} 
= \qty(-1)^n \eval{\frac{d^{2n}g}{d\omega^{2n}}}_{\omega = \omega_c}.
\ee
The odd-$n$ terms, after taking the phase factor of $\abs{t}$ into account, are
\be
\int \frac{d\omega d t}{2\pi} \, g[\omega] \abs{t}^{2n+1} e^{i (\omega - \omega_c)t} 
= \qty(-1)^{n} \int \frac{d\omega d t}{2\pi} \, e^{i (\omega - \omega_c)t} \qty[i \, \text{sgn}(t)] \frac{d^{2n+1}}{d\omega^{2n+1}} g[\omega],
\ee
with $\text{sgn}(x)$ denoting the sign function.
For small positive $\epsilon$, the integral above is equivalent to 
\be
\qty(-1)^n \int \frac{d\omega}{2\pi} \, \qty[\pdv[2n+1]{\omega} g[\omega]] \int d t\, i e^{i(\omega - \omega_c)t - \epsilon\abs{t}} \text{sgn}(t) 
= \qty(-1)^n \frac{d^{2n+1}}{d\omega_{c}^{2n+1}} \mathrm{H}(g)[\omega_c],
\ee
in which 
\be
\mathrm{H}(g)[\omega_c] \defeq \PV\!\int \frac{d\omega}{\pi} \, \frac{g[\omega]}{\omega_c - \omega}
\ee
denotes the Hilbert transform of the test function $g[\omega]$.
Thus, for a Cauchy--Lorentz distribution narrow compared to the test function's width, i.e.~$\kappa \ll m_a v_0^2$ (and implicitly $\abs{\omega_c - m_a} \gg \kappa$) for the standard halo model as the test function, we can approximate the Cauchy--Lorentz distribution as
\be
\chi[\omega] \simeq \sum_{n=0}^{\infty} \frac{\qty(-1)^n \qty(\kappa/2)^{2n}}{(2n)!} \left( \frac{d^{2n}}{d\omega^{2n}}\delta(\omega - \omega_c) + \frac{\kappa/2}{2n+1} \frac{d^{2n+1}}{d\omega^{2n+1}} \delta(\omega - \omega_c) \circ \mathrm{H}[\omega_c] \right)\!.
\ee
Here, the notation $f[\omega] \circ \mathrm{H}[\omega_c]$ indicates that when $\chi[\omega]$ is applied to a function, the Hilbert transform acts first; for example, $(f[\omega] \circ \mathrm{H}[\omega_c]) g[\omega] = f[\omega] \mathrm{H}(g)[\omega_c]$.

While evaluating the Hilbert transform of the full DM energy distribution is challenging, we know that it is well approximated in the nonrelativistic regime.
Thus,
\be
p[\omega] \simeq \eval{ \frac{2}{\sqrt{\pi} m_a v_0 v_\odot} \exp{-\frac{v^2 + v_\odot^2}{v_0^2}} \sinh(\frac{2v v_\odot}{v_0^2}) \Theta(\omega - m_a)}_{v = \sqrt{2(\omega/m_a - 1)}}
\ee
Then, its Hilbert transform can be evaluated with $d\omega = m_a v d v$ and $v_c = \sqrt{2(\omega_c/m_a - 1)}$ as
\bea
\mathrm{H}(p)[\omega_c] 
&= \frac{2}{\pi^{3/2} m_a v_0 v_\odot} \PV \int_0^\infty d v\, \frac{v}{v_c^2 - v^2} \qty( \exp{-\frac{(v-v_\odot)^2}{v_0^2}} - \exp{-\frac{(v + v_\odot)^2}{v_0^2}} ) \\
&= \frac{2}{\pi m_a v_0 v_\odot} \qty[ \mathrm{F}\qty(\frac{v_c - v_\odot}{v_0}) - \mathrm{F}\qty(\frac{v_c + v_\odot}{v_0}) ],
\eea
where $\mathrm{F}(x)$ denotes the Dawson integral.
This allows us to approximate
\be
\begin{multlined}
\int d\omega\, \chi[\omega] p[\omega] \\
\simeq p[\omega_c] + \frac{2\kappa}{m_a^2 \pi v_0^3 v_c v_\odot} \qty[ \qty(v_c -v_\odot) \mathrm{F}\qty(\frac{v_c - v_\odot}{v_0}) - \qty(v_c + v_\odot) \mathrm{F}\qty(\frac{v_c + v_\odot}{v_0}) ] + \order{\kappa^2},
\end{multlined}
\ee
justifying Eq.~\eqref{eq:GammaSig_narrowcavity}.

%%%%%%%%%%%%%%%%%%%%%%%%%%%%%%%%%
\subsection{Relation between Discrete and Continuous Frequency Modes}
\label{app:PSDfromDiscreteOp}
%%%%%%%%%%%%%%%%%%%%%%%%%%%%%%%%%

In this appendix, we derive Eq.~\eqref{eq:bfrak_to_bcont} that relates the discrete-frequency mode operator $\db[\omega_k]$ defined in Sec.~\ref{subsec:higherOrderIO} to its continuum version $b[\omega]$.
The specific claim is that for a small sampling time $\Delta t$ and a long integration time $T$, the two are related by $\db[\omega_k] \simeq b[\omega_k]/\sqrt{T}$.
To show this, first from Eqs.~\eqref{eq:discrete_b_in} and \eqref{eq:DFT} we can rewrite the discrete mode as follows
\be
\sqrt{T} \db[\omega_k] = \sqrt{\Delta t} \sum_{n=0}^{N-1} e^{i \omega_k t_n} \db(t_n)
= \sum_n e^{i \omega_k t_n} \int_{t_n}^{t_{n+1}} d t' \, b(t')
\ee
in which $t_n = t_0 + n \Delta t$, $T = N\Delta t$, and $\omega_k = 2\pi k /T$.
Now, as $\Delta t$ approaches zero whilst $T$ approaches infinity, the right-hand side of the equation above tends to $b[\omega_k]$ as we now show.
Using $b(t) = \int d \omega/(2\pi)\, e^{-i \omega t} b[\omega]$, we have,
\bea
\sqrt{T} \db[\omega_k] 
&= \int \frac{d \omega}{2\pi} \, b[\omega] \sum_n e^{i \omega_k t_n} \int_{t_n}^{t_{n+1}} d t' \, e^{-i \omega t'} \\
&= \int \frac{d \omega}{2\pi} \, b[\omega] \left[e^{-i \omega \Delta t/2} \sinc(\frac{\omega \Delta t}{2}) \right] \sum_n \Delta t \,e^{i (\omega_k - \omega)t_n}.
\eea
As $\Delta t \to 0$, the term in brackets approaches $1$ while the discrete sum becomes an integral over time.
As the integration time $T$ increases, the time integral becomes a delta function, yielding
\be
\lim_{\substack{\Delta t \to 0 \\ T \to \infty}} \sqrt{T} \, \db[\omega_k] = \int \frac{d \omega}{2\pi} \, b[\omega] \, 2\pi \delta(\omega - \omega_k) = b[\omega_k]
\ee
exactly as claimed.

The factor of $1/\sqrt{T}$ is a consequence of the discrete and continuous modes having different normalizations: $\comm{\db[\omega_j]}{\db^\dagger[\omega_k]} = \delta_{j,k}$ and $\comm{b[\omega]}{b^\dagger[\omega']} = 2\pi \delta(\omega - \omega')$.
This is the temporal analog of the matching for the discrete vs.~continuous (momentum) modes in Eq.~\eqref{eq:acomm_condis}, where here the total integration time $T$ plays the role of the quantization volume ${\cal V}$.
This also justifies why Eq.~\eqref{appeq:frequencyModes} holds for stationary states of the transmission-line modes.
Just as $N_\bk$ can be interpreted as the occupancy per unit phase-space volume, $N_{\omega}$ can be interpreted either as the occupancy of the discrete-frequency mode in that frequency bin or as the number flux per frequency at $\omega$, both of which are dimensionless.
%!TEX root = main.tex

%%%%%%%%%%%%%%%%%%%%%%%%%%%%%%%%%
\section{Measurement-Record Calculations}
%%%%%%%%%%%%%%%%%%%%%%%%%%%%%%%%%

%%%%%%%%%%%%%%%%%%%%%%%%%%%%%%%%%
\subsection{General Number Distribution for a Gaussian State}
\label{app:single_bin_PGF}
%%%%%%%%%%%%%%%%%%%%%%%%%%%%%%%%%

In Sec.~\ref{subsec:number_measurements} we considered the distribution for observing $n$ counts in a bin of length $\tau$ for the virialized Gaussian hypothesis.
In Eqs.~\eqref{eq:Geom_dist} and \eqref{eq:NB_dist} we showed the analytic results assuming $\tau \ll \tau_\tout$ and $\tau \gg \tau_\tout$, respectively, and we plotted the distribution for an arbitrary $\tau$ in Fig.~\ref{fig:p_n_comparison}.
Here we compute that general distribution.

We begin with the raw form of the distribution in Eq.~\eqref{eq:single_bin_prob}
\be
p(n) = \int d\bm{\alpha}\, P_a(\bm{\alpha})\, \text{Pois}\bigg(n, \kappa_m \int_0^\tau dt\, |\gamma(t)|^2\bigg).
\ee
Working in the discrete-mode notation, $\gamma(t)$ can be written
\be
\gamma(t) = \sum_\bk \chi_\bk \alpha_\bk e^{-i\omega_\bk t},
\ee
and the joint $P$-function for Gaussian DM is given in Eq.~\eqref{eq:VGH},
\be
P_a(\bm{\alpha}) = \prod_{\bk} \frac{1}{\pi N_{\bk}}\exp{-\frac{|\alpha_{\bk}|^2}{N_{\bk}}} = \frac{1}{\det(\pi C)}\exp{-\bm{\alpha}^\dag C^{-1}\bm{\alpha}},
\ee
where $C=\text{diag}(N_{\bk})$.
Expanding out the definition of $\gamma(t)$ in the integral of the Poisson distribution, we can rewrite the mean parameter as
\be \label{eq:lambda_M_def}
\lambda\equiv \kappa_m \int_0^\tau dt\, |\gamma(t)|^2 = \bm{\alpha}^\dag M \bm{\alpha}, \quad M_{\bk,\bq} \equiv \kappa_m \chi^*_\bk \chi_\bq \int_0^\tau dt\,e^{i(\omega_\bk -\omega_\bq )t} .
\ee

We can compute the probability generating function (PGF) of the full distribution as follows.
First, the PGF of the Poisson distribution is
\be
G_\text{Pois}(z) \equiv \sum_n \text{Pois}(n,\lambda) z^n = e^{\lambda(z-1)},
\ee
so that the PGF of the complete number distribution is given by 
\be
\begin{split}
G(z) &= \la e^{\lambda(z-1)}\ra_P \\
&= \int d\bm{\alpha} \, \frac{1}{\det(\pi C)}\exp{-\bm{\alpha}^\dag \big[ C^{-1} + (1-z)M\big]\bm{\alpha}} \\
&= \frac{1}{\det(\mathds 1+(1-z)CM)}\\
&= \frac{1}{\det(\mathds 1+(1-z)C^{1/2}MC^{1/2})}.
\end{split}
\ee
In the last equality we used that the determinants on each line are equal through a similarity relation (conjugation) with $C^{1/2}$ and $C^{-1/2}$.
We can then expand this determinant in terms of the eigenvalues $\mu_j$ of the matrix $H\equiv C^{1/2}MC^{1/2}$, so that
\be\label{eq:VGS_PGF}
G(z) = \prod_j \frac{1}{1+(1-z)\mu_j} .
\ee
This is the exact PGF for the number distribution $p(n)$, so we can recover the distribution once we have the eigenvalues $\mu_j$.
Also notice that $1 / [1+ (1-z)\mu]$ is exactly the PGF of a geometric distribution with mean $\mu$, so the above implies that $p(n)$ is exactly given by the sum of independent geometric random variables with means $\mu_j$.
(The PGF of a sum of independent variables is the product of each variable's PGF.)
The $\mu_j$ should not be confused as being proportional to the DM mode occupancies $N_\bk$, since that would only occur if $M\propto \mathds 1$, which from Eq.~\eqref{eq:lambda_M_def} it is not.
So, our problem now reduces to finding these eigenvalues of $H$.

To do this, first note that we can write the matrix elements of $H$ as
\be
H_{\bk,\bq} = \kappa_m \chi^*_\bk \chi_\bq \sqrt{N_\bk N_\bq }\int_0^\tau dt\, e^{i(\omega_\bk -\omega_\bq )t} 
\ee
or if we define the functions $v_\bk (t) = \sqrt{\kappa_m N_\bk }\, \chi_\bk e^{-i\omega_\bk t}$, then we can write these matrix elements as a matrix multiplication on the space $L^2[0,\tau]$, as in
\be
H_{\bk,\bq} = \int_0^\tau dt\,\, v^*_\bk (t)v_\bq (t)
\ee
or in shorthand, $H = V^\dag V$, with $V_{t\bk }\equiv v_\bk (t)$ as the matrix element of $V$ on the $\bk$ and $t$ space.
Crucially, since $V^\dag V$ has the same set of nonzero eigenvalues as $V V^\dag$, we can instead find our $\mu_j$ as the eigenvalues of the following kernel
\be
K(t-t') \equiv \sum_\bk v_\bk (t)v^*_\bk (t') = \kappa_m \sum_\bk N_\bk |\chi_\bk |^2 e^{-i\omega_\bk (t-t')} = \kappa_m\la \gamma^*(0)\gamma(t-t')\ra_P
\ee
which is the two-point correlation function of the steady-state output field from the cavity.
In particular, the eigenvalue equation now becomes finding an eigenfunction $f_j(t)$ with eigenvalues $\mu_j$ such that 
\be
\int_0^\tau d t'\, K(t-t') f_j(t') = \mu_j f_j(t).
\ee
These eigenvalues can be found numerically by discretizing the function $K(t-t')$ on a grid on $L^2[0,\tau]$.
From here we can compute the PGF, which can then be inverted to obtain $p(n)$.

It is also useful to understand how the general distribution approaches the negative binomial distribution as the length $\tau$ increases far beyond the coherence time.
To see this, first note that if the eigenvalues $\mu_j$ in Eq.~\eqref{eq:VGS_PGF} are identical, then $G(z)$ is exactly the PGF for the sum of independent, identically distributed geometric random variables, in other words, a negative binomial distribution.
Because the kernel $K(t) \propto \la \gamma^*(0) \gamma(t) \ra_P$ decays at large $t$, as $\tau$ increases, the function value $f_j(t)$ at large $t$ contributes less and less to the integral $\int_0^\tau d t'\, K(t-t') f_j(t')$.
This means that many eigenfunctions $f_j(t)$ with different large-$t$ behaviors can share similar eigenvalues.
This explains why, as $\tau$ increases, more eigenvalues $\mu_j$ approach each other, so that the true distribution converges to the negative binomial in the long-$\tau$ limit.

%%%%%%%%%%%%%%%%%%%%%%%%%%%%%%%%%
\subsection{Joint Distribution for the Quadrature Magnitude}
\label{app:DFT_PDF}
%%%%%%%%%%%%%%%%%%%%%%%%%%%%%%%%%

Here we explicitly evaluate the integrals marginalizing over $dr_i$ in Eq.~\eqref{eq:dft_mag_marginalization} by transforming to the frequency domain, with a view to expressing the result as an integral over the DM $P$-function.
The key step in the process is to apply Parseval's theorem to the exponent of Eq.~\eqref{eq:quadrature_measurement_record}.

Since $\{r_i\}$ is real, its DFT components defined in Eq.~\eqref{eq:DFT_def}, extended to negative $k$, satisfy $\tilde r_{-k} = \tilde r_k^*$.
Taking $N$ odd, the independent variables are $\tilde r_0 \in \mathbb{R}$ together with $\tilde r_k$ for $k = 1, \ldots, (N-1)/2$, accounting for $N$ real degrees of freedom.
We write $\tilde r_k = \rho_k e^{i\phi_k}$ for $k \geq 1$.
The same reality property holds for $X_\gamma(t)$, so $\tilde X_\gamma(-k) = \tilde X_\gamma(k)^*$.

For any real sequence $f(t_i)$, the discrete Parseval identity is
\be \label{eq:parseval}
\sum_{i=1}^N \Delta t\, f(t_i)^2
= \frac{1}{T}\!\!\sum_{k=-(N-1)/2}^{(N-1)/2}\!\!|\tilde f_k|^2
= \frac{1}{T}|\tilde f_0|^2
+ \frac{2}{T}\sum_{k=1}^{(N-1)/2}|\tilde f_k|^2 ,
\ee
where the second equality uses $\tilde f_{-k} = \tilde f_k^*$.
Applying this to $f(t_i) = r_i + \sqrt{\kappa_m}\,X_\gamma(t_i)$ converts the exponent of Eq.~\eqref{eq:quadrature_measurement_record} into a sum that factorizes over $k$:
\be \label{eq:exponent_frequency_domain}
\sum_{i=1}^N \Delta t\bigl[r_i + \sqrt{\kappa_m}\,X_\gamma(t_i)\bigr]^2
= \frac{1}{T}\bigl|\tilde r_0 + \sqrt{\kappa_m}\,\tilde X_\gamma(0)\bigr|^2
+ \frac{2}{T}\sum_{k=1}^{(N-1)/2}
\bigl|\tilde r_k + \sqrt{\kappa_m}\,\tilde X_\gamma(k)\bigr|^2 .
\ee

We now change variables in Eq.~\eqref{eq:dft_mag_marginalization} from $\{r_i\}$ to $(\tilde r_0, \{\rho_k\}, \{\phi_k\})$; the DC $k=0$ mode is marginalized out in what follows.
Under this change the delta functions become trivial, fixing the new $\rho_k$ to the magnitudes $|\tilde r_k|$, and the marginal density takes the form
\be \label{eq:marginal_new_vars}
p(\{\rho_k\})
= \int d\tilde r_0 \prod_{k=1}^{(N-1)/2} d\phi_k\;\,
J\bigl(\{\rho_k\}\bigr)\;
p\bigl(\{r_i(\tilde r_0, \rho_k, \phi_k)\}\bigr) ,
\ee
with $J(\{\rho_k\})$ the Jacobian.
The map $\{r_i\}\to\{\tilde r_0, \mathrm{Re}\,\tilde r_k, \mathrm{Im}\,\tilde r_k\}$ has matrix $D$ with $DD^T = (\Delta t)^2\,\mathrm{diag}(N, N/2, \ldots, N/2)$ by DFT orthogonality, so $|\det D| = (\Delta t)^N\,N^{N/2}\,2^{-(N-1)/2}$.
The polar change $(\mathrm{Re}\,\tilde r_k, \mathrm{Im}\,\tilde r_k)\to(\rho_k,\phi_k)$ contributes an additional factor $\rho_k$ per mode, giving
\be \label{eq:jacobian_dft_polar}
J(\{\rho_k\})
= \frac{2^{(N-1)/2}}{(\Delta t)^N\,N^{N/2}}
\prod_{k=1}^{(N-1)/2}\rho_k .
\ee

Substituting Eq.~\eqref{eq:quadrature_measurement_record} into Eq.~\eqref{eq:marginal_new_vars} with the exponent rewritten via Eq.~\eqref{eq:exponent_frequency_domain}, the $\tilde r_0$ integral is a real Gaussian: $\int d\tilde r_0\, \exp{-(\tilde r_0 + \sqrt{\kappa_m}\,\tilde X_\gamma(0))^2/T} = \sqrt{\pi T}$.
Combining this with the prefactor $(\Delta t/\pi)^{N/2}$ of Eq.~\eqref{eq:quadrature_measurement_record} and using $T = N\Delta t$, we arrive at
\be \label{eq:joint_mag_phase}
\begin{split}
p(\{\rho_k\})
= &\int d\bm{\alpha}\, P_{a}(\bm{\alpha})
\int \prod_{k=1}^{(N-1)/2} d\phi_k\,
\frac{2\rho_k}{\pi T} \\
& \times 
\exp{
-\frac{2}{T}\Bigl[
\rho_k^2 + \kappa_m\,|\tilde X_\gamma(k)|^2
+ 2\sqrt{\kappa_m}\,\rho_k\,|\tilde X_\gamma(k)|\cos(\phi_k - \arg \tilde X_\gamma(k))
\Bigr]} .
\end{split}
\ee
Performing the $\phi_k$ integrals using the Bessel function identity $\int_0^{2\pi}\! d\phi\, e^{z\cos\phi} = 2\pi I_0(z)$, we find
\be \label{eq:phase_integral}
\int_0^{2\pi}\! d\phi_k\,
\exp{
-\frac{4\sqrt{\kappa_m}\,\rho_k\,|\tilde X_\gamma(k)|}{T}
\cos(\phi_k - \arg \tilde X_\gamma(k))}
= 2\pi\, I_0\!\left(
\frac{4\sqrt{\kappa_m}\,\rho_k\,|\tilde X_\gamma(k)|}{T}
\right) .
\ee
Substituting this into Eq.~\eqref{eq:joint_mag_phase}, the integrand reduces in each mode to the Rician distribution $R\bigl(\rho_k,\sqrt{\kappa_m}\,|\tilde X_\gamma(k)|,T/4\bigr)$, defined in Eq.~\eqref{eq:rician_def}.
This establishes Eq.~\eqref{eq:joint_dft_mag_dist}.

%%%%%%%%%%%%%%%%%%%%%%%%%%%%%%%%%
\subsection{Gaussian vs.~Squeezed State Likelihood Comparison}
\label{app:quad_likelihood_test}
%%%%%%%%%%%%%%%%%%%%%%%%%%%%%%%%%

Here we compute the expected value of the test statistic for distinguishing between the Gaussian and squeezed-state hypotheses for the DM state, given that the true state is squeezed.
This is the example considered at the end of Sec.~\ref{subsec:quad_meas_stats}.
Specifically, we assume each DM mode is in a squeezed-vacuum state with squeezing parameter $r_{\bk}$ obeying $N_{\bk} = \sinh^2r_{\bk}$.
In this case the density matrix is
\be
\rho_S = \prod_{\bk } S(\bk)\ket{0_{\bk}}\bra{0_{\bk}}S^\dag(\bk) ,
\ee
where $S(\bk)$ is the squeezing operator
\be
S(\bk) = \exp{-\frac{1}{2}r_{\bk}(a_{\bk}^2 - (a^\dag_{\bk})^2)} .
\ee
We choose $r_{\bk}$ to be real, but the same analysis follows if we add phases $e^{i\theta_{\bk}}$ as long as the direction of squeezing in each mode happens to align at some point during our measurement window.
If the phases are instead random, then the effective mode of the squeezed state appears Gaussian in any case.

Both of these possibilities for the DM state are mean-zero generalized Gaussian states, so we can use a simplified version of the expected value of the test statistic:
\be
\la \Theta \ra_S = \tr{C_G^{-1}C_S - \log(C_G^{-1}C_S)-\mathds{1}},
\ee
where $C_G,\, C_S$ are the covariance matrices for the data in the Gaussian and squeezed cases, respectively.
If we define the correlation function $G(\tau)$ as
\be
G(\tau) = \kappa_m\int \frac{d^3\bk}{(2\pi)^3}|\chi_{\bk}|^2N_{\bk}\cos(\omega_{\bk}\tau),
\ee
whose envelope decays on a timescale on the order of a coherence time $\tau_\tout$, then we can write each of the covariance matrices in our trace for $N_{\mathbf{k}} \gg 1$ as 
\be
C_G(t,t') = \frac{1}{2}\delta(t-t') + G(t-t') , \qquad C_S(t,t') = \frac{1}{2}\delta(t-t') + G(t-t') + G(t+t').
\ee
Notice that the time-domain matrix $G(t-t')$ has a banded diagonal structure; that is, it attains its maximum when $t=t'$, and begins to decay from that value once $|t-t'| \gtrsim \tau_\tout$.
The matrix $G(t+t')$, on the other hand, decays once \textit{either} $t$ or $t'$ exceeds $\tau_\tout$, meaning its matrix elements are only appreciable in the upper-left corner.

We can then evaluate the trace by coarse-graining this continuum time-domain structure into $N=T/\tau_\tout$ blocks of width $\tau_\tout$, so that both $C_G,\, C_S$ can be approximated as diagonal matrices.
In particular, if we reintroduce the quantity $\mu(\tau_\tout) = G(0)\tau_\tout$ as the mean number of photons deposited in a single coherence time, we can write each of these matrices as
\be
C_G = \frac{\mathds{1}}{2}+ \mu(\tau_\tout)\mathds{1}, \qquad C_S = \frac{\mathds{1}}{2}+ \mu(\tau_\tout)\mathds{1} + \mu(\tau_\tout)P_1 ,
\ee
where $P_1$ is the projector onto the first coarse-grained subspace of length $\tau_\tout$, i.e.~the upper-left matrix element.
With this diagonal approximation, the logarithm in the trace becomes trivial, yielding the result in Eq.~\eqref{eq:squeezed_test_stat}.

%%%%%%%%%%%%%%%%%%%%%%%%%%%%%%%%%
\subsection{The Mutual Information of Virialized Gaussian Effective Modes}
\label{app:mutual_info}
%%%%%%%%%%%%%%%%%%%%%%%%%%%%%%%%%

Here we justify the statements around Eqs.~\eqref{eq:GvsC_number_dists} and \eqref{eq:factorized_eff_prob} regarding the statistical independence of adjacent measurements in time for the virialized Gaussian state.
The tool we use for this purpose is the mutual information, defined for two random variables $x,y$ described by a joint distribution $p(x,y)$ as 
\be
I(x,y) = \int dx dy \, p(x,y)\log(\frac{p(x,y)}{p(x)\,p(y)}),
\ee
where $p(x)$ and $p(y)$ are the corresponding marginalized distributions for the single random variables.
The mutual information $I(x,y)$ measures the total correlation between $x$ and $y$, and crucially $I(x,y)=0$ if and only if $x,y$ are statistically independent, that is $p(x,y)=p(x)\,p(y)$.

To justify temporal independence, the probability distribution we use is the reduced $P$-function for the effective mode at different times, which is valid as this distribution for the virialized Gaussian is always positive, and therefore may be interpreted as a genuine probability distribution over a coherent state taken as a random variable.
In terms of the fundamental modes $\alpha_\bk$, recall from Eq.~\eqref{eq:gamma_def} that the effective mode can be written (in discrete-mode notation) as
\be
\gamma(t) = \sum_\bk\chi_\bk\alpha_\bk e^{-i\omega_\bk t},
\ee
so that we can then define the reduced $P$-function of the effective mode at different times through 
\be
P_\eff(\gamma_1, \gamma_2) = \int d\bm{\alpha}\, P_a(\bm{\alpha})\, \delta\big(\gamma_1 - \gamma(t_1)\big)\,\delta\big(\gamma_2 - \gamma(t_2)\big).
\ee

To calculate $I(\gamma_1, \gamma_2)$, we first need the full distribution $P_\eff(\gamma_1, \gamma_2)$.
This is easily obtained; it is a marginalization of a multivariate Gaussian distribution for $P_a$ onto a linear combination of its random variables, and therefore it is a Gaussian itself.
Accordingly, it suffices to calculate the mean and covariance of the distribution.
Both $\la \gamma_i\ra_P $ and $\la \gamma_i \gamma_j\ra_P $ vanish, from the properties $\la \alpha_\bk\ra_P = \la \alpha_\bk \alpha_\bq\ra_P =0 $ inherited from the virialized Gaussian. 
The only non-vanishing moments we need are given by the matrix
\be
G_{ij} = \la \gamma_i \gamma_j^* \ra_P = \sum_{\bk,\bq} \chi_\bk \chi^*_\bq \la \alpha_\bk \alpha_\bq^* \ra_P e^{-i(\omega_\bk t_i - \omega_\bq t_j)} = \sum_\bk|\chi_\bk|^2 N_\bk e^{-i\omega_\bk(t_i-t_j)}.
\ee
This is nothing but the Glauber correlation function $G^{(1)}(t_i-t_j)$ of the effective mode.
We can then write the marginalized $P$-function for the effective mode at these two times as
\be
P_\eff(\gamma_1, \gamma_2) = \frac{1}{\det(\pi G)}\exp{-\bm{\gamma}^\dag G^{-1}\bm{\gamma}}
\ee
where we have introduced $\bm{\gamma} = \{\gamma_1,\gamma_2\}$ to implement matrix expressions.
We then invoke the result that the mutual information between two jointly Gaussian variables, jointly circular and zero-mean, only depends on their normalized covariance
\be
g = \frac{G_{12}}{\sqrt{G_{11}G_{22}}}
\ee
through the expression
\be
I(\gamma_1, \gamma_2) = -\log(1-|g|^2).
\ee
The object $g$ is just the normalized Glauber correlation function $g^{(1)}(t_1-t_2)$.
And since we have already established that $G^{(1)}$ decays on the timescale $\tau_\tout$, then so does the normalized $g^{(1)}$, after which the mutual information between the two is approximately zero.
This establishes the claims of approximate independence after a coherence time used in the main text. 

\bibliographystyle{utphys3}
{\scriptsize
\bibliography{refs}}

\end{document}